\documentclass[12pt, a4paper]{article}

\usepackage[utf8]{inputenc}
\usepackage[T1]{fontenc}
\usepackage{lmodern}
\usepackage[margin=2.5cm]{geometry}
\usepackage{amsmath, amssymb, amsthm}
\usepackage{mathtools}
\usepackage{bbm}                      
\usepackage{dsfont}                   
\usepackage{hyperref}
\usepackage{cleveref}                 
\usepackage{enumitem}
\usepackage{microtype}
\usepackage{setspace}
\usepackage{authblk}                  
\usepackage{algorithm}
\usepackage{algpseudocode}
\usepackage{xcolor}
\usepackage{todonotes}                
\usepackage[numbers, sort&compress]{natbib}

\hypersetup{
	colorlinks = true,
	linkcolor  = blue!60!black,
	citecolor  = green!50!black,
	urlcolor   = blue!70!black
}

\theoremstyle{plain}
\newtheorem{theorem}{Theorem}[section]

\newtheorem{definition}[theorem]{Definition}

\newtheorem{algorithm-env}[theorem]{Algorithm}

\theoremstyle{remark}

\begin{document}
	\bibliographystyle{myunsrt}
	
	\title{Perfect sampling of Ising spin glasses with bounding chains}
	
	\author[1,2]{Lukas Péron}
	\author[1]{Louise Budzynski}
	\affil[1]{DI ENS, \'{E}cole Normale Supérieure, PSL Research University, CNRS, INRIA, Paris, France}
	\affil[2]{Département de physique, \'{E}cole Normale Supérieure, PSL Research University, Paris, France}
	
	\date{}
	
	\maketitle
	
	\begin{abstract}
		We study perfect sampling algorithms for Ising spin glasses based on coupling from the past and bounding chains, on several graph families and coupling distributions.
		For the single-site bounding chain, we establish coalescence in $O(n\log n)$ updates under Dobrushin conditions.
		On sparse random graphs, our numerical results suggest that efficient sampling extends throughout the tree-uniqueness regime, with a slowdown beyond this threshold.
		On Erd\H{o}s-Rényi random graphs $\mathcal{G}(n,d/n)$ with standard Gaussian couplings, we introduce a block bounding chain and prove its coalescence in $O(n\log n)$ block updates in the tree-uniqueness regime for sufficiently large $d$, each block update requiring $O(n)$ operations, adapting a proof in~\cite{Efthymiou_Zampetakis_2024} for the rapid mixing of block Glauber dynamics.
		On periodic lattices, our experiments instead suggest that the single-site bounding chain is limited by the looseness of its bounding envelope, whereas on the Sherrington-Kirkpatrick model, slow coalescence is accompanied by damage spreading at every inverse temperature tested.
		These observations motivate alternative bounding representations and couplings as strategies to improve perfect sampling on spin glasses.
	\end{abstract}


\section{Introduction}
\label{sec:introduction}
Spin glasses are models originating in statistical physics that describe disordered systems made of a large number of interacting variables subject to quenched random interactions.
Their properties have been widely studied in statistical physics~\cite{Edwards_Anderson_1975,Parisi_79,Mezard_Parisi_Virasoro_87} and mathematics~\cite{Guerra_Toninelli_04,Stein_Newman_2013}.
Spin glasses models and methods have found applications in several areas at the interface of mathematics and theoretical computer science, including machine learning~\cite{Gabrie_2023}, Bayesian inference (for instance with the Stochastic Block Model~\cite{Decelle_2011}), random Constraint Satisfaction Problems~\cite{Zdeborova_Krzakala_2007} and combinatorial optimization~\cite{Dembo_Montanari_Subhabrata_2017}.
We focus on the particular case of Ising models with disordered interactions. On a graph $G=(V,E)$, such a model is associated with a Gibbs measure
$$
	\mu_{G,\boldsymbol{J},\beta}(\sigma)=\frac{1}{Z_\beta(G,\boldsymbol{J})}\exp\left(\beta\sum_{\{u,v\}\in E}J_{\{u,v\}}\mathbbm{1}_{\{\sigma(u)=\sigma(v)\}}\right) \quad , \quad \sigma\in\{+1,-1\}^V
$$
where $\boldsymbol{J}=\{J_{\{u,v\}}\}_{\{u,v\}\in E}$ denotes a set of i.i.d. couplings drawn from a given probability distribution.
Two fundamental and related questions concerning these models are to characterize their equilibrium properties and to determine whether one can efficiently sample from their Gibbs distribution.

At sufficiently high temperature, the infinite-volume Gibbs measure is unique. Since uniqueness requires the decay of correlations uniformly over boundary conditions, it is expected to imply the rapid mixing of local dynamics such as Glauber dynamics.
Indeed, recent results establish rapid mixing throughout the tree-uniqueness regime for several classes of spin systems~\cite{Mossel_Sly_2013,Efthymiou_Charilaos_2019, Chen_2022, Chen_Liu_Vigoda_2023, Efthymiou_Zampetakis_2024, Chen_at_uniq_2025, Bencs2026}.
Here, mixing time is defined in the worst-case sense, as the maximum over all initial conditions.
It is tempting to associate the loss of uniqueness with the appearance of small metastable regions that may trap dynamics started from unfavorable initial conditions, while remaining invisible from a typical initialization.
However, recent results prove rapid mixing beyond the tree-uniqueness threshold on random regular graphs~\cite{Chen_beyond_uniq_2025,Gobel_2026} as well as for sparse random Ising spin glasses~\cite{Kuikui_2024,efthymiou2026samplingdilutedspinglassesunbounded}, suggesting that tree-uniqueness is not the actual threshold for worst-case mixing on typical instances.

Coupling From The Past (CFTP)~\cite{Propp_Wilson_1996} is a perfect sampling technique for distributions on large configuration spaces, with application to Ising, random-cluster and dimer models among others.
The algorithm evolves copies of a Markov chain starting from every initial configuration in the state space at time $-T$, to time $t=0$, using the same random updates for all copies.
If all trajectories have coalesced by time $0$, their common state is an exact sample from the stationary distribution.
Otherwise, the simulation is restarted further into the past, preserving the random updates already generated.
For monotone dynamics, such as the Glauber heat-bath dynamics for ferromagnetic Ising models, it is sufficient to track the trajectories starting from the minimal and maximal configurations, as their coalescence certifies the coalescence of all trajectories.
Without such a monotone property, alternative strategies include bounding chains~\cite{Haggstrom_Nelander_1999,Bubley_Dyer_1997} and the patch-based method~\cite{Chanal_Krauth_2008,Chanal_Krauth_2010,Bernard_Krauth_2010,Hukushima_Krauth_2025} developed for spin-glasses and hard spheres. 
A bounding chain maintains an envelope including all configurations reachable by the trajectories. Its collapse to a singleton therefore certifies coalescence. However, this envelope need not coincide with the set of reachable configurations, so that all trajectories may have coalesced while the bounding chain still fails to detect it.

\subsection{Results}

In this work, we study the performance of a bounding chain for Ising models with disordered interactions~\cite{Huber_1998,Haggstrom_Nelander_1999,Childs_Patterson_MacKay_2001,Huber_2004}.
We consider the Gibbs distribution~(\ref{eq:Gibbs_proba}) on Erd\H{o}s-Rényi random graphs $\mathcal{G}(n,d/n)$ and random $d$-regular graphs, and on periodic $d$-dimensional lattices. 
The edge couplings $\{J_{\{u,v\}}\}_{\{u,v\}\in E}$ are i.i.d., either standard Gaussians or sampled uniformly on $\{+1,-1\}$. 
We also consider the complete graph $K_n$ with i.i.d. couplings in $\mathcal N(0,1/n)$, corresponding to the Sherrington-Kirkpatrick model~\cite{Sherrington_Kirkpatrick_1975}.

In Section~\ref{sec:numerics}, we investigate numerically the coalescence time of the bounding chain. 
On sparse random graphs, we observe rapid coalescence below the tree-uniqueness threshold, and a pronounced slowdown above it, suggesting that the uniqueness transition of the Gibbs measure on the associated infinite tree governs the dynamics of the bounding chain and defines its algorithmic threshold.
On lattices, however, the bounding chain threshold does not appear to coincide with a known equilibrium transition.  
Moreover, the bounding chain performs less effectively than the local-patch algorithm~\cite{Chanal_Krauth_2008,Chanal_Krauth_2010}: its coalescence slows down at a lower inverse temperature than that of the local-patch algorithm, whose dynamical threshold has been conjectured to be associated with damage spreading~\cite{Bernard_Krauth_2010,Hukushima_Krauth_2025}. 
On the Sherrington-Kirkpatrick model, we observe a super-polynomial growth of the coalescence time at all inverse temperature tested.

We complement these results with a numerical study of damage spreading in Section~\ref{subsec:numerics_damage_spreading}. 
On sparse random graphs, we observe the onset of damage spreading right above the tree-uniqueness threshold, suggesting that coalescence is obstructed by the fact that coupled trajectories are drifting apart, which explains the slowdown of both the bounding chain and CFTP, under the coupling considered here.
On periodic $d$-dimensional lattices, we recover the damage spreading thresholds reported in~\cite{Bernard_Krauth_2010,Hukushima_Krauth_2025}. 
While damage spreading appears to limit the performance of the local-patch algorithm, the bounding chain slows down at a smaller inverse temperature.
This observation suggests that the bounding chain slowdown results from the fact that the envelope of the coupled trajectories is not tight, rather than from the fact that trajectories are drifting apart under the coupling considered.
On the Sherrington-Kirkpatrick model, damage spreading occurs at all inverse-temperature tested, suggesting that damage spreading persists for all $\beta>0$, as already observed in~\cite{Derrida1989}, which provides an explanation for the poor performance of the bounding chain (and for any CFTP algorithm under the coupling considered here) on this model.

In Section~\ref{sec:rapid_coalescence_Dobrushin}, we prove rapid coalescence (in $O(n\log n)$ single-site updates, with $n=|V|$ the number of Ising variables) of the bounding chain under the Dobrushin condition.
The proof controls the expected number of unresolved spins through a contraction argument.
Dobrushin condition provide a sufficient criterion for uniqueness, and therefore provides a rigorous regime of efficient approximate sampling (as it implies rapid mixing of the single site Glauber dynamics~\cite{Weitz_2005}), although it generally does not cover the entire uniqueness regime.

In Section~\ref{sec:rapid_coal_block_BC}, we define a block bounding chain for Ising spin glasses on $\mathcal{G}(n,d/n)$ with i.i.d. standard Gaussian couplings, and establish its rapid coalescence in the tree-uniqueness regime for sufficiently large $d$.
Our construction uses the block partition of Efthymiou and Zampetakis~\cite{Efthymiou_Zampetakis_2024}, which exists with high probability and places vertices with large aggregate influence deep inside their blocks.
They prove rapid mixing of the corresponding block Glauber dynamics with a path coupling argument, and then transfer this bound to single-site Glauber dynamics through a spectral comparison argument.
Our contribution is to establish contraction of a potential for the block bounding chain, derived from the weighted distance introduced in~\cite{Efthymiou_Zampetakis_2024}.
This yields a coalescence time of $O(N\log n)$ block updates, where $N\leq n$ is the number of blocks.
Their spectral comparison step however exploits the reversibility of Glauber dynamics, and therefore cannot be transferred to obtain the rapid coalescence of the single-site bounding chain. 

In~\cite{Efthymiou_Zampetakis_2024}, Efthymiou and Zampetakis prove that, under the assumptions on $d$ and $\beta$ stated below, with probability at least $1-n^{-1/4}$, the block-partition exists over $G,\boldsymbol{J}$, and couplings satisfy $|J_{e}|\geq n^{-7/3}$ for each $e\in E$.
Our next theorem establishes rapid coalescence of the block bounding chain constructed from such a partition.

\begin{theorem}
	\label{thm:coal_block_BC}
	For every $\varepsilon\in(0,1)$, there exists $d_0=d_0(\varepsilon)\geq 1$, such that for all $d\geq d_0$, for $0<\beta\leq (1-\varepsilon)\beta_c(d)$ with $\beta_c(d)=\sqrt{2\pi}/d$, the following holds, for all $n\geq n_0(\varepsilon, d,\beta)$.
	Let $G=(V,E)$ be an instance of $\mathcal{G}(n,d/n)$ and $\boldsymbol{J}=\{J_e\}_{e\in E}$ be i.i.d. standard Gaussians. 
	With probability at least $1-n^{-1/4}$ over $G,\boldsymbol{J}$, the triplet $(G,\boldsymbol{J},\beta)$ admits a $(d,\varepsilon)$-block partition $\mathcal{B}$ with $N\leq n$ blocks.
	The block bounding chain $(Y_t)_{t\geq 0}$ defined in Section~\ref{sec:rapid_coal_block_BC}, built on $\mathcal{B}$ and started from $Y_0=\star^V$, satisfies for every $k>0$,
	\begin{align}
		\mathbb{P}\left[T_{\rm coal}>\frac{7N}{\varepsilon}((6+k)\log n + \log 2)\right]\leq n^{-k}
	\end{align}
	where $T_{\rm coal}=\inf\{t: Y_t\in\{+1,-1\}^V\}$ is the coalescence time of the bounding chain, and the probability is over the randomness of the chain. 
	In particular $T_{\rm coal}=O(N\log n)=O(n\log n)$, and the CFTP implementation of the block bounding chain returns an exact sample from $\mu_{G,\boldsymbol{J},\beta}$ in $O(n \log n)$ block updates with high probability. Each block update can be computed in $O(n)$ operations.
\end{theorem} 
Note that in the large $d$ limit, the tree-uniqueness regime $\mathbb{E}[\tanh(\beta|J|/2)]<1/d$ corresponds to having $\beta<\beta_c(d)=\sqrt{2\pi}/d$.

Recent rapid-mixing results~\cite{Chen_2022, Chen_Liu_Vigoda_2023,Chen_at_uniq_2025,Chen_beyond_uniq_2025,Gobel_2026,Kuikui_2024} establish rapid mixing through spectral arguments that exploit reversibility, whereas path coupling is used in~\cite{Efthymiou_Charilaos_2019} and in the block dynamics analysis of~\cite{Efthymiou_Zampetakis_2024}.
Spectral methods control convergence to equilibrium, but do not directly control coalescence of a grand coupling approach used in CFTP or in bounding chains. 
Our results demonstrate that the coupling estimates used in the block dynamics proof~\cite{Efthymiou_Zampetakis_2024} can be adapted to establish contraction of a bounding chain potential, and in turn to prove rapid coalescence.
This motivates developing coupling-based proofs for other models, especially on sparse graphs, and investigating whether they can be adapted to bounding chains.

\subsection{Related works}
Bounding chains were introduced in~\cite{Huber_1998,Haggstrom_Nelander_1999} as a general framework for implementing CFTP without explicitly tracking all initial configurations.
Applications include graph colorings~\cite{Huber_1998}, the hard-core model and antiferromagnetic Potts model~\cite{Huber_2004}, and the ferromagnetic Potts model with a bounding chain for the Swendsen-Wang dynamics~\cite{Huber_BC_Swendsen_Wang}.
More recently, an improved bounding chain for graph coloring was shown to achieve efficient perfect sampling for $q>(2.5+o(1))\Delta$, matching the lower bound for a class of CFTP algorithms~\cite{Ding_2026}.
Note that this critical threshold is stronger than Dobrushin condition for $q$-coloring, thus giving an example where CFTP algorithms require stronger condition than rapid mixing of the underlying dynamics.
For $2$-spin systems, the summary-state construction in~\cite{Childs_Patterson_MacKay_2001} coincides with the single-site bounding update used in Section~\ref{subsubsec:single_site_BC}. Numerical experiments on the triangular lattice anti-ferromagnetic Ising model reveal that the coalescence threshold depends on the implementation of the summary states, and does not seems to be related with any thermodynamic transition. This observation provides a precedent for the distinction between bounding chain coalescence and thermodynamic phase transitions, that we observe on lattices in Section~\ref{sec:numerics}.
In~\cite{Feng_Guo_Wang_2023} monotone CFTP is applied to the weighted random cluster model and yields a perfect sampling algorithm for the ferromagnetic Ising model with consistent external fields.

In~\cite{Chanal_Krauth_2008,Chanal_Krauth_2010}, a local patch implementation of CFTP is introduced and studied for Ising spin glasses in two-dimensional and three-dimensional lattices, and for hard spheres.
Similarly to bounding chains, the algorithm maintains a super-set containing all configurations attainable under the coupled dynamics of CFTP, but with a different strategy of recording admissible configurations of overlapping patches, whose size is increased during the simulation.
In~\cite{Bernard_Krauth_2010,Hukushima_Krauth_2025}, the algorithmic threshold of this local patch algorithm is found to depend strongly on the coupling of the Markov chains, and is related with damage spreading, i.e. the drifting apart of coupled trajectories starting from nearby configurations.
In principle, the local-patch approach could also be adapted to sparse graphs, although its computational efficiency in this setting remains to be investigated. 

Apart from CFTP, there have been a number of techniques to design perfect samplers.
In~\cite{Gobel2024PerfectSampling}, a generic reduction from approximate sampling to perfect sampling is established, with applications including graph coloring, matching, and anti-ferromagnetic $2$-spin systems.
In particular, for a reversible Markov chain with absolute spectral gap $\gamma$, the resulting perfect sampler has expected running time $O\left(\frac{1}{\gamma}\log \frac{|\Omega|}{|\pi_*|}\right)$ (with $\pi_*=\min_{x\in\Omega}\pi(x))$. 
For anti-ferromagnetic $2$-spin systems, the expected running time is $O(n^2)$ on graphs of bounded maximum degree $\Delta$, in the up-to-$\Delta$ unique regime.
Compared to our work, this construction does not use the CFTP strategy.
Although Ising spin glasses on $\mathcal G(n,d/n)$ are not studied in~\cite{Gobel2024PerfectSampling}, the generic reduction could also be applied to the block Glauber dynamics introduced in~\cite{Efthymiou_Zampetakis_2024}, giving in principle a perfect sampler with $O(Nn\log n)$ expected running time.

The perfect sampler constructed in~\cite{Feng_Guo_Yin_2022} yields an expected $O(n)$ running time, for spin systems with bounded maximum degree satisfying strong mixing with a rate faster than the neighborhood growth of the underlying graph.
This result is applied to spin systems with sub-exponential neighborhood growth, such as lattices graphs, to obtain a perfect sampler whenever the Gibbs sampler mixes in $O(n\log n)$, or the system shows strong
spatial mixing.
A recursive sampling algorithm in~\cite{Anand_Jerrum_2022} produces exact samples in linear time under analogous assumptions of sub-exponential growth and strong spatial mixing, with applications on the hardcore model, the $q$-coloring problem and the ferromagnetic model on the $2$-dimensional square lattice.
For the Edwards-Anderson (spin glass) model on $2$-dimensional square lattice, a different exact sampling algorithm~\cite{Thomas_Middleton_2009,Thomas_Middleton_2013} exploits a correspondence between regular lattices and dimer covering. It applies at any temperature, without requiring strong spatial mixing, and works in $O(n^{3/2})$, although the precision required for accurate sampling increases at low temperature.

\section{Definitions}
\label{sec:definitions}
\subsection{Model}
\label{subsec:model}
Let $G=(V,E)$ be a finite, simple, undirected graph, and let $\Omega=\{-1,+1\}^{V}$ be the set of spin configuration on $G$. Let $n=|V|$ be the number of vertices.
We associate with every edge $\{u,v\}\in E$ a coupling $J_{\{u,v\}}\in\mathbb{R}$ and write $\boldsymbol{J} = \left(J_{\{u,v\}}\right)_{\{u,v\}\in E}$.
For an inverse temperature $\beta>0$, the Gibbs distribution $\mu=\mu_{G,\boldsymbol{J},\beta}$ on $\Omega$ is defined by
\begin{align} 
\label{eq:Gibbs_proba}
\mu(\sigma) = \frac{1}{Z_{\beta}(G,\boldsymbol{J})} \exp\left( \beta \sum_{\{u,v\}\in E} J_{\{u,v\}}\, \mathbbm{1}_{\{\sigma(u)=\sigma(v)\}} \right), \qquad \sigma\in\Omega,
\end{align}
where
\begin{align*}
Z_\beta(G,\boldsymbol{J}) = \sum_{\sigma\in\Omega} \exp\left( \beta\sum_{\{u,v\}\in E} J_{\{u,v\}}\mathbbm{1}_{\{\sigma(u)=\sigma(v)\}} \right)
\end{align*}
is the partition function.
Notice that
$\mathbbm{1}_{\{\sigma(u)=\sigma(v)\}} =(1+\sigma(u)\sigma(v))/2$,
so that, up to a multiplicative constant, the above parametrization corresponds to the Ising interaction $(J_{uv}/2)\sigma(u)\sigma(v)$. 

\paragraph{\it Edge-influence.} 
For every edge $\{u,v\}\in E$, define its influence by
\begin{align}
\Gamma_{\{u,v\}} &= \frac{\left|1-e^{\beta J_{\{u,v\}}}\right|} {1+e^{\beta J_{\{u,v\}}}}  = \tanh\left(\frac{\beta|J_{\{u,v\}}|}{2}\right).
\end{align}
The edge-influence admits the following probabilistic interpretation. 
Define $N(v)$ as the set of neighbors of $v$.
Let $u,v$ be two adjacent vertices, and consider two configurations $\eta,\eta'\in\{+1,-1\}^{N(v)}$ disagreeing only on $u$. 
We have that
\begin{align}
	\label{eq:influence_ineq}
	\lVert \mu_v^\eta - \mu_v^{\eta'}\rVert_{\rm TV}\leq \Gamma_{\{u,v\}}
\end{align}
where $\mu_v^\eta$ is the marginal of $\mu$ at $v$ conditional on the configuration $\eta$: 
\begin{align}
\mu_v^\eta(x) = \mu\left( \sigma(v)=x \,\middle|\, \sigma(N(v))=\eta \right).
\end{align}

\paragraph{\it Aggregate-influence.} 
For every vertex $u\in V$, let $N(u)=\{v\in V: \{u,v\}\in E\}$ denote its neighborhood. 
The aggregate influence at $u$ is defined by
\begin{align}
A(u)=\sum_{v\in N(u)}\Gamma_{\{u,v\}}.
\end{align}

\subsubsection{Dobrushin criterion}
For $u,v\in V$, define the influence of $u$ on $v$ by:
\begin{align}
	\rho_{u\to v} &= \max_{\eta, \eta'\in\{+1,-1\}^{N(v)}}\left\{\lVert \mu_v^\eta - \mu_v^{\eta'}\rVert_{\rm TV} : \text{ $\eta, \eta'$ disagreeing only on $u$}\right\}
\end{align}
(with $\rho_{u\to v}=0$ whenever $u\notin N(v)$).
The Dobrushin coefficient is
\begin{align}
\label{eq:Dobrushin_coefficient}
\alpha = \max_{v\in V} \left\{\sum_{u\in N(v)} \rho_{u\to v}\right\}
\end{align}
The Dobrushin criterion $\alpha<1$ implies that the single-site random-update Glauber dynamics is rapidly mixing~\cite{Weitz_2005}.
More precisely, the mixing time is
\begin{align}
t_{\rm mix}(\delta) = O\left(\frac{n}{1-\alpha}
\log\left(\frac{n}{\delta}\right)\right)
\end{align}
In the corresponding infinite-volume setting, the Dobrushin criterion also implies the uniqueness of the Gibbs measure $\mu_{G,\boldsymbol{J},\beta}$~\cite{Dobrushin_1970}. 
From the above inequality~\eqref{eq:influence_ineq}, the condition on the aggregate influences $\max_{v\in V}A(v)<1$ is sufficient for the Dobrushin condition $\alpha<1$ to hold.
For ferromagnetic couplings with $J_{e}=1 \ \forall e\in E$ and spin glass couplings i.i.d. uniform in $\{+1,-1\}$, one has $A(v)=\deg(v)\tanh(\beta/2)$ for all $v\in V$, and the condition $\max_{v\in V}A(v)<1$ gives 
\begin{align}
	\label{eq:Dobruhsin_threshold}
	\begin{cases}
		\max_{v\in V}\{\deg(v)\} \tanh(\beta/2)<1 \quad &\text{on $\mathcal{G}(n,d/n)$} \\
		d \tanh(\beta/2)<1 \quad &\text{for random $d$ regular graphs} \\
		2d\tanh(\beta/2)<1 \quad &\text{for $d$-dimensional lattices}
	\end{cases}
\end{align}
The Dobrushin criterion is sufficient for uniqueness, but often only yields a lower bound on the uniqueness threshold $\beta_{uni}$.

\subsubsection{Uniqueness and spin-glass thresholds}
\label{subsubsec:uniqueness_and_SG}
As the inverse temperature $\beta$ increases, spin-glass models can undergo a {\it spin-glass phase transition} at $\beta_{\rm sg}$ from a high-temperature paramagnetic phase to a low temperature spin-glass phase~\cite{Sherrington_Kirkpatrick_1975,MezardParisi01,BaityJesi_2013}.
This transition is characterized by the onset of long-range spin-glass correlations between variables, and is associated with a non-trivial decomposition of the Gibbs measure into pure states~\cite{Parisi_Mezard_Virasoro_Sourlas_1984}.
This transition is associated with a critical slowing down of local dynamics, preventing the rapid equilibration of Monte Carlo Markov Chains aiming at sampling from the Gibbs measure~\cite{MontanariSemerjian06b,krzakala2007gibbs}.
For locally tree-like, sparse models, the spin-glass transition is closely related to the reconstruction problem on the corresponding infinite tree~\cite{MezardMontanari06}.

At sufficiently high temperature, the infinite-volume Gibbs measure is unique. More precisely, the {\it uniqueness regime} $\beta<\beta_{\rm uni}$ is the range of inverse temperature for which the Dobrushin-Lanford-Ruelle equations admit a unique Gibbs measure~\cite{Georgii_2000}.
Equivalently, local marginals on a vertex $i\in V$ become asymptotically insensitive to the choice of boundary condition, as the boundary is taken infinitely far away from $i$.
This is a worst-case correlation decay condition, since the insensitivity is required to hold for arbitrary boundary conditions~\cite{krzakala2007gibbs}.
While uniqueness implies extremality of the free Gibbs measure, the converse need not hold: there may exist several Gibbs measures while the free Gibbs measure remains extremal, leading to $\beta_{\rm uni}\leq\beta_{\rm sg}$.
For models on sparse random graphs (e.g. random $d$-regular graphs and $\mathcal{G}(n,d/n)$), we refer to the \emph{tree-uniqueness threshold} as the uniqueness threshold of the Gibbs measure on the infinite tree in the local weak limit of the graph ensemble~\cite{Lyons_1989}, occurring at 
\begin{align}
\label{eq:tree_uniqueness}
\operatorname{br}(T)\times\mathbb{E}\left[\tanh\left(\frac{\beta}{2}|J|\right)\right]=1
\end{align}
where $\operatorname{br}(T)$ is the branching number of the associated infinite tree, and the average is taken over the coupling distribution.
By contrast, the spin-glass (or reconstruction) transition is characterized by~\cite{MezardMontanari06} 
\begin{align}
\operatorname{br}(T)\times\mathbb{E}[\tanh^2(\beta |J|/2)]=1
\end{align}
which occurs at a strictly larger inverse temperature than the tree-uniqueness threshold.

On $d$-dimensional periodic lattices, the Ising model with i.i.d. spin glass couplings uniform in $\{+1,-1\}$ is also called the Edwards-Anderson (EA) model~\cite{Edwards_Anderson_1975}.
The nature of the low temperature phase is still an active field of research in both theoretical physics and mathematics, see for instance~\cite{Chatterjee_2023, Newman_Stein_2024,Newman_Read_Stein_2023} and references therein. 
In~\cite{DeSantis_2001}, it is shown that the uniqueness regime is strictly larger than the region of uniqueness for the related ferromagnetic system.
For the EA model on the square lattice, this yields a lower bound $\beta_{\rm uni}^{\rm EA}(d=2)>\beta_{\rm uni}^{\rm ferro}(d=2)=\log(1+\sqrt{2})\simeq0.8814$~\cite{Onsager_1944}. 
For the cubic lattice, a numerical estimation of the ferromagnetic uniqueness threshold yields $\beta_{\rm uni}^{EA}(d=3)>\beta_{\rm uni}^{\rm ferro}(d=3)\simeq0.443309252$~\cite{Ferrenberg_2018}.
Numerical studies support a zero-temperature spin-glass transition in dimension two~\cite{Toldin_2010}, and a finite temperature spin-glass transition in dimension three~\cite{BaityJesi_2013} with $\beta_{\rm sg}^{EA}(d=3)\simeq1,8150$.

\subsection{Perfect sampling algorithms}
In this section, we recall perfect sampling methods based on the Coupling From The Past (CFTP) approach~\cite{Propp_Wilson_1996}. These algorithms are built from the single-site Glauber dynamics that we define below.

\subsubsection{Single-site Glauber dynamics}
\label{subsubsec:single_site_Glauber}
For every $v\in V$ and $\sigma\in\Omega$, let $m_v(\sigma) = \mu_v^{\sigma}(+1)$ (with $\mu_v^\sigma$ the marginal of $\mu$ at $v$ conditional on the configuration $\sigma$). For $v\in V, r\in [0,1]$, we define the deterministic map $F_{v,r}:\Omega\to\Omega$ by
\begin{align}
F_{v,r}(\sigma)(u) = \begin{cases} \sigma(u) & u\neq v,\\
+1 & u=v\text{ and }r\leq m_v(\sigma),\\
-1 & u=v\text{ and }r>m_v(\sigma).
\end{cases}
\label{eq:glauber_random_map}
\end{align}
Let $(v_t)_{t\in\mathbb{Z}}$ and $(r_t)_{t\in\mathbb{Z}}$ be two mutually independent families of random variables uniformly distributed respectively on $V$ and on $[0,1]$.
The single-site heat-bath Glauber dynamics $(X_t)_{t\geq 0}$ is defined by
\begin{align}
X_{t+1}
=
F_{v_t,r_t}(X_t).
\label{eq:single_site_glauber_update}
\end{align}
Equivalently, at time $t$, the spin at the uniformly chosen vertex $v_t$ is re-sampled from its conditional distribution under $\mu$, given the spins at its neighbors, while all other spins remain unchanged. 
This Markov chain is reversible with respect to $\mu$ and has $\mu$ as its stationary distribution.

\subsubsection{Coupling from the past}
\label{subsubsec:CFTP}
Coupling from the past (CFTP), introduced by Propp and Wilson~\cite{Propp_Wilson_1996}, constructs an exact sample from $\mu$ by simultaneously coupling Glauber dynamics started from every configuration in $\Omega$.
For integers $s<t$, we define the composition of the random update maps by
\begin{align}
F_{s,t} = F_{v_{t-1},r_{t-1}} \circ\cdots\circ F_{v_s,r_s},
\label{eq:composition_random_maps}
\end{align}
with $F_{t,t}$ equal to the identity map. 
Thus, $F_{s,t}(\sigma)$ is the state at time $t$ of a Glauber chain started from $\sigma$ at time $s$ and updated using the random variables $(v_s,r_s),\ldots,(v_{t-1},r_{t-1})$.
For a time horizon $T\geq 1$, start one Markov chain from every configuration in $\Omega$ at time $-T$, updating all chains using the same sequence of random maps. 
Consider the set
\begin{align}
\mathcal{X}_T = \left\{F_{-T,0}(\sigma): \sigma\in\Omega \right\}.
\label{eq:cftp_image_set}
\end{align}
of resulting configurations at time $t=0$.
If $\mathcal{X}_T$ is a singleton, then all the chains have coalesced by time $0$. 
In this case, the unique element of $\mathcal{X}_T$ is returned.
If $\mathcal{X}_T$ is not a singleton, the time horizon is increased, for example by replacing $T$ with $2T$, and the procedure is repeated.
In order to obtain an exact sampler, the random maps that have already been generated should be reused: only the additional maps corresponding to times $-2T,\ldots,-T-1$ are sampled. 
If the procedure terminates, its output is distributed exactly according to $\mu$~\cite{Propp_Wilson_1996}.

\subsubsection{Single-site bounding chain}
\label{subsubsec:single_site_BC}
Running all the $|\Omega|=2^{n}$ Glauber chains explicitly is generally infeasible. 
A bounding chain~\cite{Huber_1998, Haggstrom_Nelander_1999} provides a compact strategy for determining whether $\mathcal{X}_T$ is a singleton.
Let $\Omega_\star=\{-1,+1,\star\}^{V}$, where the symbol $\star$ represents an unresolved spin.
For $\tau\in\Omega_\star$, define the set of configurations in $\Omega=\{+1,-1\}^V$ compatible with $\tau$ by
\begin{align}
\mathcal{C}(\tau)=\left\{\sigma\in\Omega: \sigma(u)=\tau(u) \text{ for every } u\in V \text{ such that }\tau(u)\neq \star \right\}
\end{align}
In particular, $\mathcal{C}(\star^V)=\Omega$, whereas if $\tau\in\Omega$, then $\mathcal{C}(\tau)=\{\tau\}$.
For $v\in V$, $\tau\in\Omega_\star$, define
\begin{align}
\label{eq:BC_proba}
p_v^{(-)}(\tau) = \min_{\sigma\in\mathcal{C}(\tau)} \mu_v^\sigma(+1) \quad \text{ and }\quad
p_v^{(+)}(\tau) = \max_{\sigma\in\mathcal{C}(\tau)} \mu_v^\sigma(+1)
\end{align}
For $v\in V$ and $r\in[0,1]$, define the bounding update map $G_{v,r}:\Omega_\star\to\Omega_\star$ by
\begin{align}
	G_{v,r}(\tau)(u)=\begin{cases}
	\tau(u) & u\neq v , \\
	+1  & u=v \text{ and } r\leq p_v^{(-)}(\tau)	,\\
	-1  & u=v \text{ and } r> p_v^{(+)}(\tau)	,\\
	\star , & u=v \text{ and } p_v^{(-)}(\tau)<r\leq p_v^{(+)}(\tau) .
	\end{cases}
\end{align}
The single-site bounding chain $(Y_t)_{t\geq 0}$ is the Markov chain on $\Omega_\star$, defined by 
\begin{align}
	Y_{t+1} = G_{v_t,r_t}(Y_t)
    \label{eq:single_site_bounding_chain}
\end{align}
\paragraph{\it Bounding chain inclusion.}
For every $\tau\in\Omega_\star$, $v\in V$, and $r\in[0,1]$, the Glauber and bounding update maps satisfy
\begin{align}
\left\{ F_{v,r}(\sigma): \sigma\in\mathcal{C}(\tau) \right\} \subseteq \mathcal{C}\left(G_{v,r}(\tau)\right).
\label{eq:bounding_property}
\end{align}
Indeed, if $r\leq p_v^{(-)}(\tau)$, then $r\leq m_v(\sigma)$ for every $\sigma\in\mathcal{C}(\tau)$, and all
compatible Glauber chains assign spin $+1$ to $v$. 
Similarly, if $r>p_v^{(+)}(\tau)$, all compatible chains assign spin $-1$ to $v$. 
In the remaining case, the chains may assign different values to $v$, and the bounding chain assigns $\star$.

For a CFTP horizon $T$, initialize the bounding chain at time $-T$ by $Y_{-T}^{(T)}=\star^V$, and use the same random draws $(v_t,r_t)_{-T\leq t< 0}$ used by the Glauber chains of the CFTP algorithm, i.e. $Y_{t+1}^{(T)} = G_{v_t,r_t}\left(Y_t^{(T)}\right)$ for $-T\leq t<0$.
Iterating~\eqref{eq:bounding_property} yields
\begin{align}
\mathcal{X}_T \subseteq \mathcal{C}\left(Y_0^{(T)}\right).
\label{eq:cftp_bounding_inclusion}
\end{align}
Therefore, if $Y_0^{(T)}$ contains no unresolved spins, then $\mathcal{C}(Y_0^{(T)})=\{Y_0^{(T)}\}$, and all Glauber chains started at time $-T$ have coalesced by time $0$. 
The algorithm returns $Y_0^{(T)}$, which is an exact sample from $\mu$.
If unresolved spins remain, the horizon is increased while reusing all previously generated random maps.

Note that the relation~\eqref{eq:cftp_bounding_inclusion} can be a strict inclusion, or in other words, the CFTP algorithm might have coalesced to a singleton without being detected by the bounding chain. This is what we observe numerically on periodic $d$-dimensional lattices (see Section~\ref{sec:numerics}).

\paragraph{\it Coalescence time.}
When the bounding chain is run forward from $Y_0=\star^V$, we define its coalescence time by
\begin{align}
T_{\rm coal} = \inf\left\{ t\geq0: Y_t\in\Omega \right\}.
\label{eq:bounding_coalescence_time}
\end{align}

\paragraph{\it Ferromagnetic Ising models.} For ferromagnetic models, the single site bounding chain is equivalent to tracking the two extremal trajectories under the single-site Glauber dynamics of Section~\ref{subsubsec:single_site_Glauber}. 
Let $X_t^{(+)},X_t^{(-)}$ be the two trajectories starting from the all-minus and all-plus configurations, and evolving under the same random updates. 
Monotonicity ensures that $X_t^{(-)}(v)\leq X_t^{(+)}(v)$ on each site $v\in V$ at every time, and that starting from $Y_0=\star^V$, the single-site bounding chain satisfies
$$
Y_t(v)=\begin{cases}
	X_t^{(+)}(v) & \text{if } X_t^{(-)}(v)=X_t^{(+)}(v)\\
	\star & \text{if } X_t^{(-)}(v)\neq X_t^{(+)}(v)
\end{cases}$$
Indeed, the minimal and maximal configurations compatible with $Y_t$ are $X_t^{(-)}$ and $X_t^{(+)}$.
Therefore the update probabilities are $p_v^{(-)}(\tau)=\mu_v^{X_t^{(-)}}(+1)$ and $p_v^{(+)}(\tau)=\mu_v^{X_t^{(+)}}(+1)$. 
This is easy to check for the first update, and since the Glauber dynamics preserve the ordering of the configurations (monotonicity is preserved), it remains true at all steps. 
Hence for ferromagnetic models, the single-site bounding chain coalesces exactly when the two extremal trajectories meet, at which point all trajectories of the CFTP algorithm of Section~\ref{subsubsec:CFTP} have coalesced.
\vspace{0.2cm}

This observation provides useful intuition for the behavior of the single-site bounding chain on spin-glass models on trees.
On any tree, a gauge transformation maps the model to a ferromagnetic model with positive couplings $|J_{\{u,v\}}|$.
Therefore, the single-site bounding chain coalesces exactly when the two gauge-transformed extremal trajectories meet.
On the infinite tree with ferromagnetic interactions, the tree-uniqueness threshold marks the onset of two distinct plus and minus infinite-volume Gibbs measure~\cite{Lyons_1989}.
Above this threshold, the two extremal trajectories will fail to meet in an accessible time scale, causing the slowdown of the coalescence of the bounding chain.
However, this argument should be taken with caution when considering sparse random graphs of finite size, where the presence of frustrated cycles forbid a gauge transform to a ferromagnetic model.

\section{Numerical Results}
\label{sec:numerics}
In this section and in Appendix~\ref{app:numerics_additional}, we present numerical experiments on several families of interaction graphs: Erd\H{o}s--R\'enyi random graphs $\mathcal{G}(n,d/n)$, random $d$-regular graphs, and periodic $d$-dimensional lattices. Couplings are either i.i.d. uniform in $\{+1,-1\}$ or i.i.d. standard Gaussian. We also consider the complete graph $K_n$ with i.i.d. couplings in $\mathcal N(0,1/n)$, corresponding to the Sherrington-Kirkpatrick model.
In this section, the target distribution is the Gibbs measure
\begin{equation}
	\label{eq:Gibbs_numerics}
	\mu(\sigma) \propto \exp\Big(\beta' \sum_{\{u,v\}\in E} J_{\{u,v\}}\,\sigma(u)\sigma(v)\Big),
	\qquad \sigma\in\{-1,+1\}^V,
\end{equation}
With this convention, the single-site Glauber update at site $v$ sets $\sigma(v)=+1$ with probability $\big(1+e^{-2\beta' h_v}\big)^{-1}$, where $h_v=\sum_{u\in N(v)}J_{\{u,v\}}\sigma(u)$ is the local field. 
This normalization differs from the one used in Sections~\ref{sec:introduction},\ref{sec:definitions}. and \ref{sec:rapid_coal_block_BC}. (and in~\cite{Efthymiou_Zampetakis_2024}) by a factor of two: an inverse temperature $\beta'$ in~\eqref{eq:Gibbs_numerics} corresponds to $\beta'=\beta/2$ in the definition (\ref{eq:Gibbs_proba}) of Section~\ref{sec:definitions}. 
We adopt this convention here so that our results can be compared directly with the physics literature, e.g. the damage-spreading threshold $\beta_{\rm ds}\simeq 0.58$ reported in~\cite{Bernard_Krauth_2010,Hukushima_Krauth_2025} for the two-dimensional Edwards-Anderson model.


\subsection{Coalescence time}
\label{subsec:numerics_coal_time}

\paragraph{\it Experimental setup.}
We run the forward single-site bounding chain dynamics (defined in Section~\ref{subsubsec:single_site_BC}) forward in time from $Y_0=\star^V$, until either coalescence happens or a fixed time horizon $T_{\rm hor}$ (measured in single-site updates) is reached.
For each graph family, coupling distribution, size $n$ and inverse temperature $\beta$, we generate an independent realization (edge set and couplings) for every run. 
We record the coalescence time $T_{\rm coal}$~(\ref{eq:bounding_coalescence_time}) for each run, and report its empirical mean and standard deviation over $n_{\rm runs}$ independent realizations, together with the fraction of runs that coalesced before $T_{\rm hor}$.

\paragraph{\it Fraction of coalesced runs.}
\begin{figure}[h!]
	\centering
	\includegraphics[width=0.45\textwidth]{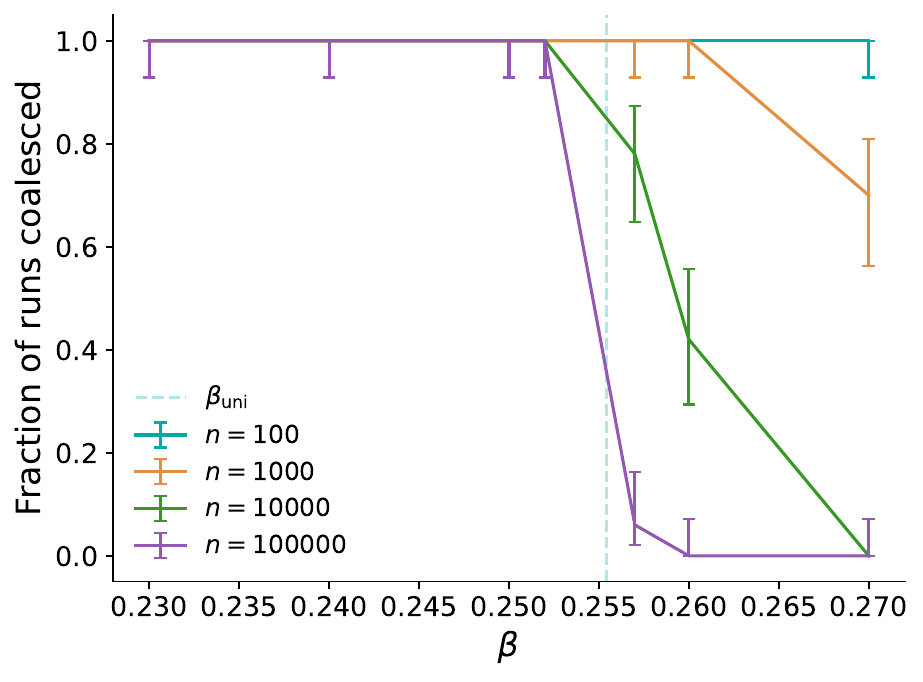}
	\includegraphics[width=0.45\textwidth]{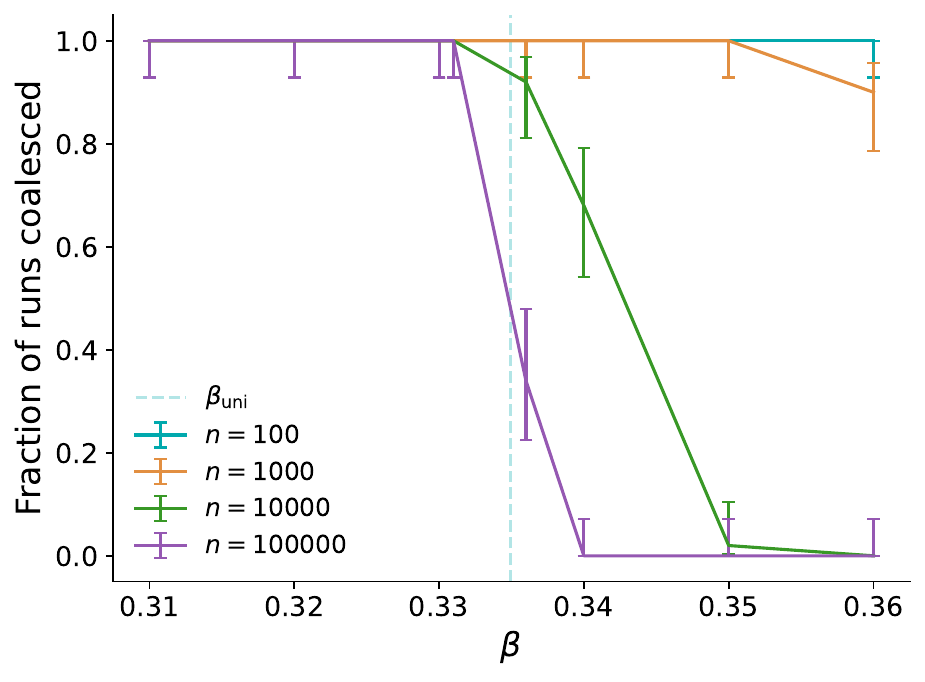}
	\caption{Fraction of coalesced runs on $\mathcal{G}(n,d/n)$ with $d=4$, i.i.d. uniform couplings in $\{+1,-1\}$ (left panel) and i.i.d. standard Gaussian couplings (right panel). We set $T_{\rm hor}=10^9$ single-site updates, and $n_{\rm runs}=50$. The corresponding tree-uniqueness threshold is $\beta_{\rm uni}\simeq0.2554$ (left) and $\beta_{\rm uni}\simeq0.3349$ (right).}
	\label{fig:coal_frac}
\end{figure}
As a first observation, Figure~\ref{fig:coal_frac}. shows the fraction of coalesced runs as a function of $\beta$, for several system sizes, together with $95\%$-Wilson confidence intervals, for Erd\H{o}s Rényi random graphs $\mathcal{G}(n,d/n)$ with $d=4$ under i.i.d. couplings both uniform in $\{+1,-1\}$ and standard Gaussian. 
Below the threshold $\beta_{\rm uni}$, all runs coalesce for all sizes tested.
The fraction drops sharply for $\beta>\beta_{\rm uni}$ as the size increases, which is consistent with a transition in the scaling of the coalescence time $T_{\rm coal}$ of the single-site bounding chain dynamics.
Although this first experiment depends on the horizon time $T_{\rm hor}$, and is not sufficient to conclude for a change in the scaling law of $T_{\rm coal}$, it already indicates that the bounding chain stops coalescing within an accessible time window above the tree-uniqueness threshold on sparse random graphs.

\paragraph{\it Scaling of the coalescence time.}
\begin{figure}[h!]
	\centering
	\includegraphics[width=0.45\textwidth]{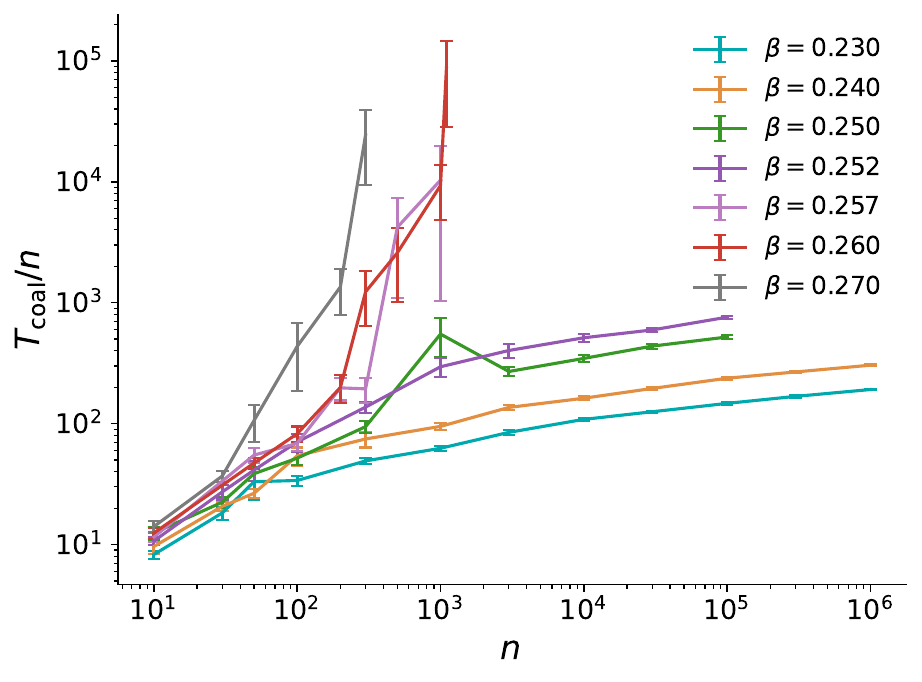}
	\includegraphics[width=0.45\textwidth]{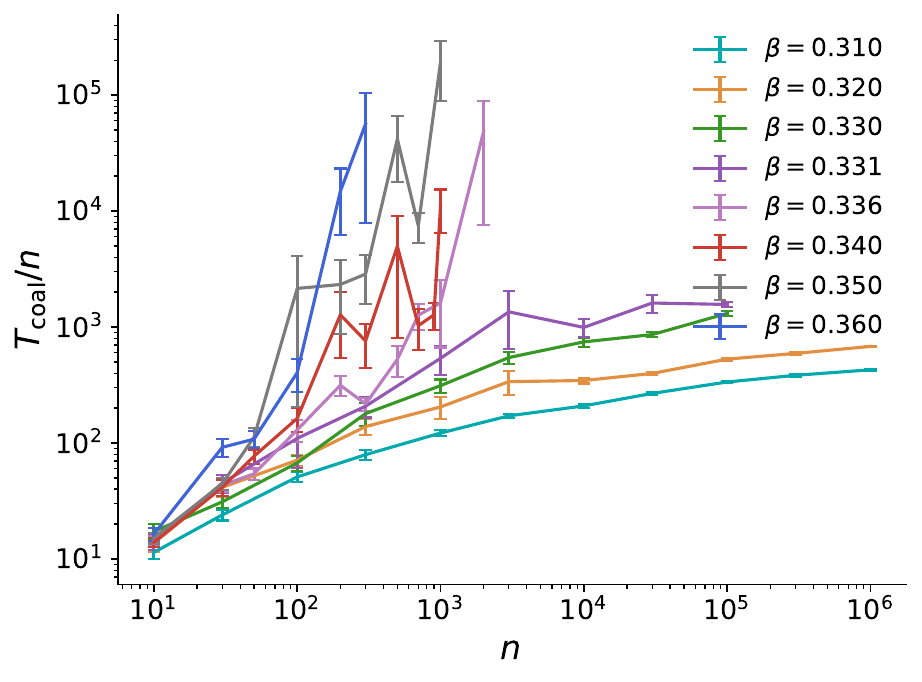}
	\caption{Empirical mean of the rescaled coalescence time $T_{\rm coal}/n$ as a function of $n$ on logarithmic axes, for $\mathcal{G}(n,d/n)$ with $d=4$. Couplings are i.i.d., drawn uniformly from $\{+1,-1\}$ (left) or from the standard Gaussian distribution (right).}
	\label{fig:coal_loglog}
\end{figure}
Figure~\ref{fig:coal_loglog} shows how the mean rescaled coalescence time $T_{\rm coal}/n$ depends on the system size $n$, for $\mathcal{G}(n,d/n)$.
We only keep the values of $\beta$ for which all $n_{\rm runs}$ runs coalesced within the horizon time $T_{\rm hor}$, in order to avoid averages computed from only a subset of the runs.
Corresponding results for random $d$-regular graphs, periodic $d$-dimensional lattices and the Sherrington-Kirkpatrick model are shown in Appendix~\ref{subsec_app:numerics_coaltime}. 
These results suggest a change in the scaling law of $T_{\rm coal}$, occurring at a value $\beta_{\rm bc}$ depending on the model. 
For $\beta<\beta_{\rm bc}$, the observed growth of $T_{\rm coal}/n$ is sub-polynomial.
For $\beta>\beta_{\rm bc}$, within the accessible system sizes, the growth is consistent with a super-polynomial scaling.
\vspace{0.1cm}

\begin{figure}[h!]
	\centering
	\includegraphics[width=0.45\textwidth]{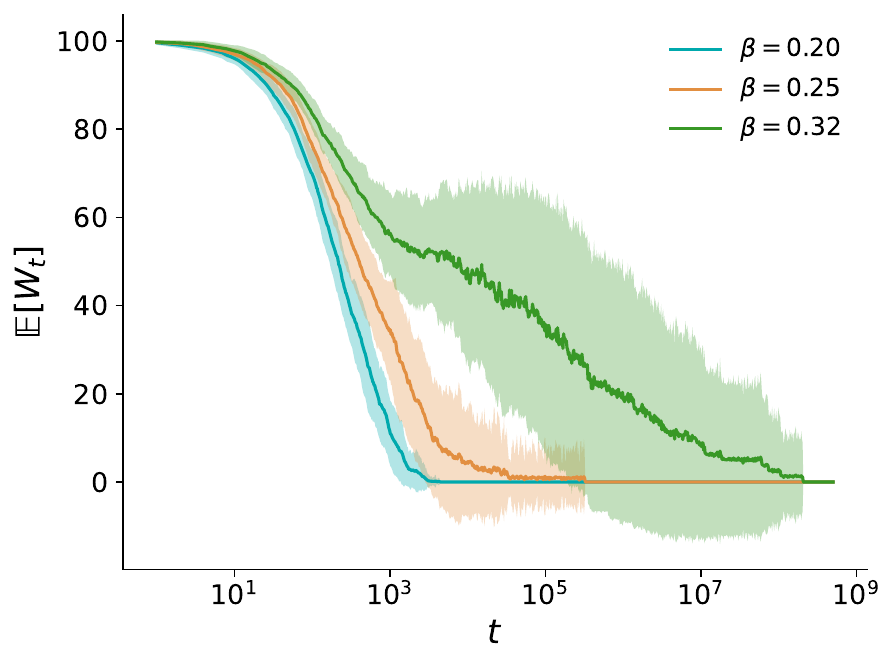}
	\includegraphics[width=0.45\textwidth]{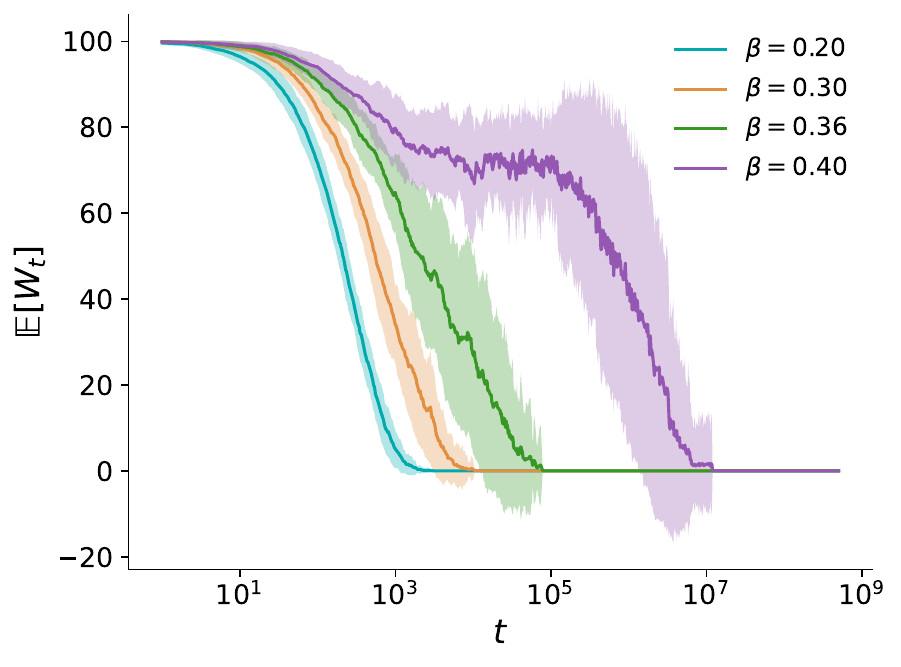}
	\caption{Number of unresolved (star) sites against iterations, averaged over $n_{\rm runs}=50$ runs, for $\beta$ values below and above $\beta_{\rm uni}$ on small system sizes ($n=100$), on $\mathcal{G}(n,4/n)$ (left panel, $\beta_{\rm uni}\simeq0.255413$) and on random $4$-regular graphs (right panel, $\beta_{\rm uni}\simeq0.346574$) with i.i.d. couplings uniform in $\{+1,-1\}$.}
	\label{fig:nb_star_evol_ER_aboveuni}
\end{figure}

{\it On Erd\H{o}s-Rényi random graphs} $\mathcal{G}(n,4/n)$, the tree-uniqueness threshold occurs at $\beta_{\rm uni}\simeq0.2554$ for $\pm 1$ couplings, and $\beta_{\rm uni}\simeq 0.3349$ for standard Gaussian couplings (see equation~(\ref{eq:tree_uniqueness})).
In both cases, these values are consistent with the inverse temperature found for $\beta_{\rm bc}$.
Above $\beta_{\rm bc}$, the coalescence time exhibits larger fluctuations between runs.
Figure~\ref{fig:nb_star_evol_ER_aboveuni} illustrates the mechanism underlying this behavior: the number of unresolved spins initially decreases, and then fluctuates around a positive plateau for a long time, before eventually reaching zero.
This suggests that coalescence in this regime requires a rare fluctuation, leading to a long waiting time varying from run to run.
Similar results for random $d$-regular graphs are shown in appendix~\ref{subsec_app:numerics_coaltime}, Figure~\ref{fig_app:coal_loglog_RR} (left panels) for various values of $d$.
Here again, the observed change in the scaling of $T_{\rm coal}$ is consistent with the tree-uniqueness threshold $\beta_{\rm uni}=\operatorname{atanh}(1/(d-1))$.
\vspace{0.1cm}

{\it On periodic $d$-dimensional lattices}, the estimated thresholds are $\beta_{\rm bc}\simeq0.36$, $0.20$, and $0.14$ for $d=2$, $3$, and $4$, respectively; see the right panels of Figure~\ref{fig_app:coal_loglog_RR} in Appendix~\ref{subsec_app:numerics_coaltime}.
These values do not coincide with any known equilibrium transition of the Edwards-Anderson model.
They are also significantly smaller than the threshold $\beta_{\rm patch}\simeq 0.58$ for $d=2$ and $\beta_{\rm patch}\simeq0.25$ for $d=3$ reported in the study of the local-patch algorithm~\cite{Bernard_Krauth_2010,Hukushima_Krauth_2025}.
A possible explanation for this discrepancy is the greater precision of the envelope maintained by the local-patch algorithm: while the single-site bounding chain only records whether each site is resolved, the patch algorithm store sets of admissible configurations on overlapping regions (patches).
Compatibility checks between patches allows to eliminate locally admissible configurations that cannot be combined consistently.
As the size of patches increase along iterations, this strategy allows to retain information on larger distances, while the single-site bounding chain discards this information.
This suggests that finding a strategy to retain correlations between unresolved sites could improve the performance of the bounding chain. 
A possible improvement could be to update blocks of spins at each time steps, as in the block bounding chain introduced in Section~\ref{sec:rapid_coal_block_BC} for Erd\H{o}s-Rényi random graphs.

Figure~\ref{fig_app:coal_loglog_RR} also compares periodic $d$-dimensional lattices with random regular graphs of the same degree.
The estimated values of $\beta_{\rm bc}$ are close, although the square lattice exhibits a slightly larger threshold.
In particular, they lie close to the tree-uniqueness threshold for the $2d$-regular tree.
This suggests that even on lattices, tree-uniqueness is also a meaningful threshold for the single-site bounding chain dynamics.
However, this is not an equilibrium phase transition, as rigorous results place the uniqueness threshold for the Edwards-Anderson model strictly above this value (see~\cite{DeSantis_2001} and discussion in Section~\ref{subsubsec:uniqueness_and_SG}).
One possible explanation is that the single-site bounding chain fails to exploit correlations and compatibility constraints associated with short loops, although establishing whether this mechanism fully explains the observed thresholds would require further analysis.
\vspace{0.1cm}

{\it On the Sherrington-Kirkpatrick model}, the coalescence time grows rapidly with system size even at the smallest inverse temperature investigated (see Appendix~\ref{subsec_app:numerics_coaltime}, Figure~\ref{fig_app:coal_loglog_SK}).
The observed growth of $T_{\rm coal}$ appears to be consistent with a super-polynomial scaling, suggesting that $\beta_{\rm bc}=0$ for this model.
One possible explanation is the accumulation of uncertainty for the many weak interactions: when many neighboring spins remain unresolved, the single-site bounding chain allows their contribution to the local field to align, producing a wide interval of possible conditional probabilities.
In turn, an updated spin can remain unresolved with a high probability.
This mechanism again illustrates that the single-site bounding chain does not retain useful information given by correlations between unresolved spins.

However, the damage spreading results in Figure~\ref{fig:coal_DS_loglog} suggests that the performance of the single-singe bounding chain is not solely affected by its envelope representation.
Disagreement between coupled trajectories persist even at small positive $\beta$, consistently with earlier numerical observations for the Sherrington-Kirkpatrick model with i.i.d. couplings uniform in $\{+1/\sqrt{n},-1/\sqrt{n}\}$~\cite{Derrida1989}.
Therefore, the slow coalescence of the underlying coupling dynamics should limit the performance of any CFTP algorithm, under the coupling considered here (which uses the same random update for each trajectory).
Investigating alternative couplings is therefore a natural direction for future work.

In order to explore the connection with sparse models, Figure~\ref{fig_app:coal_loglog_SK} also presents results for random $d$-regular graphs with i.i.d. couplings $J_{\{u,v\}}\sim \mathcal{N}(0,1/d)$.
This normalization keeps the sum of coupling variances at each vertex of order one, in such a way that increasing $d$, this model gets closer to the Sherrington-Kirkpatrick model.
The tree-uniqueness threshold (\ref{eq:tree_uniqueness}) associated with this model tends to $0$ as $d$ increases.
If the single-site bounding chain continues to have its threshold $\beta_{\rm bc}(d)$ close to the tree-uniqueness threshold $\beta_{\rm uni}(d)$, this would provide a heuristic explanation for the absence of rapid-coalescence regime on the Sherrington-Kirkpatrick model at any positive inverse temperature.
Our estimates for $\beta_{\rm bc}(d)$ are consistent with $\beta_{\rm uni}(d)$ for $d=5$ and $d=100$, and lies slightly above $\beta_{\rm uni}(d)$ for $d=20$, which could be due to finite-size effects.

\paragraph{\it Candidate scaling laws for $T_{\rm coal}$.} 
\begin{figure}[h!]
	\centering
	\includegraphics[width=0.9\textwidth]{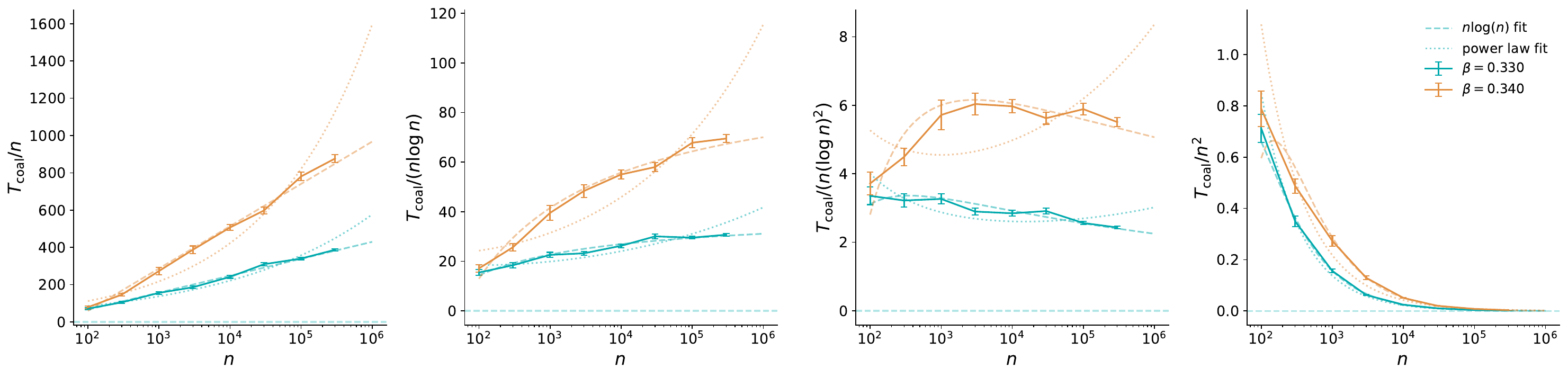}
	\includegraphics[width=0.9\textwidth]{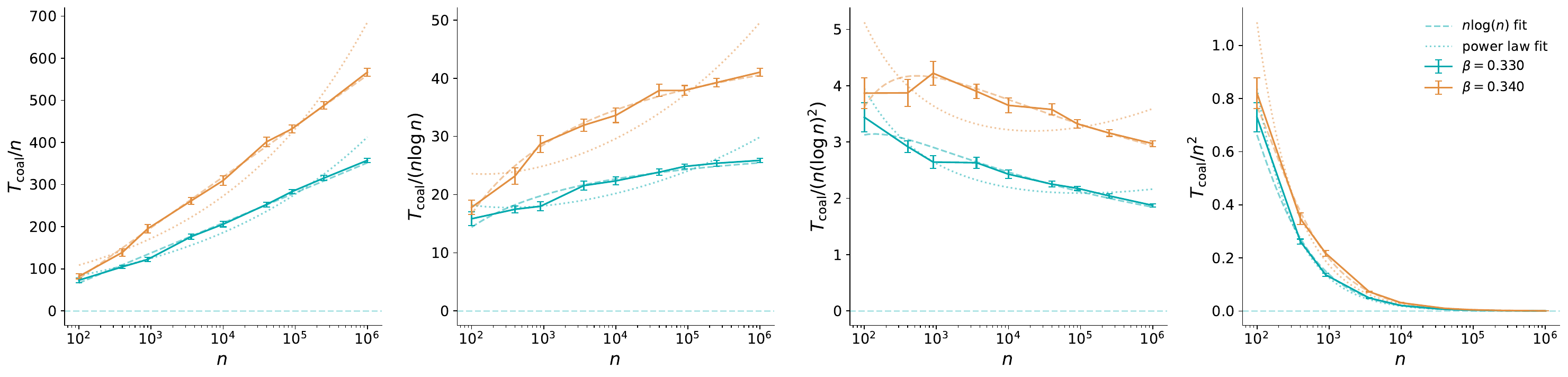}
	\caption{Rescaled empirical mean coalescence time as a function of $n$, on random $4$-regular graphs (top panels) and on periodic $2$-dimensional lattices (bottom panels). Couplings are i.i.d. drawn from $\{+1,-1\}$.
	The curves show the corresponding rescalings of two fits: $T_{\rm coal}=an\log n+bn$ (dashed line) and $T_{\rm coal}=cn^\delta$ (dotted line)}
	\label{fig:coal_scaling}
\end{figure}
We next examine which scaling forms are consistent with the observed coalescence times.
The $O(n\log n)$ bound established under the Dobrushin conditions in Section~\ref{sec:rapid_coalescence_Dobrushin} provides a natural candidate scaling, although Dobrushin conditions do not hold throughout the whole inverse temperature range considered here.
Note however that on Erd\H{o}s-Rényi random graphs $\mathcal{G}(n,d/n)$, the presence of large star sub-graphs w.h.p. affects the mixing scaling, as discussed in~\cite{Mossel_Sly_2013}, resulting in a $O(n^{1+\Theta(1/(n\log n))})$ mixing time for ferromagnetic models. This power-law scaling should transpose to the coalescence time on such model.

We therefore focus on the scaling $n\log n$, and compare it with linear growth, $n(\log n)^2$, and a generic power law $n^\delta$.
Figure~\ref{fig:coal_scaling} shows $T_{\rm coal}/n$, $T_{\rm coal}/(n\log n)$, $T_{\rm coal}/(n(\log n)^2)$ and $T_{\rm coal}/n^2$ as functions of $n$ for several values of $\beta < \beta_{\rm uni}$ on random $d$-regular graphs and on periodic square lattices.
Corresponding plots for other graph families and coupling distributions are shown in Appendix~\ref{app:numerics_additional}, Figure~\ref{fig_app:coal_scaling}.

Over the range of sizes considered, the ratio $T_{\rm coal}/n$ increases, while $T_{\rm coal}/n^2$ decreases towards zero, suggesting a scaling intermediate between linear and quadratic.
The ratio $T_{\rm coal}/(n\log n)$ increases slowly and could approach a plateau, while $T_{\rm coal}/(n(\log n)^2)$ decreases, which supports a scaling as $n(\log n)^\alpha$, with an exponent between $1$ and $2$.
These observations are therefore compatible with logarithmic corrections to linear growth, but do not determine their precise form.
In particular, the slow increase of $T_{\rm coal}/(n\log n)$ does not exclude an $n\log n$ leading term with finite-size corrections.

To compare logarithmic and power-law corrections more directly, we perform two complementary fits of the mean coalescence time: $T_{\rm coal}=an\log n+bn$ and $T_{\rm coal}=cn^\delta$ for each value of $\beta$.
Plotting these two fits against the data, we observe that the power-law fit does not capture the curvature of the rescaled data in the $T/(n\log n)$ and $T/(n(\log n)^2)$ plots.
This comparison reinforce the numerical evidence from Figure~\ref{fig:coal_loglog} in favor of a sub-polynomial scaling of $T_{\rm coal}/n$.
Note that similar observations are made on $\mathcal{G}(n,d/n)$ on the sizes tested, for which a scaling in $O(n\log n)$ is not expected for this graph family.
This last observation suggests that while all models display a scaling compatible with $T_{\rm coal}=O(n^{1+o(1)})$, the precise scaling of the correction cannot be assessed precisely within the range of sizes tested.

\paragraph{\it Evolution of the number of unresolved sites.}
\begin{figure}[h!]
	\centering
	\includegraphics[width=1.0\textwidth]{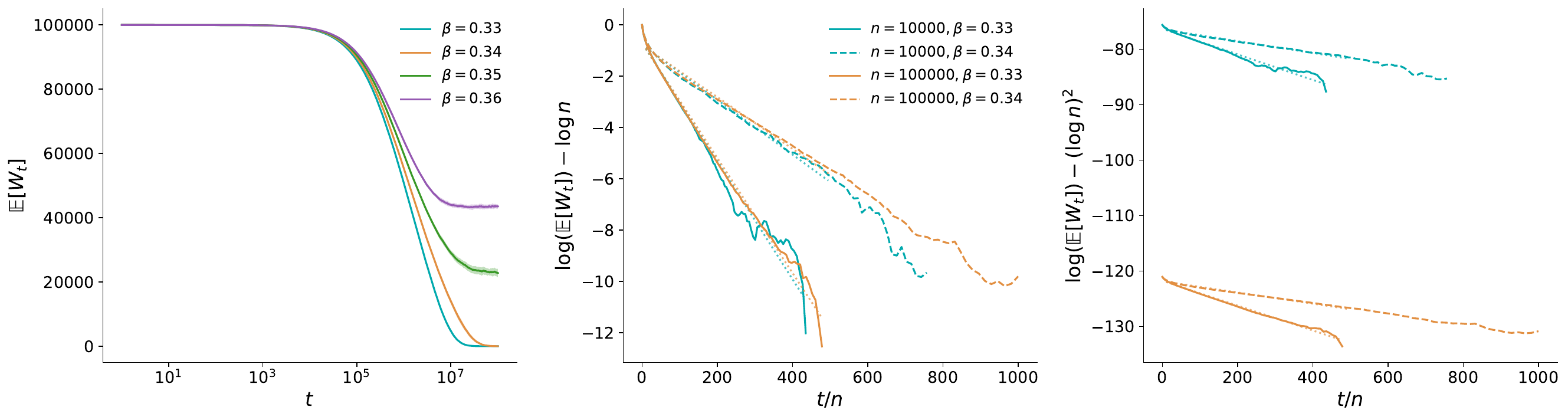}
	\caption{Evolution of the mean number of unresolved sites $\mathbb{E}[W_t]$ on random $d$-regular graphs with $d=4$, i.i.d. couplings uniform in $\{+1,-1\}$, averaged over $50$ independent runs.
	Left panel: $\mathbb{E}[W_t]$ as a function of the number of single-site updates, for $n=10^5$ and horizon time $T_{\rm hor}=10^8$. Middle and right panels: linear fits to test the scalings $W_t=Kne^{-\delta t/n}$ and $W_t=Ke^{(\log n)^2-\delta t/n}$}
	\label{fig:nb_star_evol_ER}
\end{figure}
Let $W_t=\sum_{u\in V}\mathbbm{1}_{\{Y_t(u)=\star\}}$ be the number of unresolved (star $\star$) sites as a function of $t$ (the number of single-site updates).
In Figure~\ref{fig:nb_star_evol_ER}, we show the empirical mean of $W_t$, averaged over $n_{\rm runs}=50$ independent runs on random $d$-regular graphs with i.i.d. couplings uniform in $\{+1,-1\}$.
Additional plots for other graph families are shown in Appendix~\ref{subsec_app:numerics_unresolved_sites}, Figure~\ref{fig_app:nb_star_evol}.
Consistently with Figure~\ref{fig:coal_frac}, the number of unresolved sites decays to zero for $\beta<\beta_{\rm uni}\simeq 0.346574$, and remains near a positive value above $\beta_{\rm uni}$, within the simulated time window $t<T_{\rm hor}$.

The analysis in Section~\ref{sec:rapid_coalescence_Dobrushin} provides the upper bound $\mathbb{E}[W_t]\leq ne^{-\delta t/n}$, under the Dobrushin conditions.
Motivated by this bound, we examine the scaling form $\mathbb{E}[W_t]\simeq Kn e^{-\delta t/n}$, with $K$ and $\delta$ independent of $n$ at fixed $\beta$.
In the middle panel, the curves of $\log(\mathbb{E}[W_t]/n)$ plotted against $t/n$ exhibit a linear decay after an initial transient, and approximately collapse across the two system sizes.
These observations support an exponential decay with a rate proportional to $1/n$.
Since this scaling reaches $\mathbb{E}[W_t]=O(1)$ at a characteristic time of order $n\log n$, it favors the hypothesis of a coalescence time having this scaling law.
For comparison, the right panel tests the alternative form $\mathbb{E}[W_t]=e^{(\log n)^2-\delta t/n}$.
The curves $\log(\mathbb{E}[W_t]) - (\log n)^2$ plotted against $t/n$  do not collapse across sizes, disfavoring this particular ansatz, which has an associated characteristic time of order $n(\log n)^2$.

In Figure~\ref{fig:nb_star_evol_ER_aboveuni}, we show the evolution of $W_t$ for smaller system size $n=10^2$, for which coalescence remain numerically accessible even above the threshold $\beta_{\rm bc}$.
Below the threshold, the mean number of unresolved sites decreases approximately exponentially.
Above it, the initial decrease is followed by a long plateau, before an eventual drop to zero.
This observation is consistent with individual runs fluctuating around a positive number of unresolved sites, until a rare fluctuation triggers coalescence.

\subsection{Damage spreading}
\label{subsec:numerics_damage_spreading}
\paragraph{\it Experimental set-up.}
We simulate two copies of the single-site Glauber dynamics (defined in Section~\ref{subsubsec:single_site_Glauber}), using the same update site and uniform random number at each time step, starting from two configurations sampled independently and uniformly at random in $\{+1,-1\}^V$.
For each graph family, coupling distribution, system-size $n$ and inverse temperature $\beta$, we generate an independent realization (edge set and couplings) for every run. 
For each run, we record the meeting time $T_{\rm meet}$ at which the Markov chains have met into the same configuration (after which the two trajectories remain identical).
We report its the empirical mean and standard deviation over $n_{\rm runs}$ independent realizations, together with the fraction of runs for which $T_{\rm meet}<T_{\rm hor}$. 

\paragraph{\it Scaling of the meeting time.}
\begin{figure}[h!]
	\centering
	\includegraphics[width=0.45\textwidth]{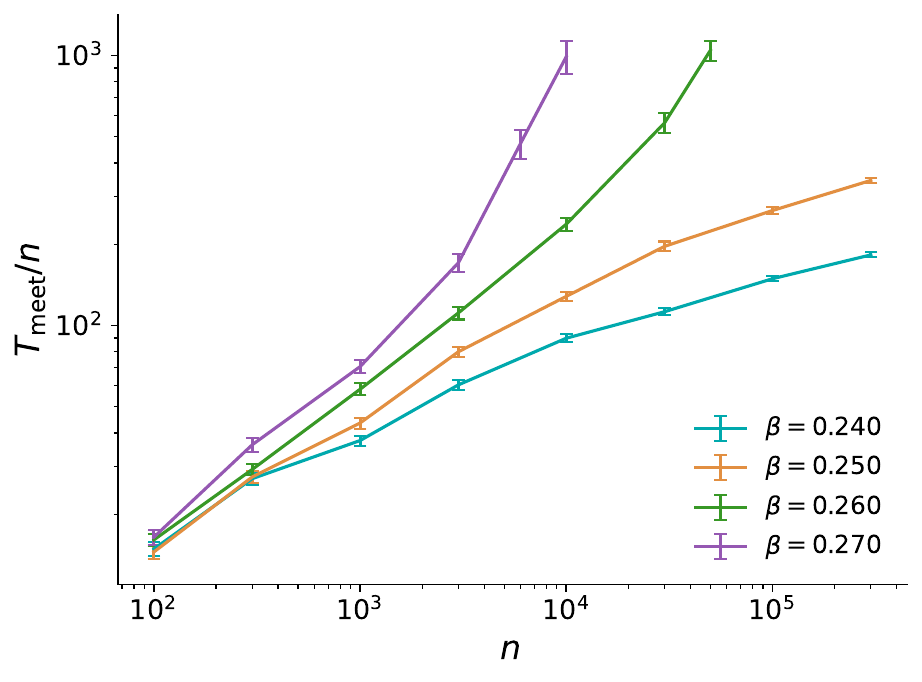}
	\includegraphics[width=0.45\textwidth]{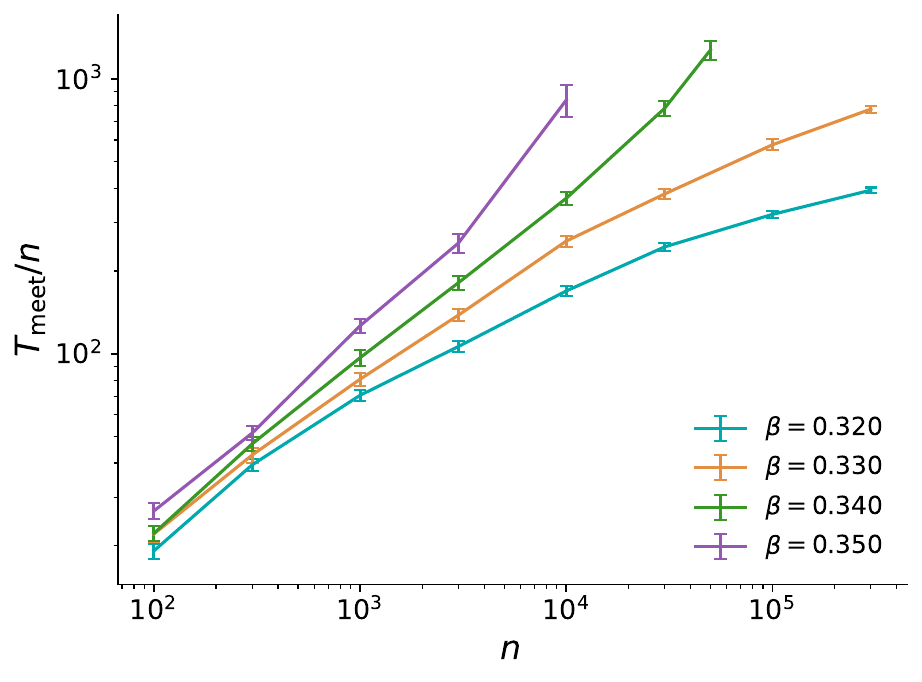}\\
	\includegraphics[width=0.45\textwidth]{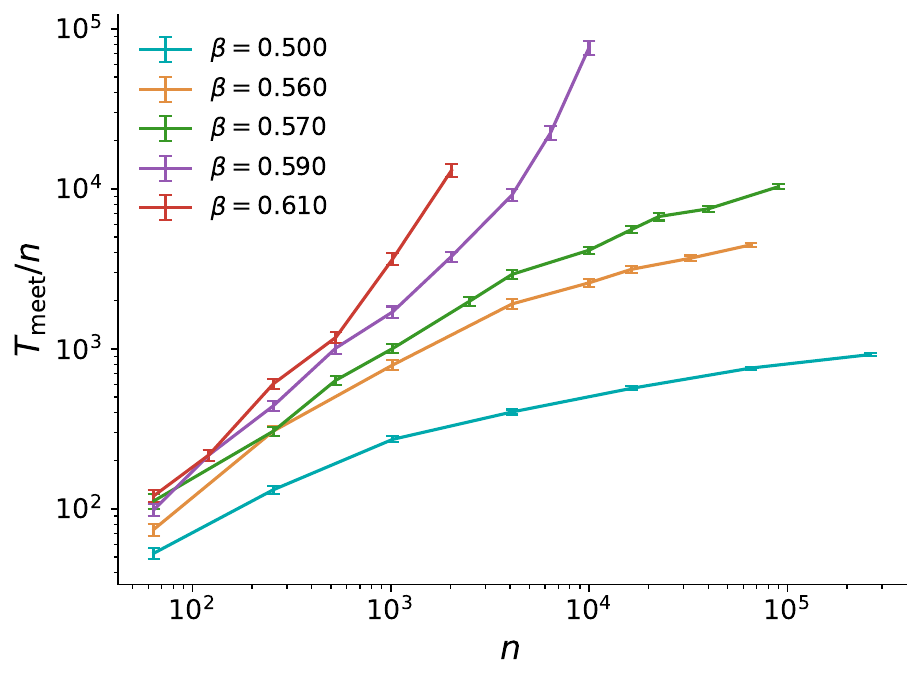}
	\includegraphics[width=0.45\textwidth]{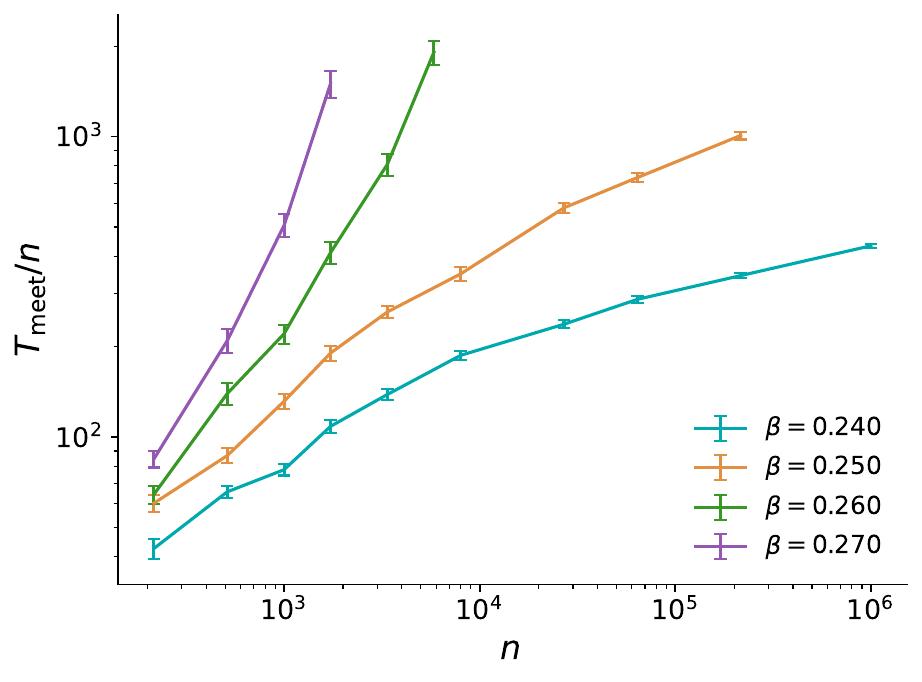}\\
	\includegraphics[width=0.45\textwidth]{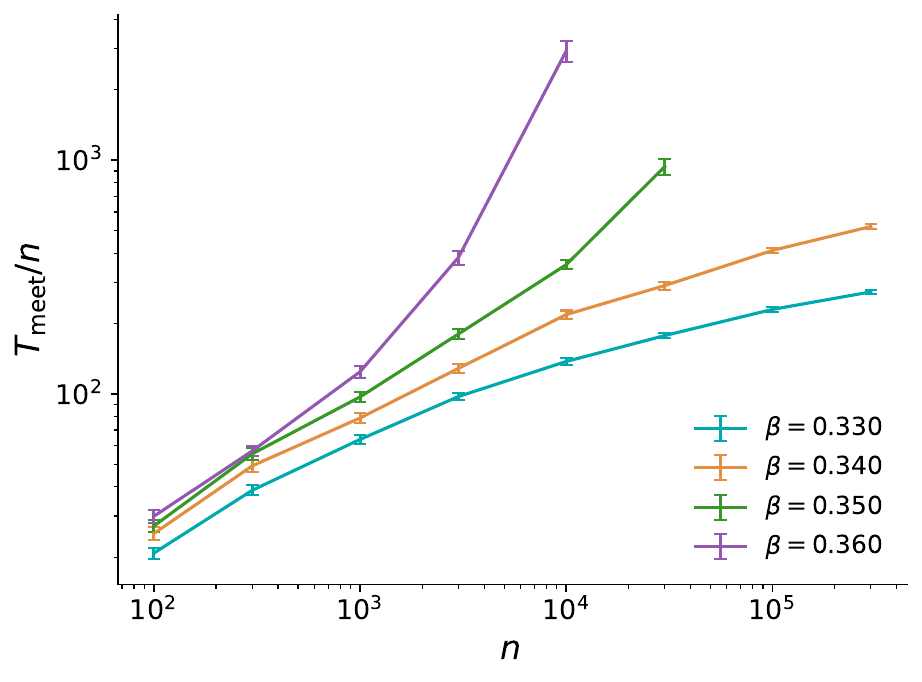}
	\includegraphics[width=0.45\textwidth]{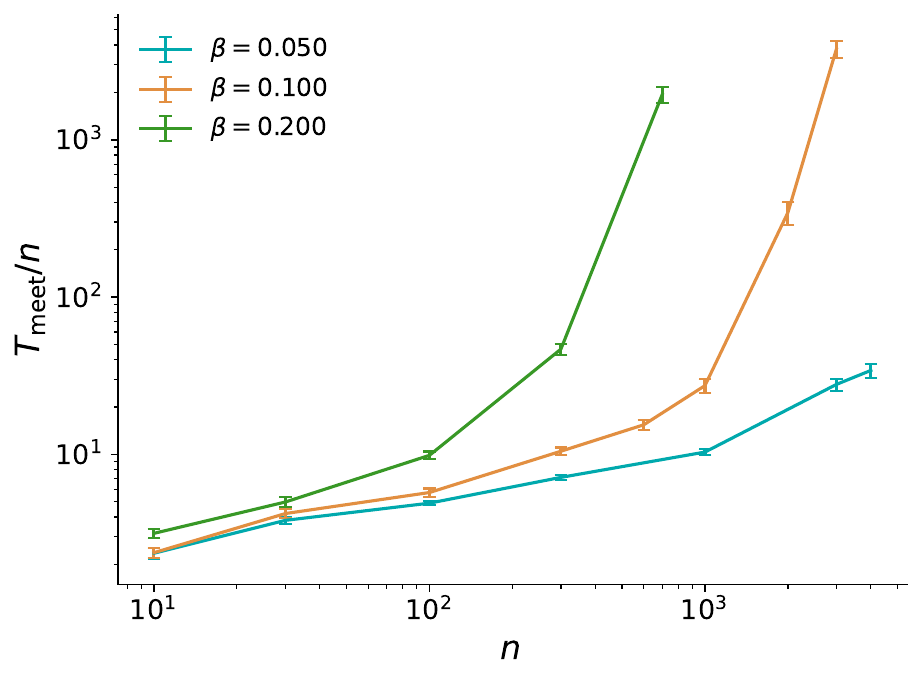}
	\caption{Empirical mean of the rescaled meeting time $T_{\rm meet}/n$ as a function of $n$ on logarithmic axes.
	Top panels: on $\mathcal{G}(n,d/n)$ with $d=4$, i.i.d couplings uniform in $\{+1,-1\}$ (top-left) and standard Gaussian (top-right).
	Middle panels: on periodic $d$-dimensional lattices with i.i.d couplings uniform in $\{+1,-1\}$, with $d=2$ (middle-left) and $d=3$ (middle-right).
	Bottom-left panel: on random $4$-regular graphs with i.i.d couplings uniform in $\{+1,-1\}$.
	Bottom-right panel: on the Sherrington-Kirkpatrick model.}
	\label{fig:coal_DS_loglog}
\end{figure}
Figure~\ref{fig:coal_DS_loglog} shows $T_{\rm meet}/n$ against $n$ on various graph families and coupling distributions. 
One observes a change in the scaling of $T_{\rm meet}$, occurring at a value $\beta_{\rm ds}$ depending on the model. 
For $\beta<\beta_{\rm ds}$, $T_{\rm meet}/n$ grows slowly, consistently with a sub-polynomial growth. 
Above it, the growth is substantially faster and compatible with a super-polynomial scaling.

{\it On sparse graphs} (Erd\H{o}s-Rényi $\mathcal{G}(n,d/n)$ and random $d$-regular graphs), the estimated damage spreading threshold $\beta_{\rm ds}$ is consistent with both the tree-uniqueness threshold $\beta_{\rm uni}$ and with the bounding chain threshold $\beta_{\rm bc}$.
This observation suggests that the slow coalescence of the underlying coupled trajectories is responsible for the poor performance of the single-site bounding chain above tree-uniqueness.
Indeed, coalescence of the CFTP algorithm requires that all trajectories coalesce, including the two trajectories followed in the damage-spreading experiment.
If this pair has not met by the time $T_{\rm hor}$, then neither the CFTP algorithm described in Section~\ref{subsubsec:CFTP}, nor its bounding chain envelope can have coalesced by that time.
Therefore, the performance of the single-site bounding chain is limited by the choice of the coupling itself, and cannot be improved by tightening the bounding representation only.

{\it On periodic $d$-dimensional lattices}, the situation is quite different, as shown in the middle panels of Figure~\ref{fig:coal_DS_loglog}.
The damage spreading threshold is estimated to be $\beta_{\rm ds}\simeq 0.58$ for $d=2$, and $\beta_{\rm ds}\simeq 0.25$ for $d=3$, consistently with the values reported in~\cite{Bernard_Krauth_2010,Hukushima_Krauth_2025}.
Both values are significantly larger than the corresponding bounding chain thresholds: $\beta_{\rm bc}\simeq0.36$ for $d=2$ and $\beta_{\rm bc}\simeq0.20$ for $d=3$.
This separation suggests that damage spreading does not explain the deterioration of the single-site bounding chain.
Indeed, numerical studies of the local-patch CFTP algorithm~\cite{Chanal_Krauth_2008,Chanal_Krauth_2010} indicate a rapid coalescence up to the damage spreading threshold.
In the interval $\beta_{\rm bc}<\beta<\beta_{\rm ds}$, the deterioration of the performance of the single-site bounding chain may instead be explained by its failure to retain correlations between unresolved sites.
Unlike the local-patch algorithm, its envelope can therefore remain too loose to detect coalescence of the underlying trajectories.

{\it On the Sherrington-Kirkpatrick model} (Figure~\ref{fig:coal_DS_loglog}, bottom-right panel), the meeting time appears to grow super-polynomially at every positive temperature tested.
This suggests that damage spreading persists for all $\beta>0$, consistently with numerical observations in~\cite{Derrida1989}.
As on sparse graphs, this behavior limits CFTP implementation based on the coupling using the same random updates for all trajectories.
Contrarily to what is observed on lattices, improving the bounding envelope alone would therefore be insufficient to obtain rapid coalescence.
A natural direction is to investigate alternative couplings, potentially including non-Markovian constructions inspired by~\cite{Griffeath1975}, although their efficient implementation in a perfect sampling algorithm remains an open question.

\section{Rapid coalescence under the Dobrushin condition}
\label{sec:rapid_coalescence_Dobrushin}
In this section, we show that the single-site bounding chain $(Y_t)_{t\geq 0}$ defined in Section~\ref{subsubsec:single_site_BC} coalesces rapidly whenever the Dobrushin coefficient~\eqref{eq:Dobrushin_coefficient} is strictly smaller than one.
Throughout this section, we therefore assume that $\alpha<1$ independently of $n$.

\begin{theorem}
	Let $\delta = 1-\alpha>0$. The single-site bounding chain $(Y_t)_{t\geq 0}$ defined in Section~\ref{subsubsec:single_site_BC} started from $Y_0=\star^V$, satisfies for every $k>0$,
	\begin{align}
		\mathbb{P}\left[T_{\rm coal}\leq \frac{(k+1)n\log n}{\delta} \right] \geq 1-n^{-k}
	\end{align}
	where $T_{\rm coal}=\inf\{t: Y_t\in\{+1,-1\}^V\}$ is the coalescence time of the bounding chain, and the probability is over the randomness of the chain. In particular $T_{\rm coal}=O(n\log n)$ with high probability, and the CFTP implementation of the single-site bounding chain returns an exact sample from $\mu_{G,\boldsymbol{J},\beta}$.
\end{theorem} 

\paragraph{\it Proof.}
We define $W(Y_t)$ as the number of unresolved states:
\begin{align}
W(Y_t)=\sum_{v\in V}\mathbbm{1}_{\{Y_t(v)=\star\}} \  ,
\end{align}
and show that $\mathbb{E}\left[W(Y_t)\right]$ decreases along the iterations.
Conditionally on $Y_t=\tau$ and on $v$ being updated, the probability that the updated spin remains unresolved is
\begin{align}
\mathbb P\left( Y_{t+1}(v)=\star \,\middle|\, Y_t=\tau,\ v_t=v \right) = p_v^{(+)}(\tau)-p_v^{(-)}(\tau)
\end{align}
with $p_v^{(\pm)}$ defined in~\eqref{eq:BC_proba}.
Let $\eta^{(-)}$ and $\eta^{(+)}$ be two configurations on $N(v)$
compatible with $\tau$ such that
\begin{align}
p_v^{(-)}(\tau) = \mu_v^{\eta^{(-)}}(+1), \qquad p_v^{(+)}(\tau) = \mu_v^{\eta^{(+)}}(+1).
\end{align}
These two configurations agree at each vertex $u\in N(v)$ for which $\tau(u)\neq\star$, and disagree on some of the unresolved neighbors. 
Let 
$$D_v(\tau)=\left\{u\in N(v):\eta^{(+)}(u)\neq \eta^{(-)}(u)\right\}=(u_1,\dots,u_k)$$
and choose a sequence of configurations $(\eta^{(0)}=\eta^{(-)},\eta^{(1)},\ldots,\eta^{(k)}=\eta^{(+)})$ such that $\eta^{(i-1)}$ and $\eta^{(i)}$ differ only on $u_i$.
Telescoping and applying the triangular inequality give
\begin{align}
\begin{aligned}
p_v^{(+)}(\tau)-p_v^{(-)}(\tau)&=\sum_{i=1}^k\mu_v^{\eta^{(i)}}(+1)-\mu_v^{\eta^{(i-1)}}(+1) \leq \sum_{i=1}^k\lvert\mu_v^{\eta^{(i)}}(+1)-\mu_v^{\eta^{(i-1)}}(+1)\rvert\\
&=\sum_{i=1}^k\lVert\mu_v^{\eta^{(i)}}-\mu_v^{\eta^{(i-1)}}\rVert_{\rm TV} \leq \sum_{i=1}^k\rho_{u_i\to v}\leq \sum_{u\in N(v)}\rho_{u\to v}\mathbbm{1}_{\{\tau(u)=\star\}}
\end{aligned}
\label{eq:unresolved_probability_Dobrushin}
\end{align}
For every $v\in V$ and $t\geq0$, we define $q_t(v)=\mathbb P\left(Y_t(v)=\star\right)$, where the bounding chain is initialized from $Y_0=\star^V$.
With probability $1-1/n$, the vertex $v$ is not updated at time $t$, and its state remains unchanged. 
Thus, using~\eqref{eq:unresolved_probability_Dobrushin}, we
obtain
\begin{align}
q_{t+1}(v) &= \left(1-\frac{1}{n}\right)q_t(v) + \frac{1}{n} \mathbb E\left[ p_v^{(+)}(Y_t)-p_v^{(-)}(Y_t) \right] \\
&\leq \left(1-\frac{1}{n}\right)q_t(v) + \frac{1}{n} \sum_{u\in N(v)} \rho_{u\to v} q_t(u).
\end{align}
Let $q_t^{\max} = \max_{v\in V}q_t(v)$, in such a way that
\begin{align}
q_{t+1}(v) \leq \left( 1-\frac{1}{n} + \frac{\alpha}{n} \right) q_t^{\max} 
= \left( 1-\frac{\delta}{n} \right) q_t^{\max}.
\end{align}
where we recall the expression of the Dobrushin coefficient $\alpha = \max_{v\in V} \left\{\sum_{u\in N(v)} \rho_{u\to v}\right\}$.
Taking the maximum over $v$ yields $q_{t+1}^{\max} \leq \left( 1-\frac{\delta}{n} \right) q_t^{\max}$, and since $q_0^{\max}=1$ one gets
\begin{align}
\begin{aligned}
q_t^{\max} &\leq \left( 1-\frac{\delta}{n} \right)^t \leq \exp\left(-\frac{\delta t}{n}\right) \quad \text{and therefore}\\
\mathbb E[W(Y_t)] &= \sum_{v\in V}q_t(v) \leq n q_t^{\max} \leq n\exp\left(-\frac{\delta t}{n}\right).
\end{aligned}
\end{align}
Since $W(Y_t)$ is a non-negative integer-valued random variable, we obtain by Markov inequality
\begin{align}
\mathbb P(T_{\rm coal}>t) &= \mathbb P(W(Y_t)\geq1) \leq \mathbb E[W(Y_t)] \leq \exp\left( -\frac{\delta t}{n}+\log n \right).
\end{align}
Therefore, for every $C>0$, $t\geq \frac{n}{\delta}\left(\log n +C\right)$ implies $\mathbb{P}(T_{\rm coal}>t)\leq e^{-C}$.
In particular, taking $C=k\log n$ with $k$ a constant of $n$ leads to
\begin{align}
\mathbb P\left( T_{\rm coal} \leq \frac{(k+1)n\log n}{\delta} \right) \geq 1-n^{-k}.
\end{align}
Thus, if $\delta=1-\alpha$ is bounded below by a positive constant
independent of $n$, the single-site bounding chain coalesces in
$O(n\log n)$ steps with high probability.
\hfill$\square$

\section{Rapid coalescence below the uniqueness threshold}
\label{sec:rapid_coal_block_BC}
In this Section, we introduce the block bounding chain, a variant of the single-site bounding chain defined in Section~\ref{subsubsec:single_site_BC}, based on a vertex partition introduced in~\cite{Efthymiou_Zampetakis_2024}.
Section~\ref{subsec:block_partition} introduce such vertex partition and recall the Theorem 4.2 in~\cite{Efthymiou_Zampetakis_2024} proving its existence with high probability on $\mathcal G(n,d/n)$ with i.i.d. standard Gaussian couplings.
Section~\ref{subsec:block_Glauber} reminds the block Glauber dynamics introduced in~\cite{Efthymiou_Zampetakis_2024} and the coupling estimate given in Theorem 7.1 of~\cite{Efthymiou_Zampetakis_2024} for proving its rapid mixing (in $O(n\log n)$ block updates).
Section~\ref{subsec:block_BC_trees} first introduces a block bounding chain valid when every block in the vertex partition is a tree, and proves its rapid coalescence.
Section~\ref{subsec:block_BC_unicyclic} adapts the construction of the block bounding chain of Section~\ref{subsec:block_BC_trees} to the case where blocks are uni-cyclic, and gives the proof of Theorem~\ref{thm:coal_block_BC} for the rapid coalescence of the block bounding chain in the general case.

\subsection{Block partition} 
\label{subsec:block_partition}
Let $\mathcal{B}=\{B_1,\ldots,B_N\}$ be a vertex partition of $V$. In the definitions below, we fix $\varepsilon\in(0,1)$.
For $B\in\mathcal{B}$, let $\partial_{\rm out}B$ be the set of vertices outside $B$ which have a neighbor inside the block.
We recall the vertex and path weighting scheme introduced in~\cite{Efthymiou_Zampetakis_2024}.
\paragraph{\it Vertex weight.}For any $u\in V$, define its ($d,\varepsilon$)-vertex weight by
\begin{align}
M(u)=\begin{cases}
1-\frac{\varepsilon}{4} &\text{ if } A(u)\leq 1-\frac{\varepsilon}{2}\\
d\cdot A(u) & \text{ otherwise.}
\end{cases}
\end{align}
\paragraph{\it Path weight.} For any path $P=(w_0,\ldots,w_\ell)$ in $G$, the ($d,\varepsilon$)-path weight is obtained by multiplying the weights of its vertices
\begin{align}
	M(P)=\prod_{i=0}^\ell M(w_i)
\end{align}
\paragraph{\it Block vertex.} A vertex $u$ in $G$ is called a ($d,\varepsilon$)-block vertex, if for every path $P$ of length at most $\log n$ emanating from $u$, the ($d,\varepsilon$)-weight of $P$ is less than $1$, i.e. $M(P)<1$.
\vspace{0.2cm}

From these definitions, we can see that any path connecting a block-vertex to a heavy vertex (with $A(v)>1$) must contain a large number of light vertices.
We recall now the properties of the block partition of Definition~4.1 in~\cite{Efthymiou_Zampetakis_2024}.
\begin{definition} [($d,\varepsilon$)-Block Partition~\cite{Efthymiou_Zampetakis_2024}] A vertex partition $\mathcal{B}=\{B_1,\dots,B_N\}$ is called a $(d,\varepsilon)$-block partition if for every $B\in\mathcal{B}$ the following is true:
	\begin{enumerate}
		\item block $B$ is a tree with at most one extra edge,
		\item if $B$ is a multi-vertex block, then we have the following:
		\begin{enumerate}
			\item every $w\in\partial_{\rm out} B$ is a $(d,\varepsilon)$-block vertex,
			\item every $w\in\partial_{\rm out} B$ has exactly one neighbor in $B$,
			\item if $B$ contains a cycle, this is a short one, i.e. its length is at most $4\frac{\log n}{\log^4 d}$,
			\item the distance of the short cycle from the boundary of its block is at least $\log^5 d$.
		\end{enumerate}
		\item if $B$ is a single-vertex block, then this vertex is ($d,\varepsilon$)-block.
	\end{enumerate}
\end{definition}
This construction guarantees that the heavy vertices are hidden deep inside the blocks, a property that is used in the path coupling argument.
Such a partition exists with high probability for Erd\H{o}s R\'{e}nyi random graphs $\mathcal G(n,d/n)$ and i.i.d. standard Gaussian couplings $\boldsymbol{J}$, for sufficiently large $d$. 
More precisely, for $\varepsilon>0$, there exists $d_0=d_0(\varepsilon)\geq 1$ such that for any $d\geq d_0$, and for any $0<\beta\leq (1-\varepsilon)\beta_c(d)$, $G$ admits a ($d,\varepsilon$)-block partition with probability at least $1-n^{-1/4}$ (see Theorem 4.2 in~\cite{Efthymiou_Zampetakis_2024}).

\subsection{Block Glauber dynamics}
\label{subsec:block_Glauber}
For a block $B\in\mathcal B$ and an exterior configuration $\eta\in\{+1,-1\}^{V\setminus B}$, we denote by $\mu_B^\eta$ the conditional Gibbs distribution on $\{+1,-1\}^B$
\begin{align}
\mu_B^\eta(\tau) &= \mu(\sigma(B)=\tau\mid \sigma(V\setminus B)=\eta) \ , \quad \tau\in\{+1,-1\}^B
\end{align}
The block Glauber dynamics is the Markov chain $(X_t)_{t\geq 0}$ on $\Omega = \{+1,-1\}^V$ whose transition from $X_{t}$ to $X_{t+1}$ is performed as follows:
\begin{enumerate}
	\item Select a block $B_{t}\in\mathcal{B}$ uniformly at random.	
	\item Keep the configuration outside $B_{t}$ unchanged: $X_{t+1}(V\setminus B_{t})=X_{t}(V\setminus B_{t})$.
	\item Sample the configuration on $B_{t}$ from the Gibbs distribution conditional on the configuration outside $B_{t}$: $X_{t+1}(B_{t})\sim \mu_{B_{t}}^{\,X_{t}(V\setminus B_{t})}$.
\end{enumerate}

\paragraph{\it Distance metric.}
The path-coupling proof of the rapid mixing of the block Glauber dynamics in~\cite{Efthymiou_Zampetakis_2024} is based on a weighted Hamming distance adapted to the block partition $\mathcal{B}$.
For each vertex $z\in V$ let $B_z$ be the block that $z$ belongs to, and define the aggregate influence of $z$ on its neighbors outside $B_z$
\begin{align}
A_{\rm out}(z)=\sum_{w\in N(z), \ w\notin B_z}\Gamma_{\{z,w\}}
\end{align}
Let $W\subseteq V$ be the subset of vertices $z\in V$ such that $A_{\rm out}(z)>0$. Since $\lvert J_e\rvert>0$ for each edge (see Theorem 4.2 in~\cite{Efthymiou_Zampetakis_2024}), $A_{\rm out}(z)=0$ implies that $z$ has no neighbors outside $B_z$. Define the metric used in~\cite{Efthymiou_Zampetakis_2024} for the path coupling argument. For any two configurations $\tau,\tau'\in\Omega$, let:
\begin{align}
\label{eq:distance_Efthymiou}
\operatorname{dist}(\tau,\tau')=\sum_{z\in V\setminus W}\mathbbm{1}_{\{\tau(z)\neq \tau'(z)\}} + n^4\sum_{z\in W}A_{\rm out}(z)\mathbbm{1}_{\{\tau(z)\neq \tau'(z)\}}
\end{align}
In this weighted Hamming metric, a disagreement at an internal vertex $z\notin W$ has weight $1$, whereas a disagreement at an external vertex $z\in W$ has weight $n^4A_{\rm out}(z)$.

\paragraph{\it Coupling construction.} 
Consider two copies of the block Glauber dynamics $(X_t)_{t\geq0}$ and $(Y_t)_{t\geq0}$. 
Suppose that at time $t\geq0$, $X_t$ and $Y_t$ differ at a single vertex $u^*$. 
We recall the following theorem, corresponding to Theorem~7.1 of~\cite{Efthymiou_Zampetakis_2024}.
\begin{theorem}~\cite{Efthymiou_Zampetakis_2024}
\label{thm:contraction}
	Suppose that $u^*\in W$. For any $B\in\mathcal{B}$ such that $u^*\in\partial_{\rm out}B$, there exists a coupling between $X_{t+1}$ and $Y_{t+1}$ such that 
	\begin{align}
		\mathbb{E}\left[\operatorname{dist}(X_{t+1},Y_{t+1})-\operatorname{dist}(X_{t},Y_{t})\mid X_t,Y_t, B_t=B\right]\leq \Gamma_{\{u^*,z\}}\cdot n^4\cdot (1-\varepsilon/6)
	\end{align}
	where $z\in B$ is the unique neighbor of $u^*$ in $B$.
\end{theorem}
This theorem controls the expected increase in the distance, when the disagreement at the exterior vertex $u^*$ propagates into the updated block $B$.
To construct the coupling when $B$ is updated, since the configurations outside $B$ are left unchanged, one only needs to couple the two conditional Gibbs distributions $\mu_B^{X_t(V\setminus B)}$ and $\mu_B^{Y_t(V\setminus B)}$.
These two boundary conditions differ only at $u^*$, which has a unique neighbor $z\in B$. 
One first uses a maximal coupling of the marginals at $z$. 
The probability of creating a disagreement at $z$ is therefore bounded by
\begin{align}
\mathbb{P}\left( X_{t+1}(z)\neq Y_{t+1}(z) \mid X_t,Y_t,B_{t}=B \right) \leq \Gamma_{{u^*,z}}.
\end{align}
If the two chains agree at $z$, their conditional distributions on $B\setminus{z}$ coincide, since $u^*$ interacts with $B$ only through $z$, and the identity coupling is used on the remainder of the block. 
If they disagree at $z$, the disagreement is propagated recursively away from $z$: on every tree branch, the conditional maximal coupling of Lemma~7.2 in~\cite{Efthymiou_Zampetakis_2024} is applied across successive edges. 
If $B$ is uni-cyclic and the disagreement reaches its unique cycle, the specialized coupling constructed in the proof of Lemma~7.8 is used across the cyclic part, after which the recursive coupling continues along the remaining tree branches. 

\subsection{Block bounding chain for tree blocks}
\label{subsec:block_BC_trees}
In the following two sections, we now define the block bounding chain constructed from the block partition, and prove its rapid coalescence below $\beta_c(d)$.
We assume that $\varepsilon\in(0,1)$, that $d\geq d_0(\varepsilon)$ and that $0<\beta<(1-\varepsilon)\beta_c(d)$ is fixed independently of $n$.

Suppose that every block $B\in\mathcal B$ is a tree. In this case, a gauge transformation maps the conditional Gibbs measure $\mu_B^\eta$ on each block to a ferromagnetic Ising measure, which allows us to construct a monotone coupling of the block Glauber update.

\subsubsection{Block bounding chain algorithm}
\label{subsubsec:block_BC_tree_construction}
Fix a block $B$, and choose once for all an arbitrary root $r_B$. 
For each vertex $v\in B$, we define a sign $g^B_v\in\{+1,-1\}$ recursively along the rooted tree as follows.
We set $g^B_{r_B}=+1$, and for any $w\in B\setminus\{r_B\}$ with parent $v$ on the rooted tree, we define $g^B_w=g^B_v\cdot \operatorname{sgn}\left(J_{\{v,w\}}\right)$.
Every boundary vertex $u\in\partial_{\rm out}B$ has only one neighbor inside $B$ (see definition 4.1 of~\cite{Efthymiou_Zampetakis_2024}), which we denote $z_B(u)$. 
We can therefore extend the gauge signs to $\partial_{\rm out} B$ by setting $g^B_u=g^B_{z_B(u)}\cdot \operatorname{sgn}\left(J_{\{u,z_B(u)\}}\right)$.
For a block configuration $\sigma\in\{+1,-1\}^B$ and a boundary configuration $\eta\in\{+1,-1\}^{\partial_{\rm out} B}$, we define the gauge transformed configurations $s(v)=g^B_v\sigma(v)$ for $v\in B$ and $b(u)=g^B_u\eta(u)$ for $u\in\partial_{\rm out} B$.
Up to terms independent of $\sigma$, the conditional Gibbs distribution on $B$ can be written as
\begin{align}
	\mu_B^\eta(\sigma)=\frac{1}{Z_B^\eta}\exp\left[\frac{\beta}{2}\left(\sum_{\{v,w\}\in E_B}J_{\{v,w\}}\sigma(v)\sigma(w)+\sum_{u\in\partial_{\rm out}B}J_{\{u,z_B(u)\}}\eta(u)\sigma(z_B(u))\right)\right] \ ,
\end{align} 
where $E_B$ is the edge set of the sub-graph induced by $B$. 
Note that by Theorem 4.2 in~\cite{Efthymiou_Zampetakis_2024}, each coupling is non-zero: $|J_e|>n^{-7/3}$ w.h.p.
Under the gauge transformation, all interactions are non-negative and this measure becomes a ferromagnetic Ising measure
\begin{align}
	\hat{\mu}_B^\eta(s)=\frac{1}{\widehat{Z}_B^b}\exp\left[\frac{\beta}{2}\left(\sum_{\{v,w\}\in E_B}\lvert J_{\{v,w\}}\rvert s(v)s(w)+\sum_{u\in\partial_{\rm out}B}\lvert J_{\{u,z_B(u)\}}\rvert b(u)s(z_B(u))\right)\right] \ .
\end{align}
One can therefore build a monotone sampler~\cite{Propp_Wilson_1996,Haggstrom_Nelander_1999} as follows.
We denote by $\preceq$ the partial order on $\{+1,-1\}^B$: $\sigma\preceq\sigma'$ when $\sigma(v)\leq\sigma'(v)$ for each $v\in B$ (and similarly on $\partial_{\rm out} B$). 
We define $U=(U_v)_{v\in B}$ with $U_v\sim \operatorname{Unif}([0,1])$ i.i.d. and construct the deterministic map
$F_{B,U}:\{+1,-1\}^{\partial_{\rm out} B}\longrightarrow \{+1,-1\}^B$ such that for every boundary condition $b\in\{+1,-1\}^{\partial_{\rm out} B}$, one has $F_{B,U}(b) \sim \hat{\mu}_B^b$, and for every fixed realization of $U$
\begin{align}
 b\preceq b' \implies F_{B,U}(b)\preceq F_{B,U}(b')
\end{align}
The map $F_{B,U}$ can be constructed recursively on the tree. We order the vertices in $B$ as $(v_0=r_B, v_1,\ldots,v_{|B|-1})$ in such a way that every parent precedes its children. 
\begin{enumerate}
	\item First, we compute the root marginal $\hat{m}_{v_0}(b)=\hat{\mu}_B^b(s(v_0)=+1)$ using the Belief-Propagation messages from the leaves to the root. 
	We set 
	\begin{align}
	F_{B,U}(b)(v_0)=\begin{cases}
	+1 & U_{v_0}\leq \hat{m}_{v_0}(b) \\
	-1 & U_{v_0}> \hat{m}_{v_0}(b) 	
	\end{cases}
	\end{align}
	\item The remaining spins are sampled recursively from the root towards the leaves. Once the spin of the parent $v_i$ has been sampled, we compute the conditional marginal $\hat{\mu}_{B}^b(s(v_i)=+1\mid s(v_0),\ldots,s(v_{i-1}))$, and use $U_{v_i}$ to sample $F_{B,U}(b)(v_i)$.
\end{enumerate}
We now define the bounding-chain update. One can define the two extreme gauge-transformed boundary conditions $b^{(+)}, b^{(-)}\in\{+1,-1\}^{\partial_{\rm out} B}$ by
\begin{align}
 b^{(\pm)}(u)=\begin{cases}
 g^B_u Y_t(u)  & \text{ if } Y_t(u)\neq \star \\
 \pm 1 & \text{ if } Y_t(u)=\star
 \end{cases}
\end{align}
Using the same random vector $U$ for both boundary conditions, set $s^{(+)}=F_{B,U}\left(b^{(+)}\right)$ and $s^{(-)}=F_{B,U}\left(b^{(-)}\right)$.
By monotonicity, if $\eta$ is a boundary configuration compatible with $Y_t$, and $b$ is its gauge transformation, then we have $b^{(-)}\preceq b\preceq b^{(+)}$, which implies that $s^{(-)}\preceq F_{B,U}(b)\preceq s^{(+)}$.
Therefore, the bounding chain update is
\begin{align}
 	Y_{t+1}(v)=\begin{cases}
 	Y_t(v) & \text{ if } v \notin B\\
	g^B_v s^{(+)}(v) & \text{ if } v\in B \text{ and } s^{(+)}(v)=s^{(-)}(v)\\
	\star & \text{ if }  v\in B \text{ and } s^{(+)}(v)\neq s^{(-)}(v)	
 	\end{cases}
\end{align}
With this definition, the bounding chain update simultaneously bounds all block Glauber updates associated with every boundary configuration compatible with $Y_t$. 
Note that once $Y_t$ has no $\star$-sites, then $b^{(-)}=b^{(+)}$ on each block, and no $\star$-site can be created on any block at time $t+1$: coalescence is achieved.
Note also that the block Glauber dynamics does not depend on the choice of the root $r_B$ and the vertex ordering of each block, but the monotone coupling and the resulting bounding chain may depend on these choices.
Composing these random bounding maps from the past yields a valid CFTP algorithm~\cite{Propp_Wilson_1996}, provided that the choice of the root and of the ordering of each block is fixed in advance and used consistently in each update of the CFTP algorithm.

\paragraph{\it Remark on the computation time of each block update.}
The two extreme boundary conditions $b^{(\pm)}$ can be computed in $O(|\partial_{\rm out}B|)$ operations, and $\hat{m}_{r_B}(b^{(\pm)})$, the marginals at the root $r_B$ can be computed with Belief Propagation in $O(|B|)$ operations.
For each remaining site $v_i$, the conditional marginals $\hat{\mu}_{B}^{b^{(\pm)}}(s(v_i)=+1\mid s(v_0),\ldots,s(v_{i-1}))$ depend only on the parent of $v_i$ on the tree, and can be computed in $O(1)$ operations.
Therefore, sampling the full block configuration $Y_{t+1}(B)$ can be done in $O(|\partial_{\rm out}B|+|B|)=O(n)$ operations.

\subsubsection{Rapid coalescence below $\beta_c(d)$}
We now prove that the block bounding chain defined above coalesces in $O(N\log n)$ steps with high probability. Since the number of blocks is $N\leq n$ and each block update requires $O(n)$ operations, this implies a $O(n^2\log n)$ coalescence time.

\begin{theorem}
	If every block in the block partition $\mathcal B$ is a tree, then the block bounding chain $(Y_t)_{t\geq0}$ defined in Section~\ref{subsubsec:block_BC_tree_construction}, built on $\mathcal B$ and started from $Y_0=\star^V$, satisfies for every $k>0$,
	\begin{align}
		\mathbb{P}\left[T_{\rm coal}>\frac{6N}{\varepsilon}\left((6+k)\log n + \log 2\right)\right]\leq n^{-k}
	\end{align}
	where $T_{\rm coal}=\inf\{t: Y_t\in\{+1,-1\}^V\}$ is the coalescence time of the bounding chain, and the probability is over the randomness of the chain. 
	In particular $T_{\rm coal}=O(N\log n)=O(n\log n)$, and the CFTP implementation of the block bounding chain returns an exact sample from $\mu_{G,\boldsymbol{J},\beta}$ in $O(n \log n)$ block updates with high probability.
\end{theorem}

\paragraph{\it Proof.} For a configuration $Y_t\in\Omega_\star$, we define the potential
\begin{align}
	\Phi(Y_t) = \sum_{u\in V} c(u) \mathbbm{1}_{\{Y_t(u)=\star\}} \text{ , with } c(u)=\begin{cases}
	1 & \text{ if } u\notin W\\
	n^4\cdot A_{\rm out}(u)& \text{ if } u\in W
	\end{cases}
\end{align}
More generally, let $\Phi_\Lambda(Y_t)=\sum_{u\in \Lambda} c(u) \mathbbm{1}_{\{Y_t(u)=\star\}}$ for any $\Lambda\subseteq V$. 
Note that $c(u)$ is precisely the weight assigned to a disagreement in the distance metric~\eqref{eq:distance_Efthymiou}.
This is the analogue, for the weighted metric distance~\eqref{eq:distance_Efthymiou}, of the number of unresolved spins used in the single-site bounding chain analysis in Section~\ref{sec:rapid_coalescence_Dobrushin}.
Here again, we will show that $\mathbb{E}[\Phi(Y_t)]$ decreases along the iterations.
Let $B\in\mathcal{B}$ be the block updated at time $t+1$, and define $\mathcal D_B(Y_t)=\{u\in\partial_{\rm out} B : Y_t(u)=\star\}$ as the set of unresolved variables at the boundary of $B$. Enumerate this set as $\mathcal{D}_B(Y_t)=(u_1,\dots u_k)$. 
Starting from the two extreme gauge-transformed boundary configurations $b^{(-)},b^{(+)}$, we build a monotone sequence of boundary configurations $\left(b^{(0)}=b^{(-)}, b^{(1)},\ldots,b^{(k)}=b^{(+)}\right)$ where $b^{(i-1)},b^{(i)}$ differ only on site $u_i$.
Note that each $u_i\in \partial_{\rm out} B \subset W$ is an external vertex.
For $v\in B$ and a gauge transformed boundary configuration $b\in\{+1,-1\}^{\partial_{\rm out}B}$, let $\hat{m}_v(b)=\hat{\mu}_B^b(s(v)=+1)$.
Recall that the two extreme block configurations are generated using the same random vector $U$: $s_t^{(-)}=F_{B,U}\left(b_t^{(-)}\right)$ and $s_t^{(+)}=F_{B,U}\left(b_t^{(+)}\right)$.
Taking expectations over $U$ and using the fact that the two maps $F_{B,U}\left(b^{(\pm)}\right)$ have the correct marginals gives, for every $v\in B$,
\begin{align}
	\begin{aligned}
 \mathbb{P}\left[Y_{t+1}(v)=\star\mid Y_t, B_t=B\right] &= \mathbb{P}\left[s^{(+)}(v)\neq s^{(-)}(v)\right] \\
 &=\mathbb{P}\left[\{s^{(+)}(v)=+1\} \setminus \{s^{(-)}(v)=+1\}\right]\\
 &= \hat{m}_v(b^{(+)}) - \hat{m}_v(b^{(-)})
	\end{aligned}
\end{align}
with $\{s^{(-)}(v)=+1\}\subseteq \{s^{(+)}(v)=+1\}$ by monotonicity of the block bounding chain.
By telescoping along the sequence of boundary configurations
\begin{align}
\mathbb{P}\left[Y_{t+1}(v)=\star\mid Y_t, B_t=B\right]=\sum_{i=1}^k \hat{m}_v(b^{(i)}) - \hat{m}_v(b^{(i-1)}) = \sum_{i=1}^k \lVert \hat{\mu}_{B,v}^{b^{(i)}} - \hat{\mu}_{B,v}^{b^{(i-1)}}\rVert_{\rm TV}
\end{align}
where each term in the sum of the first equality is non-negative by monotonicity of the ferromagnetic gauge-transformed measure, and where $\hat{\mu}_{B,v}^b$ denotes the marginal of $\hat{\mu}_B^b$ at $v$.
We now bound the weighted sum 
\begin{align}
\mathbb{E}\left[\Phi_B(Y_{t+1})\mid Y_t,B\right] = \sum_{v\in B}c(v)\mathbb{P}\left[Y_{t+1}(v)=\star\mid Y_t, B_t=B\right]= \sum_{i=1}^k\sum_{v\in B}c(v)\lVert \hat{\mu}_{B,v}^{b^{(i)}} - \hat{\mu}_{B,v}^{b^{(i-1)}}\rVert_{\rm TV}
\end{align}
For each $i\in\{1,\ldots,k\}$, we transform back $b^{(i-1)}, b^{(i)}$ to the original boundary configurations $\eta^{(i-1)},\eta^{(i)}$, which differ only at $u_i$.
We extend them to two configurations $X_t^{(i-1)},X_t^{(i)}$ on $\Omega$, that differ only at $u_i$ and agree everywhere else.
The two block update distributions for $X_t^{(i-1)}$ and $X_t^{(i)}$ are the gauge transforms of $\widehat{\mu}_B^{b^{(i-1)}}$ and $\widehat{\mu}_B^{b^{(i)}}$.
Let $Q_i$ be the coupling of these block updates provided by Theorem~7.1 of~\cite{Efthymiou_Zampetakis_2024}. 
The coupling inequality for total variation distance gives for each $v\in B$
\begin{align}
\lVert \hat{\mu}_{B,v}^{b^{(i)}} - \hat{\mu}_{B,v}^{b^{(i-1)}}\rVert_{\rm TV} = \lVert \mu_{B,v}^{\eta^{(i)}} - \mu_{B,v}^{\eta^{(i-1)}}\rVert_{\rm TV} \leq \mathbb{P}_{Q_i}\left(X_{t+1}^{(i)}(v)\neq X_{t+1}^{(i-1)}(v)\right)
\end{align}
Because $u_i\in\partial_{\rm out} B$, the initial disagreement at $u_i$ remains unchanged when $B$ is updated, and all other vertices outside $B$ agree in the two chains, therefore
\begin{align}
\operatorname{dist}\left(X_{t+1}^{(i)},X_{t+1}^{(i-1)}\right)-\operatorname{dist}\left(X_t^{(i)},X_t^{(i-1)}\right)=\sum_{v\in B}c(v)\mathbbm{1}_{\{X_{t+1}^{(i)}(v)\neq X_{t+1}^{(i-1)}(v)\}}.
\end{align}
Therefore, applying the Theorem 7.1 of~\cite{Efthymiou_Zampetakis_2024} gives
\begin{align}
\begin{aligned}
\sum_{v\in B}c(v)\left(\hat{m}_v(b^{(i)})-\hat{m}_v(b^{(i-1)})\right)&\leq \sum_{v\in B}c(v)\mathbb{P}_{Q_i}\left[X_{t+1}^{(i)}(v)\neq X_{t+1}^{(i-1)}(v)\mid X_t^{(i)}, X_t^{(i-1)}, B_t=B\right] \\
&=\mathbb{E}_{Q_i}\left[\operatorname{dist}(X_{t+1}^{(i)},X_{t+1}^{(i-1)}) - \operatorname{dist}(X_{t}^{(i)},X_{t}^{(i-1)}) \mid X_t^{(i)}, X_t^{(i-1)}, B_t=B\right] \\
&\leq \Gamma_{\{u_i, z_B(u_i)\}}n^4\left(1-\varepsilon/6\right)
\end{aligned}
\end{align}
where $z_B(u_i)$ is the unique neighbor of $u_i$ in $B$.
In turn 
\begin{align}
\label{eq:contraction_PhiB_tree_blocks}
\begin{aligned}
\mathbb{E}&\left[\Phi_B(Y_{t+1})\mid Y_t, B_t=B\right] =\sum_{v\in B}c(v)\mathbb{P}\left[Y_{t+1}(v)=\star\mid Y_t, B_t=B\right] \\
&\leq n^4\left(1-\varepsilon/6\right)\sum_{i=1}^k\Gamma_{\{u_i, z_B(u_i)\}} =n^4\left(1-\varepsilon/6\right)\sum_{u\in\partial_{\rm out} B} \Gamma_{\{u, z_B(u)\}}\mathbbm{1}_{\{Y_t(u)=\star\}}
\end{aligned}
\end{align}
When block $B$ is updated in the bounding chain update, all previous unresolved variables are overwritten, while variables outside $B$ remain unchanged, therefore
\begin{align}
\begin{aligned}
	\mathbb{E}\left[\Phi(Y_{t+1})\mid Y_t, B_t=B\right] &=\Phi(Y_t)-\Phi_B(Y_t) + \mathbb{E}\left[\Phi_B(Y_{t+1})\mid Y_t, B_t=B\right] \\
	&\leq \Phi(Y_t)-\Phi_B(Y_t) + n^4\left(1-\varepsilon/6\right)\sum_{u\in\partial_{\rm out} B} \Gamma_{\{u, z_B(u)\}}\mathbbm{1}_{\{Y_t(u)=\star\}}
\end{aligned}
\end{align}
Averaging over the uniformly chosen block $B$ gives
\begin{align}
	\mathbb{E}\left[\Phi(Y_{t+1})\mid Y_t\right] \leq \left(1-\frac{1}{N}\right)\Phi(Y_t) + \frac{1}{N}n^4\left(1-\varepsilon/6\right)\sum_{B\in\mathcal{B}}\sum_{u\in\partial_{\rm out}B}\Gamma_{\{u,z_B(u)\}}\mathbbm{1}_{\{Y_t(u)=\star\}}
\end{align}
We now reorganize the boundary contribution. 
Since for each $B$, any $u\in\partial_{\rm out} B$ has only one neighbor $z_B(u)$ in $B$, the last sum can be re-written 
\begin{align}
\begin{aligned}
n^4\sum_{B\in\mathcal{B}}\sum_{u\in\partial_{\rm out}B}\Gamma_{\{u,z_B(u)\}}\mathbbm{1}_{\{Y_t(u)=\star\}} &= n^4\sum_{u\in W}\mathbbm{1}_{\{Y_t(u)=\star\}}\sum_{v\in N(u) , v\notin B_u}\Gamma_{\{u,v\}}\\
&=\sum_{u\in W} n^4 A_{\rm out}(u)\mathbbm{1}_{\{Y_t(u)=\star\}} = \Phi_W(Y_t)
\end{aligned}
\end{align}
And we obtain the contraction inequality
\begin{align}
\label{eq:contraction_tree_blocks}
\mathbb{E}\left[\Phi(Y_{t+1})\mid Y_t\right]\leq \left(1-\frac{1}{N}\right)\Phi_{V\setminus W}(Y_t) + \left(1-\frac{\varepsilon}{6N}\right)\Phi_W(Y_t)\leq \left(1-\frac{\varepsilon}{6N}\right)\Phi(Y_t)
\end{align}
Starting the bounding chain from the completely unresolved configuration $Y_0=\star^V$, we have
\begin{align}
\Phi(Y_0)=\sum_{v\notin W} 1 + \sum_{v\in W}n^4 A_{\rm out}(v)\leq n + n^4\sum_{v\in V}\operatorname{deg}(v)\leq n + n^6 \leq 2n^6
\end{align}
Iterating~\eqref{eq:contraction_tree_blocks} we obtain
\begin{align}
 \mathbb{E}\left[\Phi(Y_t)\right]\leq \left(1-\frac{\varepsilon}{6N}\right)^t\Phi(Y_0)\leq 2n^6 e^{-\varepsilon t/6N}
\end{align}
Note that $\Phi(Y_t)$ is a non-negative random variable taking values either $0$ or $\geq 1$, indeed, for $u\notin W: c(u)=1$, and for $u\in W$, $c(u)= n^4A_{\rm out}(u) \geq  n^4 \operatorname{tanh}(\beta n^{-7/3}/2) \geq (\beta/4) n^{5/3} \geq  1$ for sufficiently large
$n$, using $\operatorname{tanh}x \geq x/2$ on $[0,1]$, and Theorem 4.2 in~\cite{Efthymiou_Zampetakis_2024}.
One obtains by Markov inequality
\begin{align}
\mathbb{P}\left[T_{\rm coal}>t\right]=\mathbb{P}\left[\Phi(Y_t)\geq 1\right]\leq \mathbb{E}\left[\Phi(Y_t)\right]\leq 2n^6e^{-\varepsilon t/6N}
\end{align}
For every $k> 0$, taking $t_k=\frac{6N}{\varepsilon}\left((6+k)\log n + \log 2\right)$ gives $\mathbb{P}\left[T_{\rm coal}> t_k\right]\leq n^{-k}$.
Therefore, for fixed $\varepsilon, \beta$ and $k$, the coalescence time of the block bounding chain is $T_{\rm coal}=O(N\log n)$ with probability at least $1-n^{-k}$.
\hfill$\square$

\subsection{Block bounding chain for uni-cyclic blocks}
\label{subsec:block_BC_unicyclic}

Suppose that some blocks in $\mathcal B$ are uni-cyclic. A gauge transformation makes all interactions within such a block non-negative if and only if its unique cycle is not frustrated.
When the cycle is frustrated, no gauge transformation on the original vertices can make all interactions non-negative. Nevertheless, one can first sample one distinguished vertex on the cycle, and then apply a gauge transformation to the forest obtained by removing such vertex.
For every uni-cyclic block, we choose once for all a distinguished vertex $w_B$ on its unique cycle (we use the shorthand notation $w_B=w$ whenever it is clear from the context).
This choice is kept fixed throughout the CFTP construction.
Note that by Definition 4.1 in~\cite{Efthymiou_Zampetakis_2024}, the cycle is at distance at least $\log^5 d$ from $\partial_{\rm out} B$. In particular $w\notin W$ is an internal vertex and $N(w) \subseteq B$.

\subsubsection{Block bounding chain algorithm}
Let $B$ be the block updated at time $t+1$, and assume its unique cycle is $C=(w=x_0,x_1,\dots,x_k)$.
Define the sign of the cycle by $\operatorname{sgn}(C)=\prod_{i=1}^{k+1}\operatorname{sgn}\left(J_{\{x_{i-1},x_i\}}\right)$, with $x_{k+1}=x_0=w$.
If $\operatorname{sgn}(C)=+1$, the cycle is not frustrated and a gauge transformation can be chosen so that every interaction in $B$ is non-negative.
The tree-block bounding chain described in Section~\ref{subsec:block_BC_trees} therefore applies without modification.
Hence, we focus on the case where the cycle is frustrated: $\operatorname{sgn}(C)=-1$.

\paragraph{\it Step 1. Sampling the variable at $w$.}
Let
\begin{align}
\label{eq:compatible_boundaries}
\mathcal{C}_{\partial_{\rm out}B}(Y_t) = \left\{ \eta\in\{-1,+1\}^{\partial_{\rm out}B}:	\eta(u)=Y_t(u)	\text{ whenever }Y_t(u)\neq\star \right\}
\end{align}
be the set of boundary configurations compatible with $Y_t$. 
For every $\eta\in\mathcal{C}_{\partial_{\rm out}B}(Y_t)$, define  $m_w(\eta)=\mu_{B}^\eta(\sigma(w)=+1)$. We introduce the two extreme marginal probabilities
\begin{align}
	p_w^{(+)}=\max_{\eta\in\mathcal C_{\partial_{\rm out} B}(Y_t)}m_w(\eta) \quad , \quad p_w^{(-)}=\min_{\eta\in\mathcal C_{\partial_{\rm out} B}(Y_t)}m_w(\eta)
\end{align}
Define the random variable $U=\left(U_v\right)_{v\in B}$ with $U_v\sim \operatorname{Unif}([0,1])$ i.i.d. for each $v\in B$.
For a fixed boundary configuration $\eta$ compatible with $Y_t$, the corresponding block update at $w$ sets $+1$ if $U_w\leq m_w(\eta)$, and $-1$ otherwise. 
The bounding chain update at $w$ is therefore defined by
\begin{align}
	Y_{t+1}(w)=\begin{cases}
	+1 & \text{ if } U_w\leq p_w^{(-)} \\
	\star & \text{ if } p_w^{(-)} < U_w\leq p_w^{(+)} \\
	-1 & \text{ if } U_w > p_w^{(+)}
	\end{cases}
\end{align}
{\it Remark:} A simple strategy allows to compute $p_w^{(\pm)}$ in $O(n)$ operations, see Appendix~\ref{app:sampling_at_w}. 

\paragraph{\it Step 2. Gauge transformation conditional on the spin at $w$.}
The bounding update on the remaining vertices $B\setminus\{w\}$ must simultaneously contain the conditional updates associated with every compatible pair $(\eta,\eta(w))$ where $\eta\in\mathcal{C}_{\partial_{\rm out} B}(Y_t)$, and where $\eta(w)$ is sampled from $m_w(\eta)$ and $U_w$.
Since $B$ is uni-cyclic and $w\in C$, the graph induced by $T=B\setminus\{w\}$ is a forest.
A separate gauge transform can therefore be introduced on each of its connected components, after treating the incidences with $w$ as boundary interactions. 
For each vertex $v\in N(w)$, we introduce a fictive vertex $w_v$, connected only to $v$ and carrying coupling $J_{\{w,v\}}$.
In other words, the edge $\{w,v\}$ is replaced, for the purpose of sampling on $T$, by the boundary edge $\{w_v,v\}$.
We use the shorthand notation $w_1=w_{x_1}$ and $w_k=w_{x_k}$.
Let $T_C$ be the connected component of $T$ containing the path $x_1,x_2,\ldots,x_k$.
The augmented component $T_C\cup\{w_1,w_k\}$ is also a tree, and we root it at $w_1$.
For every $v\in N(w)\setminus\{x_1,x_k\}$, let $T_v$ be the connected component of $T$ containing $v$. The augmented tree $T_v\cup\{w_v\}$ is rooted at $w_v$.

Fix an exterior boundary configuration $\eta\in\mathcal{C}_{\partial_{\rm out}B}(Y_t)$, and a given value $\eta(w)\in\{+1,-1\}$ for the spin at $w$.
Under the constraint that $\eta(w)=\eta(w_v)$ for each $v\in N(w)$, the conditional distribution $\mu_{T}^\eta$ can be written, for $\sigma\in\{+1,-1\}^T$
\begin{align}
\mu^\eta_{T}(\sigma)=\frac{e^{\frac{\beta}{2}\left[\sum_{\{u,v\}\in E_{T}}J_{\{u,v\}}\sigma(u)\sigma(v)+\sum_{u\in\partial_{\rm out} B}J_{\{u,z_B(u)\} } \sigma(z_B(u))\eta(u)+ \sum_{v\in N(w)}J_{\{v,w\}}\eta(w_v)\sigma(v)\right]}}{Z^\eta_{T}}
\end{align}
where $E_T$ is the edge-set of the forest induced by $T$.
We now define the gauge signs.
On the component $T_C$, we set $g_{w_1}=+1$ and propagate the signs from $w_1$ along the tree. 
More precisely, if $q$ is a child of $p$, we set $g_q=g_p\operatorname{sgn}\left(J_{\{p,q\}}\right)$.
Since the cycle $C$ is frustrated, one obtains $g_{w_k}=g_{w_1}\operatorname{sgn}(C)=-1$.
On every other component $T_v$ with $v\in N(w)\setminus\{x_1,x_k\}$, we set $g_{w_v}=+1$ and propagate the signs from $w_v$.
Finally, for every $u\in\partial_{\rm out} B$, we define $g_u=g_{z_B(u)}\operatorname{sgn}\left(J_{\{u,z_B(u)\}}\right)$.

For $\sigma\in\{+1,-1\}^T$, we introduce the gauge-transformed spins $s(v)=g_v\sigma(v)$ for $v\in T$. 
The transformed boundary conditions are $b(u)=g_u\eta(u)$ for each $u\in\partial_{\rm out}B$ and $b(w_v)=g_{w_v}\eta(w)$ for each $v\in N(w)$.
The gauge-transformed conditional distribution, denoted $\hat{\mu}_T^b$	 have only ferromagnetic interactions
\begin{align}
	\hat{\mu}_{T}^b(s)=\frac{e^{\frac{\beta}{2}\left[\sum_{\{u,v\}\in E_{T}}\lvert J_{\{u,v\}}\rvert s(u)s(v)+\sum_{u\in\partial_{\rm out} B}\lvert J_{\{u,z_B(u)\} }\rvert  s(z_B(u))b(u)+ \sum_{v\in N(w)}\lvert J_{\{v,w\}}\rvert b(w_v)s(v)\right]}}{\widehat{Z}^b_{T}}
\end{align}
The equality constraint $\eta(w_v)=\eta(w)$ for each $v\in N(w)$ in the original spin variables now becomes $b(w_v)=\eta(w)$ for $v\in N(w)\setminus \{x_k\}$ and $b(w_k)=-\eta(w)$.

\paragraph{\it Step 3. Monotone samplers.} 
Conditional on the value assigned to $w$, the connected components of $T=B\setminus\{w\}$ are independent, we can therefore build for each of them a monotone sampler.
For each $v\in N(w)\setminus\{x_1,x_k\}$, define $\partial_{\rm out}T_v=\{u\in \partial_{\rm out}B : z_B(u)\in T_v\}$ as the set of vertices in $\partial_{\rm out} B$ having one neighbor in $T_v$. 
Similarly, define $\partial_{\rm out}T_C=\{u\in \partial_{\rm out}B : z_B(u)\in T_C\}$.
For each component $T_v$, let $U_{T_v}=\left(U_q\right)_{q\in T_v}$ and define the deterministic map
$G_{v,U_{T_v}}: \{+1,-1\}^{\partial_{\rm out}T_v\cup \{w_v\}}\longrightarrow\{+1,-1\}^{T_v}$.
The map is constructed recursively on the rooted tree $T_v$: first sample the marginal at $v$, using $U_v$ and the marginal $\hat{m}_v(b)=\hat{\mu}_{T_v}^b(s(v)=+1)$. Then sample its descendants recursively, using the random vector $U_{T_v}$ and the conditional marginals $\hat{\mu}_{T_v}^b(s(v_i)=+1\mid s(v_0),\dots,s(v_{i-1}))$.
We define a similar map $G_{C,U_{T_C}}:\{+1,-1\}^{\partial_{\rm out}T_C\cup\{w_1,w_k\}}\longrightarrow\{+1,-1\}^{T_C}$ for $T_C$.
For each $u\in\partial_{\rm out} B$, we define the boundary spins
\begin{align}
\label{eq:boundary_spins}
b_{Y_t}^{(+)}(u)=\begin{cases}
g_uY_{t}(u) & \text{ if } Y_{t}(u)\neq \star \\
+ 1 & \text{ if } Y_t(u)=\star 	
\end{cases} \quad , \quad  
b_{Y_t}^{(-)}(u)=\begin{cases}
g_uY_{t}(u) & \text{ if } Y_{t}(u)\neq \star \\
- 1 & \text{ if } Y_t(u)=\star 	
\end{cases}
\end{align}

{\it Step 3(a). Block bounding chain for the off-cycle components.}
Let $v\in N(w)\setminus\{x_1,x_k\}$, and define the two extreme boundary configurations $b_v^{(\pm)}$ on $\partial_{\rm out} T_v\cup\{w_v\}$ by
\begin{align}
b_v^{(\pm)}(w_v)=\begin{cases}
	Y_{t+1}(w) & \text{ if } Y_{t+1}(w)\neq \star \\
	\pm 1 & \text{ if } Y_{t+1}(w)=\star	
	\end{cases}
\end{align}
and for each $u\in \partial_{\rm out}T_v$ by $b_v^{(\pm)}(u)=b_{Y_t}^{(\pm)}(u)$ as defined in equation~(\ref{eq:boundary_spins}).
We compute $s_v^{(\pm)}=G_{v,U_{T_v}}\left(b_v^{(\pm)}\right)$.
If $\eta$ is a configuration on $\{+1,-1\}^{\partial_{\rm out}T_v\cup\{w_v\}}$, compatible with $Y_t(\partial_{\rm out} T_v)$ and $Y_{t+1}(w)$, then its gauge transformation is such that $b_v^{(-)}\preceq b\preceq b_v^{(+)}$. 
By monotonicity, $s_v^{(-)}\preceq G_{v,U_{T_v}}(b)\preceq s_v^{(+)}$.
We can therefore define the bounding chain update on each $q\in T_v$ as
\begin{align}
Y_{t+1}(q)=\begin{cases}
g_q s_v^{(+)}(q) & \text{ if } s_v^{(+)}(q)=s_v^{(-)}(q)\\
\star & \text{ if }  s_v^{(+)}(q)\neq s_v^{(-)}(q)	
\end{cases}
\end{align}

{\it Step 3(b). Block bounding chain on the cycle component $T_C$.}
In order to define the two extreme boundary configurations, we need to distinguish the two cases depending on whether $Y_{t+1}(w)$ is unresolved or not.
\begin{itemize}
	\item[--] If $Y_{t+1}(w)=x\in\{+1,-1\}$, then we define the two extreme boundary conditions $b_x^{(\pm)}$ on $\{+1,-1\}^{\partial_{\rm out}T_C\cup\{w_1,w_k\}}$ by $b_x^{(\pm)}(w_1)=x$ and $b^{(\pm)}_x(w_k)=-x$, and on each $u\in\partial_{\rm out}T_C$ by
	$b_x^{(\pm)}(u)=b_{Y_t}^{(\pm)}(u)$.
	\item[--] If $Y_{t+1}(w)=\star$, then both configurations $(b(w_1),b(w_k))=(+1,-1)$ and $(b(w_1),b(w_k))=(-1,+1)$ are compatible with $Y_{t+1}(w)$, and they are not comparable under the order $\preceq$.
	More precisely, the set of gauge-transformed compatible boundary configurations on $\partial_{\rm out} T_C\cup\{w_1,w_k\}$ is either bounded by $(b_+^{(-)},b_+^{(+)})$ or by $(b_-^{(-)},b_-^{(+)})$, where $b_x^{(\pm)}$ are the extreme configurations defined in the resolved case (i.e. with $b_x^{(\pm)}(w_1)=x$ and $b_x^{(\pm)}(w_k)=-x$).
	However, every compatible gauge-transformed boundary configurations is bounded by the artificial interval defined by $b_\star^{(+)},b_\star^{(-)}$ with
	\begin{align}
	b_\star^{(+)}(w_1)=b_\star^{(+)}(w_k)=+1 \text{ ,  }
		b_\star^{(-)}(w_1)=b_\star^{(-)}(w_k)=-1
	\end{align}
	and for each $u\in\partial_{\rm out} T_C: b_\star^{(\pm)}(u)=b_{Y_t}^{(\pm)}(u)$.
	Although $b_\star^{(-)}$ and $b_\star^{(+)}$ do not satisfy the frustration constraint, they are legitimate boundary conditions for the gauge-transformed ferromagnetic measure.
	
\end{itemize}
Let $Y_{t+1}(w)=y\in\{+1,-1,\star\}$.
We compute $s_y^{(\pm)}=G_{C,U_C}\left(b_y^{(\pm)}\right)$, and the bounding chain update for every $q\in T_C$ is
\begin{align}
Y_{t+1}(q)=\begin{cases}
g_q s_y^{(+)}(q) & \text{ if } s_y^{(+)}(q)=s_y^{(-)}(q)\\
\star & \text{ if } s_y^{(+)}(q)\neq s_y^{(-)}(q)	
\end{cases}
\end{align}

\paragraph{\it Remark on the computation time of each block update.} 
Sampling at the first vertex $w$ can be done in $O(n)$ operations (see Appendix~\ref{app:sampling_at_w}). In order to sample on the remaining sites, we need to compute the Belief Propagation marginals at each $w_v\in N(w)\setminus\{x_0,x_k\}$ with the two extreme boundary conditions $b_v^{(\pm)}$ and at $w_0,w_k$ with $b_y^{(\pm)}$, $y\in\{+1,-1,\star\}$, which can be done in $O(|B|)$ operations.
Computing the extreme boundary configurations can be made in $O(|\partial_{\rm out} B|)$ operations.
Sampling each descendant on each tree can be done in $O(1)$ operations by computing the marginal of each child spin conditioned on the value of its parent.
Therefore, updating the block $B$ can be done in $O(n)$ operations.

\subsubsection{Rapid coalescence below $\beta_c(d)$}
In this Section, we now prove the Theorem~\ref{thm:coal_block_BC} for the rapid coalescence of the block bounding chain.

\paragraph{\it Proof.} Let $r_w=p_w^{(+)}-p_w^{(-)}$ and define for $y\in\{+1,-1,\star\}$
\begin{align}
D_y=\sum_{v\in T}c(v)\mathbb{P}\left[Y_{t+1}(v)=\star\mid Y_{t+1}(w)=y, Y_t, B_{t}=B\right]
\end{align}
with $T=B\setminus \{w\}$. Then we can write
\begin{align}
 \mathbb{E}\left[\Phi_B(Y_{t+1})\mid Y_t, B_{t}=B\right] = r_w c(w) + p_w^{(-)}D_+ + \left(1- p_w^{(+)}\right)D_- + r_w D_\star
\end{align}
and since $w$ belongs to the cycle, which by Definition 4.1 in~\cite{Efthymiou_Zampetakis_2024} of the block partition, is at distance at least $\log^5 d$ from the boundary, we have $w\notin W$ and therefore $c(w)=1$. We need to upper bound each of these terms.

\paragraph{\it Step 1. Bounding $D_x$, with $x\in\{+1,-1\}$.}
Defining $\hat{m}_v(b)=\hat{\mu}_T^b(s(v)=+1)$, we have, by monotonicity of the block bounding chain, $\{s_v^{(-)}(q)=+1\}\subseteq \{s_v^{(+)}(q)=+1\}$ for each $v\in N(w)\setminus \{x_1,x_k\}, q\in T_v$, and $\{s_x^{(-)}(q)=+1\}\subseteq \{s_x^{(+)}(q)=+1\}$ for each $q\in T_C$, and therefore
\begin{align}
\begin{aligned}
D_x &= \sum_{v\in T}c(v)\mathbb{P}\left[Y_{t+1}(v)=\star\mid Y_{t+1}(w)=x, Y_t, B_{t}=B\right]\\
&=\sum_{v\in N(w)\setminus\{x_1,x_k\}}\sum_{q\in T_v}c(q)\mathbb{P}\left[s_v^{(+)}(q)=+1,s_v^{(-)}(q)=-1\mid Y_{t+1}(w)=x, Y_t, B_{t}=B\right] \\
&+ \sum_{q\in T_C}c(q)\mathbb{P}\left[s_x^{(+)}(q)=+1,s_x^{(-)}(q)=-1\mid Y_{t+1}(w)=x, Y_t, B_{t}=B\right]\\
&=\sum_{v\in N(w)\setminus\{x_1,x_k\}}\sum_{q\in T_v}c(q)\left(\hat{m}_q(b_v^{(+)})-\hat{m}_q(b_v^{(-)})\right) + \sum_{q\in T_C}c(q)\left(\hat{m}_q(b_x^{(+)})-\hat{m}_q(b_x^{(-)})\right)
\end{aligned}
\end{align}
where $b_x^{(\pm)}$ and $b_v^{(\pm)}$ are the boundary configurations defined in the block bounding chain above. 
Since $Y_{t+1}(w)=x\neq \star$ is resolved, $b_x^{(+)}$ and $b_x^{(-)}$ differ only on the vertices $u\in\partial_{\rm out} T_C$ such that $Y_t(u)=\star$.
Similarly, for each $v\in N(w)\setminus\{x_1,x_k\}$, $b_v^{(+)}$ and $b_v^{(-)}$ differ only on the unresolved vertices of $\partial_{\rm out} T_v$.

For each component $K\in\{\{T_v\}_{v\in N(w)\setminus\{x_1,x_k\}}, T_C\}$ of the forest $T$, we enumerate the unresolved boundary vertices $(u_{K,1},\dots,u_{K,\ell_K})=\{u\in\partial_{\rm out} K : Y_t(u)=\star\}$, and build a path of configurations $(b_{K,x}^{(0)},b_{K,x}^{(1)},\ldots,b_{K,x}^{(\ell_K)})$ such that $	b_{T_C,x}^{(0)}=b_x^{(-)}$, $b_{T_C,x}^{(\ell_{T_C})}=b_x^{(+)}$  and $\forall v\in N(w)\setminus\{x_1,x_k\}$:  $b_{T_v,x}^{(0)}=b_v^{(-)}$  and $b_{T_v,x}^{(\ell_{T_v})}=b_v^{(+)}$, and such that $b_{K,x}^{(i-1)}, b_{K,x}^{(i)}$ differ only on $u_{K,i}$.
One can write by telescoping the sums
\begin{align}
D_x &= \sum_{K}\sum_{i=1}^{\ell_K}\sum_{q\in K}c(q)\left(\hat{m}_q(b_{K,x}^{(i)})-\hat{m}_q(b_{K,x}^{(i-1)})\right) = \sum_{K}\sum_{i=1}^{\ell_K}\sum_{q\in K}c(q)\lVert \hat{\mu}_{T,q}^{b_{K,x}^{(i)}}-\hat{\mu}_{T,q}^{b_{K,x}^{(i-1)}}\rVert_{\rm TV} 
\end{align}
where each term in the sums are non-negative by monotonicity of the ferromagnetic gauge transformed measure.
Let $K$ be a component of $T$, and $u_{K,i}$ be a unresolved boundary vertex of $\partial_{\rm out} K$.
Transforming back $b_{K,x}^{(i-1)}, b_{K,x}^{(i)}$ to the original spin coordinates produces two Gibbs measures on $K$, whose boundary conditions agree everywhere except at $u_{K,i}$, while all fictitious copies of $w$ have the same fixed values in both measures.
Applying the recursive coupling used in the proof of Theorem 7.1 of~\cite{Efthymiou_Zampetakis_2024}, the initial disagreement enters $K$ through the edge $\{u_{K,i},z_B(u_{K,i})\}$, and reaches $z_B(u_{K,i})$ with probability at most $\Gamma_{\{u_{K,i},z_B(u_{K,i})\}}$. 
Conditional on a disagreement at parent $p$ in $K$ (with $z_B(u_{K,i})$ considered as the root), the disagreement propagates to child $q$ with probability at most $\Gamma_{\{p,q\}}$. 
Since $K$ is a tree, there is a unique path from $z_B(u_{K,i})$ to every vertex of $K$. 
The recursive decomposition and the path-weight estimates are therefore the same as those used in Lemmas~7.5 and~7.6 of~\cite{Efthymiou_Zampetakis_2024}, with every sum restricted to vertices of $K$.
More precisely, Corollaries~7.3 and~7.4, used in the proofs of Lemmas~7.5,7.6 hold for any path from $u_{K,i}$ to $v\in K$.
The set $\hat{T}$ in Lemma~7.5 of vertices at distance at most $\log d$ from $u_{K,i}$ is included in $K$, so Lemma~7.5 applies directly and gives
\begin{align}
\sum_{q\in \hat{T}\cap W}c(q)\lVert \hat{\mu}_{T,q}^{b_{K,x}^{(i)}}-\hat{\mu}_{T,q}^{b_{K,x}^{(i-1)}}\rVert_{\rm TV} \leq \Gamma_{\{u_{K,i},z_B(u_{K,i})\}}n^4(1-\varepsilon/2)
\end{align}
For $K=T_C$, the situation is simpler than in~\cite{Efthymiou_Zampetakis_2024}, since once $w$ is pinned, the remaining vertices of the cycle $x_1,\dots,x_k$ belong the the tree $T_C$, and the recursion relation (35) in Lemma 7.6 runs over all unique paths in $K$ (Lemma 7.8 and proposition 7.7 treating the unicyclic case are not needed).
The additional pinned boundary spins are identical in the two measures and hence cannot create disagreements. 
Furthermore, a disagreement reaching a fictitious copy of $w$ terminates there: it cannot cross through $w$ into another component. 
Therefore, pinning $w$ only removes non-negative terms from the recursive upper bound in Lemma~7.6, and: 
\begin{align}
\sum_{q\in (K\setminus \hat{T})\cap W}c(q)\lVert \hat{\mu}_{T,q}^{b_{K,x}^{(i)}}-\hat{\mu}_{T,q}^{b_{K,x}^{(i-1)}}\rVert_{\rm TV} \leq \Gamma_{\{u_{K,i},z_B(u_{K,i})\}}n^4d^{-\varepsilon/5}		
\end{align}
Finally, the contribution of internal sites can be bounded by $\Gamma_{\{u_{K,i},z_B(u_{K,i})\}}n$, leading to 
\begin{align}
\begin{aligned}
\sum_{q\in K}c(q)\lVert \hat{\mu}_{T,q}^{b_{K,x}^{(i)}}-\hat{\mu}_{T,q}^{b_{K,x}^{(i-1)}}\rVert_{\rm TV} &\leq \Gamma_{\{u_{K,i},z_B(u_{K,i})\}}\left(n^4(1-\varepsilon/2) +n^4d^{-\varepsilon/5} + n \right)\\
&\leq \Gamma_{\{u_{K,i},z_B(u_{K,i})\}}n^4\left(1-\varepsilon/6\right)
\end{aligned}
\end{align}
for $d\geq d_0(\varepsilon), n\geq n_0(\varepsilon)$.
Applying this inequality to each unresolved boundary vertex, we obtain (since each $u\in\partial_{\rm out}B$ has exactly one neighbor in $B$)
\begin{align}
D_x \leq n^4(1-\varepsilon/6)\sum_{u\in\partial_{\rm out}B}\Gamma_{\{u,z_B(u)\}}\mathbbm{1}_{\{Y_t(u)=\star\}}
\end{align} 

\paragraph{\it Step 2. Bounding $D_\star$.}
We have
\begin{align}
\begin{aligned}
D_\star &= \sum_{v\in T}c(v)\mathbb{P}\left[Y_{t+1}(v)=\star\mid Y_{t+1}(w)=\star, Y_t, B_{t}=B\right] \\
&= \sum_{v\in N(w)\setminus\{x_1,x_k\}}\sum_{q\in T_v} c(q)\left(\hat{m}_q(b_{v}^{(+)}) - \hat{m}_q(b_{v}^{(-)})\right) + \sum_{q\in T_C}c(q)\left(\hat{m}_q(b_\star^{(+)}) - \hat{m}_q(b_\star^{(-)})\right)
\end{aligned}
\end{align}
For each $v\in N(w)\setminus\{x_1,x_k\}$, the boundary configurations $b_v^{(\pm)}$ differ on the vertices $\{u\in\partial_{\rm out} T_v: Y_t(u)=\star\}$ and on $w_v$.
The boundary configurations $b_\star^{(\pm)}$ differ on $\{u\in\partial_{\rm out} T_C: Y_t(u)=\star\}$ and on $w_1,w_k$.
For the off-cycle components $T_v$, $v\in N(w)\setminus\{x_1,x_k\}$, we can define the following path of nearby configurations 
\begin{align}
(b_{T_v,-}^{(0)},b_{T_v,+}^{(0)},b_{T_v,+}^{(1)}\ldots,b_{T_v,+}^{(\ell_{T_v})})
\end{align}
in such a way that $b_{T_v,-}^{(0)}=b_v^{(-)}$, that $b_{T_v,+}^{(0)}$ and $b_{T_v,-}^{(0)}$ differ only on $w_v$, that for each $i\in\{1,\ldots,\ell_{T_v}\}$, $b_{T_v,+}^{(i-1)}$ and $b_{T_v,+}^{(i)}$ differ only on $u_{T_v,i}$, and that $b_{T_v,+}^{(\ell_{T_v})}=b_v^{(+)}$.
For the contribution of $T_C$, we define the path of nearby configurations
\begin{align}
(b_\star^{(-)},b_{T_C,+}^{(0)},\dots,b_{T_C,+}^{(\ell_{T_C})},b_\star^{(+)})
\end{align}
in such a way that $b_\star^{(-)}$ and $b_{T_C,+}^{(0)}$ differ only on $w_1$, that $b_{T_C,+}^{(\ell_{T_C})}$ and $b_\star^{(+)}$ differ only on $w_k$, and that for each $i\in\{1,\dots,\ell_{T_C}\}$, $b_{T_C,+}^{(i-1)}$ and $b_{T_C,+}^{(i)}$ differ only on $u_{T_C,i}$.
In particular, for each component $K$, the intermediate segment from $b_{K,+}^{(0)}$ to $b_{K,+}^{(\ell_{K})}$ corresponds to the path of boundary configurations built in {\it Step 1}, when fixing the original spin at $w$ to $x=+1$.
Telescoping the sums, we get
\begin{align}
\label{eq:D_star_decomposition}
\begin{aligned}
D_\star &= \sum_K \sum_{i=1}^{\ell_K}\sum_{q\in K}c(q)\left(\hat{m}_q(b_{K,+}^{(i)}) - \hat{m}_q(b_{K,+}^{(i-1)})\right) \\
&+ \sum_{v\in N(w)\setminus\{x_1,x_k\}}\sum_{q\in T_v}c(q) \left(\hat{m}_q(b_{T_v,+}^{(0)}) - \hat{m}_q(b_{T_v,-}^{(0)})\right)\\
&+\sum_{q\in T_C}c(q)\left[\left(\hat{m}_q(b_\star^{(+)}) - \hat{m}_q(b_{T_C,+}^{(\ell_{T_C})})\right) + \left(\hat{m}_q(b_{T_C,+}^{(0)}) - \hat{m}_q(b_\star^{(-)})\right)\right]
\end{aligned}
\end{align}
We recognize $D_+$ in the first term, which we have already bounded in {\it Step 1.}
In the second term, $b_{T_v,-}^{(0)}$ and $b_{T_v,+}^{(0)}$ are two gauge-transformed boundary configurations on $\partial_{\rm out}T_v\cup\{w_v\}$, differing only on $w_v$
In the last terms, $b_\star^{(+)}$ and $b_{T_C,+}^{(\ell_{T_C})}$ are two gauge-transformed boundary configurations on $\partial_{\rm out}T_C\cup\{w_1,w_k\}$ differing only on $w_k$, while $b_{T_C,+}^{(0)}$ and $b_\star^{(-)}$ differ only on $w_1$.

Let $K$ be a component of $T$, and let $v$ be a neighbor of $w$ in $K$.
Let $b, b'$ be two gauge-transformed boundary configurations that differ only on $w_v$. Transforming back $b,b'$ to the original spin coordinates produces two Gibbs measures on $K$, whose boundary conditions $\eta,\eta'$ agree everywhere except on $w_v$.
We use the following recursive coupling $Q_v$ between $\mu_K^{\eta}, \mu_K^{\eta'}$, on the tree $K$, starting at $v$. 
By maximally coupling at $v$, we obtain a disagreement at $v$ with probability at most $\Gamma_{\{w,v\}}$.
Along the recursion, if the two coupled spins agree at a given vertex, we use the identity coupling in every descendant sub-tree. If they disagree, we use maximal coupling for its children.
Therefore, a disagreement at any $q\in K$ requires a propagation of disagreements along the unique path from $v$ to $q$.
For the internal vertices $q\in K\setminus W$, we have $c(q)=1$ and we use the simple bound
\begin{align}
\sum_{q\in K\setminus W}\mathbb{P}_{Q_v}\left[\sigma(q)\neq \sigma'(q)\right]\leq |K|\Gamma_{\{w,v\}}\leq n\Gamma_{\{w,v\}}
\end{align}
For the contribution of the external vertices $K\cap W$, we define for each $p\in K\cup\{w_v\}$, and for each of its children $q$ on the rooted tree $K$, the quantity
\begin{align}
S_p(K_q)=\sum_{x\in K_q\cap W}c(x)\mathbb{P}_{Q_v}\left[\sigma(x)\neq \sigma'(x)\mid \sigma(p)\neq \sigma'(p)\right]
\end{align}
where $K_q$ is the sub-tree of $K$ rooted at $q$.
We can write a recursion on $S_{w_v}(K)$ as it is done in~\cite{Efthymiou_Zampetakis_2024}:
\begin{align}
\begin{aligned}
S_{w_v}(K)&\leq \Gamma_{\{w,v\}}\left(n^4A_{\rm out}(v)+ \sum_{x\in N(v)\cap K}\Gamma_{\{v,x\}}\left(n^4A_{\rm out}(x)+\sum_{y\in N(x)\cap K\setminus\{v\}}S_x(K_y)\right)\right)\\
&\leq \Gamma_{\{w,v\}}\left(n^4A_{\rm out}(v)+ \left[A(v)-A_{\rm out}(v)\right]\max_{x\in N(v)\cap K}\left(n^4A_{\rm out}(x)+\sum_{y\in N(x)\cap K\setminus\{v\}}S_x(K_y)\right)\right)
\end{aligned}
\end{align}
We can repeat the above steps, as in the proof of Lemma 7.6 in~\cite{Efthymiou_Zampetakis_2024}, and find a worst-case path in $K$ emanating from $v$, $P=(q_0=v,\ldots,q_\ell)$, such that
\begin{align}
S_{w_v}(K)\leq \Gamma_{\{w,v\}}n^4\left(\sum_{j=0}^\ell A_{\rm out}(q_j)\prod_{i=0}^{j-1}\left[A(q_i)-A_{\rm out}(q_i)\right]\right)
\end{align}
Let $L_v$ be the distance from $v$ to the closest vertex of $K$ having a neighbor outside $B$.
Since $v$ either belongs to the cycle or is adjacent to it, whereas any vertex $q$ such that $A_{\rm out}(q)>0$ has a neighbor in $\partial_{\rm out}B$, the block partition properties imply that $L_v\geq \log^5(d)-2$.
All the terms with $j<L_v$ vanish in the above sum, and hence
\begin{align}
S_{w_v}(K)\leq\Gamma_{\{w,v\}}n^4\sum_{j=L_v}^{\ell}A_{\rm out}(q_j)\prod_{i=0}^{j-1}A(q_i).
\end{align}
We now bound each non-zero term separately. Fix $j\geq L_v$ such that $A_{\rm out}(q_j)>0$. There exists a vertex $u_j\in\partial_{\rm out}B$ adjacent to $q_j$. Consider the path $(v=q_0,\ldots,q_j, u_j)$ and let $h_j$ be the number of heavy vertices on this path.
We apply Corollary 7.3 in~\cite{Efthymiou_Zampetakis_2024} to obtain
\begin{align}
 A_{\rm out}(q_j)\prod_{i=0}^{j-1}A(q_i)\leq \prod_{i=0}^{j}A(q_i) \leq \left(\frac{1-\varepsilon/2}{1-\varepsilon/4}\right)^{j+1-h_j}d^{-h_j}\leq \left(1-\frac{\varepsilon}{4}\right)^{j+1-h_j}d^{-h_j}\leq \left(1-\frac{\varepsilon}{4}\right)^{j+1}
\end{align}
where the last inequality holds for sufficiently large $d$.
We obtain finally, using $L_v\geq \log^5(d)-2$,
\begin{align}
S_{w_v}(K)\leq\Gamma_{\{w,v\}}n^4\sum_{j=L_v}^{+\infty}\left(1-\frac{\varepsilon}{4}\right)^{j+1}\leq\Gamma_{\{w,v\}}n^4\frac{4}{\varepsilon}\left(1-\frac{\varepsilon}{4}\right)^{L_v}\leq \Gamma_{\{w,v\}}n^4\frac{4}{\varepsilon}\left(1-\frac{\varepsilon}{4}\right)^{\log^5(d)-2}
\end{align}
We conclude that, for sufficiently large $d$,
\begin{align}
S_{w_v}(K)
\leq
\Gamma_{{w,v}}n^4d^{-\varepsilon/5}.
\end{align}
Combining the contribution of the internal and external vertices, we obtain for sufficiently large $n,d$
\begin{align}
\begin{aligned}
\sum_{q\in K}c(q)\lvert \hat{m}_q(b)-\hat{m}_q(b')\rvert&\leq \sum_{q\in K}c(q)\lVert\hat{\mu}_{K,q}^b-\hat{\mu}_{K,q}^{b'}\rVert_{\rm TV}\leq \sum_{q\in K}c(q)\mathbb{P}_{Q_v}\left[\sigma(q)\neq \sigma'(q)\right]\\
&\leq \Gamma_{\{w,v\}}\left(n+n^4d^{-\varepsilon/5})\right)\leq \Gamma_{\{w,v\}}n^4(1-\varepsilon/6)
\end{aligned}
\end{align}
We therefore bound the contribution in $D_\star$ of each $v\in N(w)$ in equation~\eqref{eq:D_star_decomposition}, by $n^4(1-\varepsilon/6)\Gamma_{\{w,v\}}$ and obtain
\begin{align}
D_\star \leq n^4(1-\varepsilon/6)\left(\sum_{u\in\partial_{\rm out}B}\Gamma_{\{u,z_B(u)\}}\mathbbm{1}_{\{Y_t(u)=\star\}} + A(w)\right)
\end{align}
with $A(w)=\sum_{v\in N(w)}\Gamma_{\{w,v\}}$.
Collecting all the terms in $\mathbb{E}\left[\Phi_B(Y_{t+1})\mid Y_t, B_t=B\right]$, and noting that $p_w^{(-)}+1-p_w^{(+)}+r_w=1$, one obtains 
\begin{align}
\begin{aligned}
\mathbb{E}\left[\Phi_B(Y_{t+1})\mid Y_t, B_{t}=B\right]&\leq  n^4(1-\varepsilon/6)\sum_{u\in\partial_{\rm out}B}\Gamma_{\{u,z_B(u)\}}\mathbbm{1}_{\{Y_t(u)=\star\}}\\ &\quad+r_w\left(1+A(w)n^4(1-\varepsilon/6)\right)
\end{aligned}
\end{align}

\paragraph{\it Step 3. Bounding $r_wA(w)$.}
It remains to bound the probability
\begin{align}
	r_w = p_w^{(+)}-p_w^{(-)} = \mathbb{P}\left[Y_{t+1}(w)=\star\mid Y_t, B_{t}=B\right]
\end{align}
With $m_w(\eta)=\mu_B^\eta(\sigma(w)=+1)$ and
$$
p_w^{(+)}=\max_{\eta\in\mathcal C_{\partial_{\rm out}B}(Y_t)}m_w(\eta) = m_w\left(\eta^{(+)}\right) \quad , \quad p_w^{(-)}=\min_{\eta\in\mathcal C_{\partial_{\rm out}B}(Y_t)}m_w(\eta)=m_w\left(\eta^{(-)}\right)
$$
The two configurations $\eta^{(\pm)}\in\{+1,-1\}^{\partial_{\rm out}B}$ achieving the min and the max in the above expressions coincide on each resolved vertex $u\in \partial_{\rm out} B$ such that $Y_t(u)\neq \star$, and may differ on some unresolved boundary vertices.
We enumerate such vertices as $(u_1,\dots,u_s)=\mathcal{U}_B(Y_t)=\left\{u\in\partial_{\rm out}B: \eta^{(+)}(u)\neq \eta^{(-)}(u)\right\}$, and find a path of nearby configurations $(\eta^{(0)}=\eta^{(-)}, \eta^{(1)},\ldots,\eta^{(s)}=\eta^{(+)})$ such that $\eta^{(i-1)}, \eta^{(i)}$ differ only on $u_i$.
We can write
\begin{align}
 r_w \leq \sum_{i=1}^s \lvert m_w\left(\eta^{(i)}\right)-m_w\left(\eta^{(i-1)}\right)\rvert =\sum_{i=1}^s r_{u_i\to w}
\end{align}
Let $u\in \mathcal{U}_B(Y_t)$, and let $\eta,\eta'$ be two boundary configurations differing only on $u$, in such a way that $r_{u\to w}=\lvert m_w(\eta)-m_w{(\eta')}\rvert$.
We consider the coupling $Q_u$ of the paper~\cite{Efthymiou_Zampetakis_2024} between $\mu_B^{\eta}$ and $\mu_B^{\eta'}$, the two conditional distributions on $\{+1,-1\}^B$.
If $z(u)$, the unique neighbor of $u$ in $B$, belongs to $T_v$, with $v\in N(w)\setminus\{x_1,x_k\}$, i.e. one of the tree-components in the partition of $T$, then there is a unique path $z(u)\leadsto w=(z(u)=v_0,v_1,\ldots,v_{\ell_u}=w)$ of vertices inside $B$.
Since within $Q_u$, a disagreement at $w$ can only propagate along the path $z(u)\leadsto w$, we get
\begin{align}
	r_{u\to w}\leq \mathbb{P}_{Q_u}\left[\sigma(w)\neq\sigma'(w)\right]\leq \Gamma_{\{u,z(u)\}}\prod_{i=0}^{\ell_u-1}\Gamma_{\{v_{i},v_{i+1}\}}\leq \Gamma_{\{u,z(u)\}}\prod_{i=0}^{\ell_u-1}A(v_i)
\end{align}
If $z(u)\in T_C$, with $T_C$ the component including the remaining of the cycle $x_1,\ldots,x_k$ then there are two paths from $z(u)$ to $w$.
In that case, similarly as in the proof of Lemma 7.8 in~\cite{Efthymiou_Zampetakis_2024}, we let $y$ be the first vertex in the cycle reached by any path going from $u$ to $w$.
We define $\Lambda$ as the set of vertices in $B$ (apart from $y$) that cannot be reached from $u$ by a path that do not pass through $C$. 
We define $M$ as the set of vertices which are outside $\Lambda$ and have at least one neighbor in $\Lambda$, and build the coupling between $\mu_\Lambda\left(\cdot\mid \tau\right)$ and $\mu_\Lambda\left(\cdot\mid \tau'\right)$, where $\tau,\tau'\in\{+1,-1\}^M$ are two configurations differing only on $y$, with $\tau(y)=+1,\tau'(y)=-1$. 

To do so, we introduce the modified boundary $\widehat{M}=M\cup\{y'\}$, where a fictive vertex $y'$ is added between $y$ and one of its neighbors $x$ in the cycle, and let the modified coupling be $\widehat{J}_{\{y',x\}}=J_{\{y,x\}}$. 
Let $\tau^{++},\tau^{--},\tau^{+-}\in\{+1,-1\}^{\widehat{M}}$ be three configurations differing only on $\tau(y),\tau(y')$, with $\tau^{s_y,s_{y'}}(y)=s_y$ and $\tau^{s_y,s_{y'}}(y')=s_{y'}$ for $s_y,s_{y'}\in\{+1,-1\}$, in such a way that $\mu_\Lambda(\cdot\mid \tau)=\mu_\Lambda(\cdot\mid \tau^{++})$ and $\mu_\Lambda(\cdot\mid \tau')=\mu_\Lambda(\cdot\mid \tau^{--})$.
We build the coupling $\kappa_1$ between $\mu_\Lambda(\cdot\mid \tau^{++})$ and $\mu_\Lambda(\cdot\mid \tau^{+-})$, starting from the disagreeing vertex $y'$ and propagating the disagreement in $\Lambda$. Similarly, we build the coupling $\kappa_2$ between $\mu_\Lambda(\cdot\mid \tau^{+-})$ and $\mu_\Lambda(\cdot\mid \tau^{--})$.
Let $P_{y\to w}=(y=p_0,\ldots,p_\ell=w)$ and $P'_{y\to w}=(y=p'_0,p'_1=x\ldots,p'_{\ell'}=w)$ be the two paths going from $y$ to $w$ along the cycle, with $P'_{y\to w}$ passing through $x$.
Under the coupling $\kappa_1$ (resp. $\kappa_2$), a disagreement at $w$ occurs with probability at most $\prod_{i=1}^{\ell'}\Gamma_{\{p'_{i-1},p'_i\}}$ (resp. with probability at most $\prod_{i=1}^{\ell}\Gamma_{\{p_{i-1},p_i\}}$).

We compose the couplings $\kappa_1$ and $\kappa_2$, by generating first a configuration $\sigma_A$ from $\mu_\Lambda(\cdot\mid \tau)$ and then use $\kappa_1$ to obtain a configuration $\sigma_B$ from $\mu_\Lambda(\cdot\mid \tau^{+-})$ conditional on $\sigma_A$. Then we use $\kappa_2$ to generate $\sigma_C$ from $\mu_\Lambda(\cdot\mid \tau')$ conditional on $\sigma_B$. We call $\kappa$ the resulting coupling between $\sigma_A,\sigma_C$.
Under this coupling, conditional on a disagreement at $y$, a disagreement at $w$ occurs with probability at most
$$
\prod_{i=1}^{\ell}\Gamma_{\{p_{i-1},p_i\}}+\prod_{i=1}^{\ell'}\Gamma_{\{p'_{i-1},p'_i\}}
$$
Finally, we let $R_{u\to y}=(u=r_0,\ldots,r_m=y)$ be the unique path from $u$ to $y$, and we consider the usual coupling with disagreement propagation from $u$ to $y$, to obtain the bound
\begin{align}
\begin{aligned}
r_{u\to w}&\leq \prod_{i=1}^m\Gamma_{\{r_{i-1},r_i\}}\left(\prod_{i=1}^{\ell}\Gamma_{\{p_{i-1},p_i\}}+\prod_{i=1}^{\ell'}\Gamma_{\{p'_{i-1},p'_i\}}\right)\\
&\leq \Gamma_{\{u,z(u)\}}\prod_{i=1}^mA(r_i)\left(\prod_{i=1}^{\ell-1}A(p_i)+\prod_{i=1}^{\ell'-1}A(p'_i)\right)
\end{aligned}
\end{align}
Consider a path $P_{u\to w}=(u=q_0,z(u)=q_1,\ldots,w=q_t)$, and let $h$ be the number of heavy vertices in $P_{u\to w}$, i.e. vertices $x$ with $A(x)>1-\varepsilon/2$. 
By definition of the block partition, $u$ and $z(u)$ are light, and by Corollary 7.3 in~\cite{Efthymiou_Zampetakis_2024}, we have that
\begin{align}
\prod_{i=1}^{t}A(q_i)\leq \left(\frac{1-\varepsilon/2}{1-\varepsilon/4}\right)^{t-h}d^{-h}
\end{align}
Since $w$ belongs to the cycle, we have $t\geq\log^5(d)$.
For sufficiently large $d$, we have $d^{-1}<\frac{1-\varepsilon/2}{1-\varepsilon/4}$ and therefore 
\begin{align}
\prod_{i=1}^{t}A(q_i)\leq \left(\frac{1-\varepsilon/2}{1-\varepsilon/4}\right)^{t}\leq \left(\frac{1-\varepsilon/2}{1-\varepsilon/4}\right)^{\log^5(d)}=e^{-c_\varepsilon \log^5(d)}
\end{align}
With $c_\varepsilon = -\log\left(\frac{1-\varepsilon/2}{1-\varepsilon/4}\right)$.
For sufficiently large $d$, taking $c_\varepsilon\log^5(d)\geq 12\log d$ leads to $\prod_{i=1}^{t}A(q_i)\leq d^{-12}$. 
Therefore, we obtain $r_{u\to w}A(w)\leq 2\Gamma_{\{u,z(u)\}}d^{-12}$ for each $u\in\mathcal{U}_B(Y_t)$. In turn, we obtain the following bound for $r_w\cdot A(w)$:
\begin{align}
 r_w\cdot A(w) \leq 2d^{-12}\sum_{u\in\partial_{\rm out}B}\Gamma_{\{u,z(u)\}}\mathbbm{1}_{\{Y_t(u)=\star\}}
\end{align}

\paragraph{\it Step 4. Contraction inequality.}
Let $\Delta_B(Y_t)=\sum_{u\in\partial_{\rm out}B}\Gamma_{\{u,z_B(u)\}}\mathbbm{1}_{\{Y_t(u)=\star\}}$.
Combining the bounds obtained at the previous steps leads to
\begin{align}
\begin{aligned}
\mathbb{E}\left[\Phi_B(Y_{t+1})\mid Y_t, B_t=B\right]&\leq n^4(1-\varepsilon/6)\Delta_B(Y_t)+r_wA(w)\left(\frac{1}{A(w)}+n^4(1-\varepsilon/6)\right)\\
&\leq n^4\Delta_B(Y_t)\left(1-\frac{\varepsilon}{6}+2d^{-12}\left(\frac{1}{n^4A(w)}+1-\frac{\varepsilon}{6}\right)\right)
\end{aligned}
\end{align}
By Theorem 4.2 of~\cite{Efthymiou_Zampetakis_2024}, we have that $\Gamma_{e}\geq \tanh(\beta n^{-7/3}/2)$ with probability at least $1-n^{-1/4}$. 
Therefore, since $w$ belongs to the cycle and has at least two neighbors, $A(w)\geq 2 \tanh(\beta n^{-7/3}/2)$ and $1/(n^4A(w))\leq 1/(2n^4\tanh(\beta n^{-7/3}/2))<1$ for fixed $\beta>0$ and sufficiently large $n$, giving
\begin{align}
	\begin{aligned}
		\mathbb{E}\left[\Phi_B(Y_{t+1})\mid Y_t, B_t=B\right]\leq n^4\Delta_B(Y_t)\left(1-\frac{\varepsilon}{6}+2d^{-12}\left(2-\frac{\varepsilon}{6}\right)\right)
	\end{aligned}
\end{align}
Equation~(\ref{eq:contraction_PhiB_tree_blocks}) gives when $B$ is a tree 
$$
\mathbb{E}\left[\Phi_B(Y_{t+1})\mid Y_t, B_t=B\right]\leq n^4\Delta_B(Y_t)\left(1-\frac{\varepsilon}{6}\right)\leq n^4\Delta_B(Y_t)\left(1-\frac{\varepsilon}{6}+2d^{-12}\left(2-\frac{\varepsilon}{6}\right)\right)
$$
When block $B$ us updated, all previous unresolved variables are overwritten. Therefore
\begin{align}
\begin{aligned}
\mathbb{E}\left[\Phi(Y_{t+1})\mid Y_t,B_t=B\right]&\leq \Phi(Y_t)-\Phi_B(Y_t)+\mathbb{E}\left[\Phi_B(Y_{t+1})\mid Y_t, B_{t}=B\right]\\
&\leq \Phi(Y_t)-\Phi_B(Y_t)+n^4\Delta_B(Y_t)\left(1-\varepsilon/6+2d^{-12}\left(2-\varepsilon/6\right)\right)
\end{aligned}
\end{align}
Averaging over the uniformly chosen block gives
\begin{align}
\begin{aligned}
\mathbb{E}\left[\Phi(Y_{t+1})\mid Y_t\right]&\leq \left(1-\frac{1}{N}\right)\Phi(Y_t)+\frac{n^4}{N}\left(1-\varepsilon/6+2d^{-12}\left(2-\varepsilon/6\right)\right)\sum_{B\in\mathcal{B}}\Delta_B(Y_t)\\
&\leq \left(1-\frac{1}{N}\right)\Phi(Y_t)+\frac{1}{N}\left(1-\varepsilon/6+2d^{-12}\left(2-\varepsilon/6\right)\right)\Phi_W(Y_t)\\
&\leq \left(1-\frac{1}{N}\right)\Phi_{V\setminus W}(Y_t) + \Phi_W(Y_t)\left(1-\frac{1}{N}\left(\varepsilon/6-2d^{-12}\left(2-\varepsilon/6\right)\right)\right)
\end{aligned}
\end{align}
We choose $d$ sufficiently large so that $\varepsilon/6-2d^{-12}\left(2-\varepsilon/6\right)>\varepsilon/7$, and we obtain
\begin{align}
\mathbb{E}\left[\Phi(Y_{t+1})\mid Y_t\right]\leq \left(1-\frac{\varepsilon}{7N}\right)\Phi(Y_t)
\end{align}
Just as in the case of tree-blocks, we can deduce that the coalescence time of the block bounding chain is $O(n\log n)$ with high probability.
\hfill$\square$

\section{Conclusion}
In this work, we study perfect sampling algorithms for Ising spin glass models, based on coupling from the past and bounding chains.
We investigate the range of inverse temperature over which the single-site bounding chain produces perfect samples efficiently, on various models defined on several graph families and coupling distribution.
Our numerical results suggests a threshold $\beta_{\rm bc}$ separating rapid coalescence from a regime with super-polynomial growth of the coalescence time.
On sparse random graphs, this algorithmic threshold is consistent with the tree-uniqueness threshold of the associated infinite tree model, occurring at a $\beta$ value strictly smaller than the spin-glass transition.

Under the Dobrushin conditions, we prove rapid coalescence of the single-site bounding chain, in $O(n\log n)$ single-site updates, through a contraction argument (see Section~\ref{sec:rapid_coalescence_Dobrushin}), consistently with our numerical observations.
This condition also guarantees uniqueness, but the inverse temperature range it covers is narrower than the regime $\beta<\beta_{\rm bc}$ in which we observe rapid coalescence numerically.
In particular, our simulations suggest that the sub-polynomial scaling of $T_{\rm coal}/n$ persists throughout this larger regime.

On Erd\H{o}s-Rényi random graphs $\mathcal{G}(n,d/n)$ with standard Gaussian couplings, we introduce an block bounding chain and prove that it coalesces in $O(n\log n)$ block updates in the tree-uniqueness regime for sufficiently large degree $d$, each block update requiring $O(n)$ operations (see Section~\ref{sec:rapid_coal_block_BC}).
Although this algorithm differs from the single-site bounding chain, the regime for which its rapid coalescence is proven matches the range observed in our numerical results for the single-site bounding chain.
We do not investigate the behavior of the block bounding chain beyond the tree-uniqueness threshold, so whether it remains efficient in a larger inverse-temperature regime remains an open question.

Our proof builds on~\cite{Efthymiou_Zampetakis_2024}, where rapid mixing of the corresponding block Glauber dynamics is established through a path coupling method.
We adapt the contraction estimates to a potential associated with the block bounding chain.
This suggests that coupling-based proofs of rapid mixing may provide a route to prove rapid coalescence of bounding chains, and more generally of CFTP algorithms, for other models and dynamics.
By contrast, the spectral argument used in~\cite{Efthymiou_Zampetakis_2024} to transfer rapid mixing from block to single-site dynamics cannot be directly transposed to prove rapid coalescence on the single-site bounding chain.

Our numerical study in Section~\ref{sec:numerics} suggests that, on sparse graphs, the algorithmic threshold of the single-site bounding chain coincides with the tree-uniqueness threshold. 
Recent results establish rapid mixing beyond the tree-uniqueness threshold in some models~\cite{Chen_beyond_uniq_2025,Gobel_2026,Kuikui_2024,efthymiou2026samplingdilutedspinglassesunbounded}, raising the question of whether bounding chain algorithms could also be efficient in this regime.

Our damage-spreading experiments suggest that, on sparse random graphs, the deterioration of the bounding chain is associated with the increasing meeting time of Glauber trajectories under shared-update coupling.
Improving the bounding chain envelope alone would therefore not be sufficient to overcome this obstruction.
A possible direction would be to design alternative coupling that favor coalescence, for instance based on non-Markovian coupling constructions~\cite{Griffeath1975}. 
Whether such constructions can be implemented efficiently within a CFTP algorithm remains to be investigated.

On periodic lattices, the algorithmic threshold of the single-site bounding chain does not appear to coincide with any known equilibrium phase transition and lies below the damage-spreading threshold.
The better performance of the local-patch CFTP algorithm~\cite{Chanal_Krauth_2008,Chanal_Krauth_2010} suggests that the limitation is instead due to the looseness of the bounding envelope, which fails to capture correlations between unresolved neighboring spins.
Incorporating such correlations in the envelope representation, for instance by updating blocks of variables at each iterations, could therefore improve the efficiency of bounding chain algorithms on lattices.

On the Sherrington-Kirkpatrick model, both the coalescence time of the bounding chain and the meeting time of the two coupled Glauber trajectories appear to grow super-polynomially at every positive inverse temperature tested.
These observations suggest that damage spreading limits the efficiency of CFTP under shared-update coupling for $\beta>0$, consistently with~\cite{Derrida1989}.
As for sparse graphs, improving the bounding envelope alone would not be sufficient. 
Investigating alternative coupling is a natural direction for this model.

These observations suggest that the performance of the single-site bounding chain depends on how accurately its representation with unresolved spins captures the evolution of the coupled CFTP trajectories.
On locally tree-like sparse graphs, neglecting correlations between unresolved spins may be less restrictive than on lattices or on dense models.
An interesting direction for a better understanding of the dynamics of bounding chains could be to study the propagation and extinction of unresolved sites on the graph as a percolation model, across several graph families.

\paragraph{Acknowledgments.}
We thank Ana Busic, Werner Krauth and Guilhem Semerjian for insightful discussions.
This work has received support under the 2025 PSL Young Researcher Starting Grant operated by Université PSL (grant agreement no. 2025-392).
A public GitHub repository containing an implementation of the algorithm and notebooks is available at~\cite{github_ref}.

\paragraph{Statement of AI use.} 
The authors were assisted by Fable 5.1 (Anthropic) and GPT 6 Astra (Open AI) in the steps of the proofs of Sections~\ref{sec:rapid_coalescence_Dobrushin}. and~\ref{sec:rapid_coal_block_BC}, and in the numerical simulations of Section~\ref{sec:numerics}.
The authors independently checked the resulting arguments, calculations, numerical results, and take full responsibility for the content of the present work.

	\bibliography{biblio.bib}
	
\appendix
\section{Sampling at the first vertex of an unicyclic block}
\label{app:sampling_at_w}
For each $x_i$ in the cycle, let $T_i$ be the tree hanging below $x_i$.
Let $\partial_{\rm out}T_i=\{u\in\partial_{\rm out}B: z_B(u)\in T_i\}$ be the set of vertices in $\partial_{\rm out}B$ having one neighbor in $T_i$.
For a given compatible boundary condition $\eta\in \mathcal{C}_{\partial_{\rm out}B}(Y_t)$ (with $\mathcal{C}_{\partial_{\rm out}B}(Y_t)$ defined in (\ref{eq:compatible_boundaries})), one can compute the Belief-Propagation message $h_i(\eta)$ coming from the leaves of $T_i$ towards $x_i$ in $O(|T_i|)$ operations.
The message $h_i(\eta)$ depends only on the boundary variables $\eta(\partial_{\rm out}T_i)$.
Consider a gauge transformation starting from $x_i$, with $g_{x_i}=+1$, and propagating towards the leaves of $T_i$.
Define the two gauge-transformed boundary configurations $b_i^{(+)},b_i^{(-)}$ in $\{+1,-1\}^{\partial_{\rm out}T_i}$ such that for each $u\in\partial_{\rm out}T_i$
\begin{align}
b_i^{(\pm)}(u)=\begin{cases}
g_uY_t(u) \quad \text{if} \quad Y_t(u)\neq \star \\
\pm 1 \quad \text{otherwise}
\end{cases}
\end{align}
Then, the minimum and maximum values $h_i^{(-)},h_i^{(+)}$ of $h_i(\eta)$ are attained by two boundary configurations $\eta_i^{(-)},\eta_i^{(+)}$ whose gauge transformation are precisely $b_i^{(-)},b_i^{(+)}$.
The marginal probability $\mu_B^\eta(\sigma)$, $\sigma\in\{+1,-1\}^{k+1}$ on the sites of the cycle $C$ can be written as
\begin{align}
\label{eq:joint_cycle}
\mu_B^\eta(\sigma)=\frac{\prod_{i=0}^ke^{\frac{\beta}{2} J_{\{x_i,x_{i+1}\}}\sigma(x_{i})\sigma(x_{i+1})+\frac{\beta}{2} h_i(\eta)\sigma(x_i)}}{Z_{C}}
\end{align}
And the marginal $m_w(\eta)$ can be written for each $i\in\{1,\dots,k\}$ as
\begin{align}
\begin{aligned}
m_w(\eta)&=\frac{\sum_{s_i\in\{+1,-1\}}e^{\frac{\beta}{2} h_i(\eta)s_i}G(+1,s_i)}{\sum_{s_0,s_i\in\{+1,-1\}^2}e^{\frac{\beta}{2} h_i(\eta)s_i}G(s_0,s_i)} \\ \text{with}\quad G(s_0,s_i)&=\sum_{\{s_j\}_{j\in[k]\setminus\{i\}}}\prod_{j=0}^ke^{\frac{\beta}{2} J_{\{x_j,x_{j+1}\}}s_js_{j+1}}\times\prod_{j=0, j\neq i}^k e^{\frac{\beta}{2}h_j(\eta)s_j}
\end{aligned}
\end{align}
From this expression, it is easy to see that $m_w(\eta)=\widetilde{m}_w(\{h_j\}_{j\in \{0,1,\dots,k\}})$ is a monotonous function in each of its arguments $h_0,\dots,h_k$. 
Indeed, let $y_i=e^{\beta h_i(\eta)}$, then for $i\in\{1,\dots,k\}$	
\begin{align}
\partial_{y_i}m_w(\eta)=\frac{G(+,+)G(-,-)-G(+,-)G(-,+)}{\left(y_i\left[G(+,+)+G(-,+)\right]+G(+,-)+G(-,-)\right)^2}
\end{align}
is of the sign of $G(+,+)G(-,-)-G(+,-)G(-,+)$ which does not depends on $y_i$.
And similarly one can see that $m_w(\eta)$ is an increasing function of $h_0$
\begin{align}
\begin{aligned}
m_w(\eta)&=\frac{y_0 H(+)}{y_0 H(+) + H(-)} \\
\text{with} \quad H(s)&=\sum_{s_1,\dots,s_k}\prod_{j=0}^ke^{\frac{\beta}{2} J_{\{x_j,x_{j+1}\}}s_js_{j+1}}\times\prod_{j=1}^k e^{\frac{\beta}{2}h_j(\eta)s_j} \\
\partial_{y_0}m_w(\eta)&=\frac{H(+)H(-)}{\left(y_0 H(+) + H(-)\right)^2}>0
\end{aligned}
\end{align}
Therefore, the maximum and minimum of $m_w(\eta)$ is attained by a combination of the extreme values taken by $(h_0,h_1\dots,h_k)$
\begin{align}
\begin{aligned}
p_w^{(+)}&=\max_{\eta\in\mathcal C_{\partial_{\rm out} B}(Y_t)}m_w(\eta)=\max_{\varepsilon_1,\dots,\varepsilon_k\in\{+1,-1\}^k}\left\{\widetilde{m}_w(h_0^{(+)},h_1^{(\varepsilon_1)},\dots,h_k^{(\varepsilon_k)})\right\}\\ 
p_w^{(-)}&=\min_{\eta\in\mathcal C_{\partial_{\rm out} B}(Y_t)}m_w(\eta)=\min_{\varepsilon_1,\dots,\varepsilon_k\in\{+1,-1\}^k}\left\{\widetilde{m}_w(h_0^{(-)},h_1^{(\varepsilon_1)},\dots,h_k^{(\varepsilon_k)})\right\}
\end{aligned}
\end{align}
Computing the two extreme boundary configurations $\eta_i^{(\pm)}$ via the gauge-transformation for each $i\in\{0,1,\dots,k\}$ can be done in $O(|\partial_{\rm out}B|)$ operations.
The $2(k+1)$ Belief Propagation messages $\{h_i^{(\pm)}\}_{i\in\{0,1,\dots,k\}}$ can be computed in $O(|B|)$ operations.
For each combination $(\varepsilon_1,\dots,\varepsilon_k)$, computing $\widetilde{m}(h_0^{(\pm)},h_1^{(\varepsilon_1)},\dots,h_k^{(\varepsilon_k)})$ from the expression (\ref{eq:joint_cycle}) can be done in $O(k)$ operations with transfer matrices.
It remains to perform the optimization over the sign $h_1^{(\varepsilon_1)},\dots,h_k^{(\varepsilon_k)}$, which has to be done in $O(2^k)$ operations.
Since the cycle $C$ is of length at most $4\frac{\log n}{\log^4 d}$ from the definition 4.1 in~\cite{Efthymiou_Zampetakis_2024} of a $(d,\varepsilon)$-block partition, computing the probabilities $p_w^{(\pm)}$ can be done in $O(|\partial_{\rm out}B|+|B|+k\times 2^k)=O(n)$ operations.

\section{Additional numerical results}
\label{app:numerics_additional}
\subsection{Coalescence time}
\label{subsec_app:numerics_coaltime}
\begin{figure}[h!]
	\centering
	\includegraphics[width=0.45\textwidth]{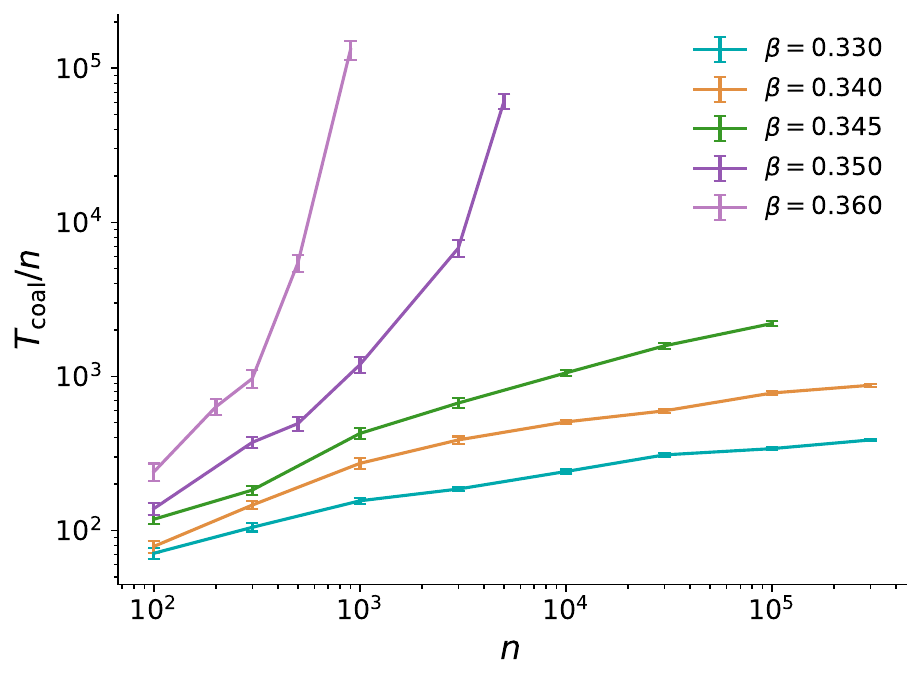}
	\includegraphics[width=0.45\textwidth]{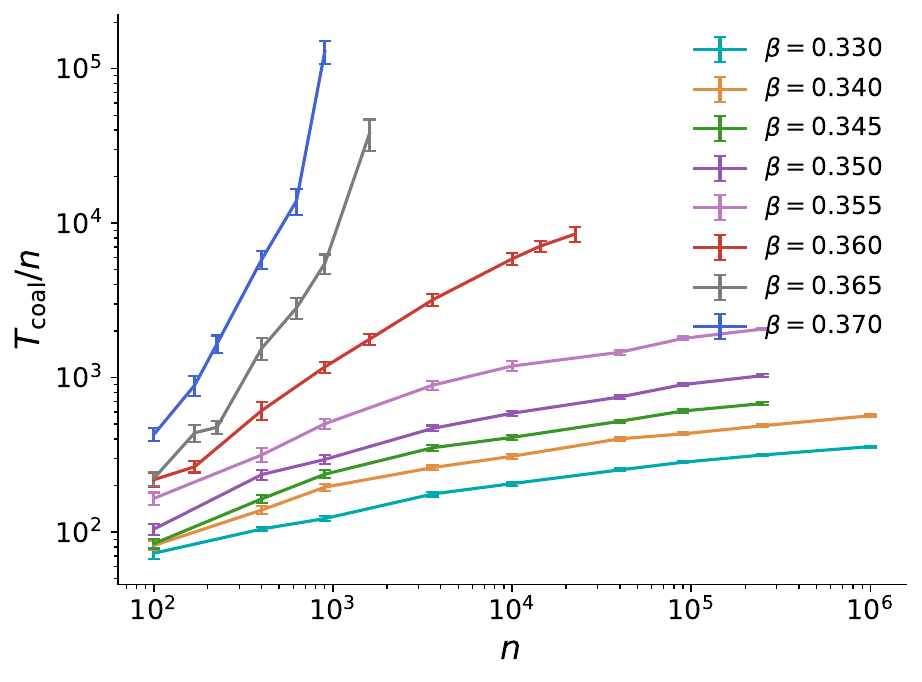}\\
	\includegraphics[width=0.45\textwidth]{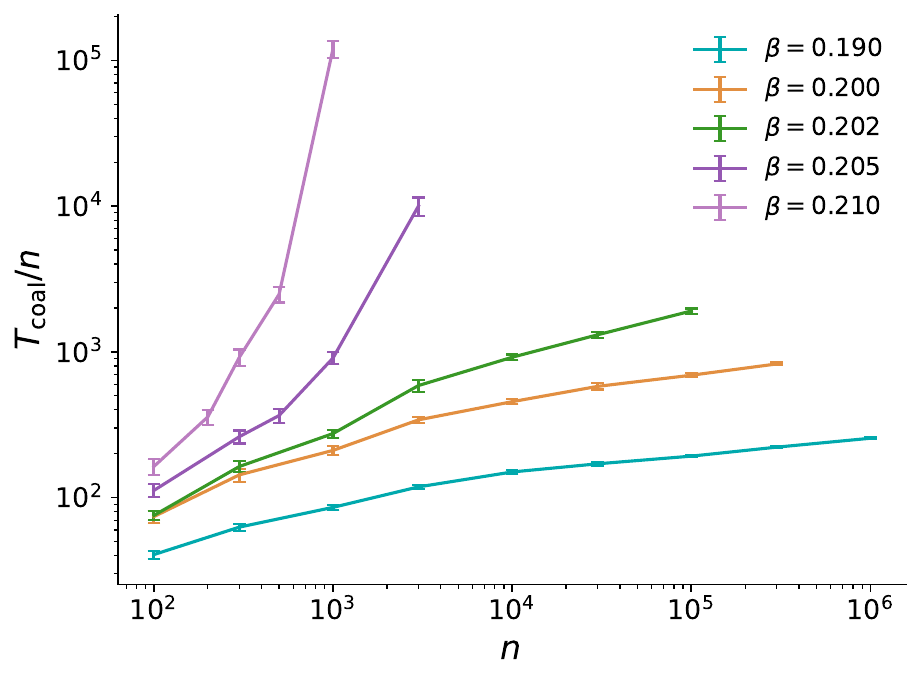}
	\includegraphics[width=0.45\textwidth]{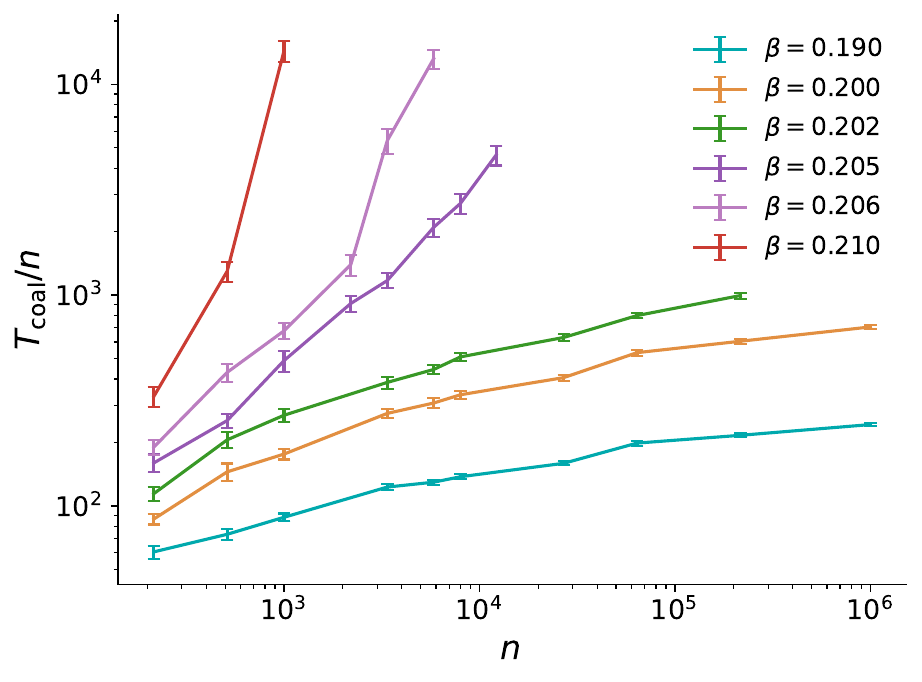}\\
	\includegraphics[width=0.45\textwidth]{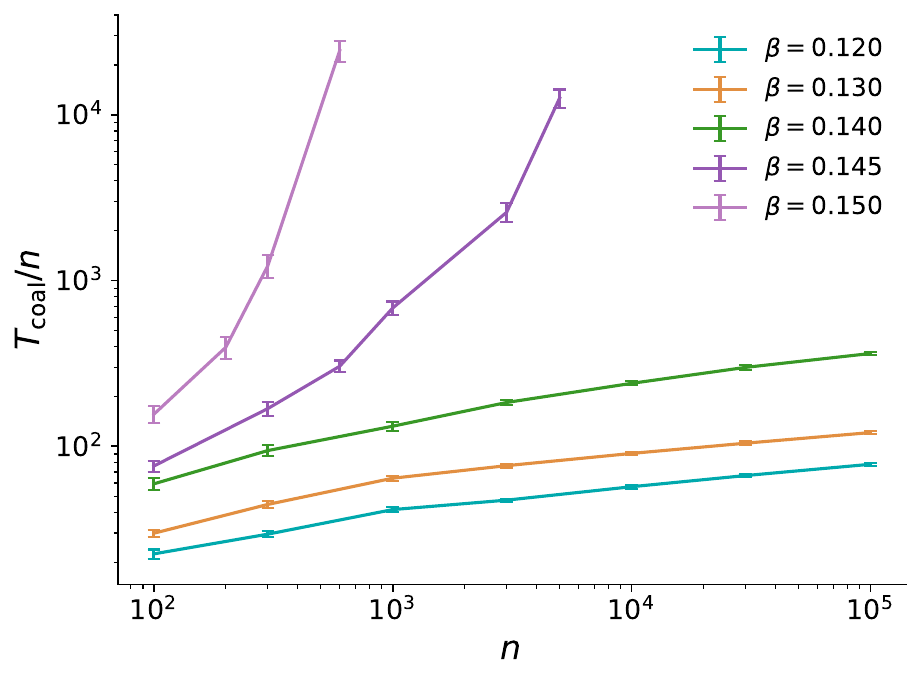}
	\includegraphics[width=0.45\textwidth]{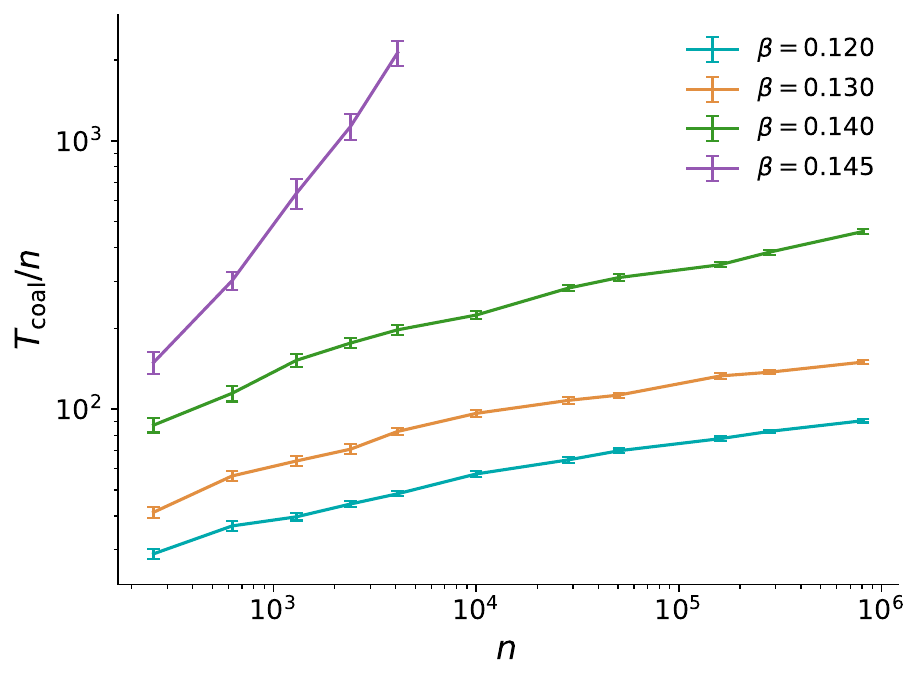}
	\caption{Empirical mean of the rescaled coalescence time $T_{\rm coal}/n$ as a function of $n$ on logarithmic axes. Left panels: on random $d$-regular graphs with i.i.d. couplings uniform in $\{+1,-1\}$, $d=4$ (top-left panel, $\beta_{\rm uni}\simeq0.346574$), $d=6$ (middle-left panel, $\beta_{\rm uni}\simeq0.202733$) and $d=8$ (bottom-left panel, $\beta_{\rm uni}\simeq0.143841$).
	Right panels: on periodic $d$-lattices with uniform i.i.d. couplings in $\{+1,-1\}$, with $d=2$ (top-right), $d=3$ (middle-right) and $d=4$ (bottom-right).}
	\label{fig_app:coal_loglog_RR}
\end{figure}

\begin{figure}[h!]
	\centering
	\includegraphics[width=0.45\textwidth]{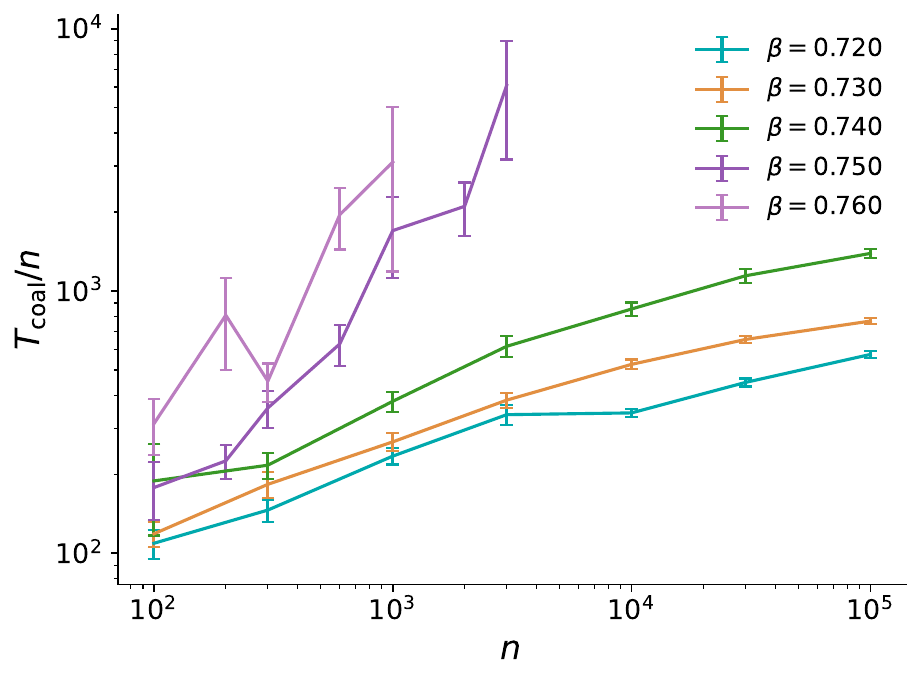}
	\includegraphics[width=0.45\textwidth]{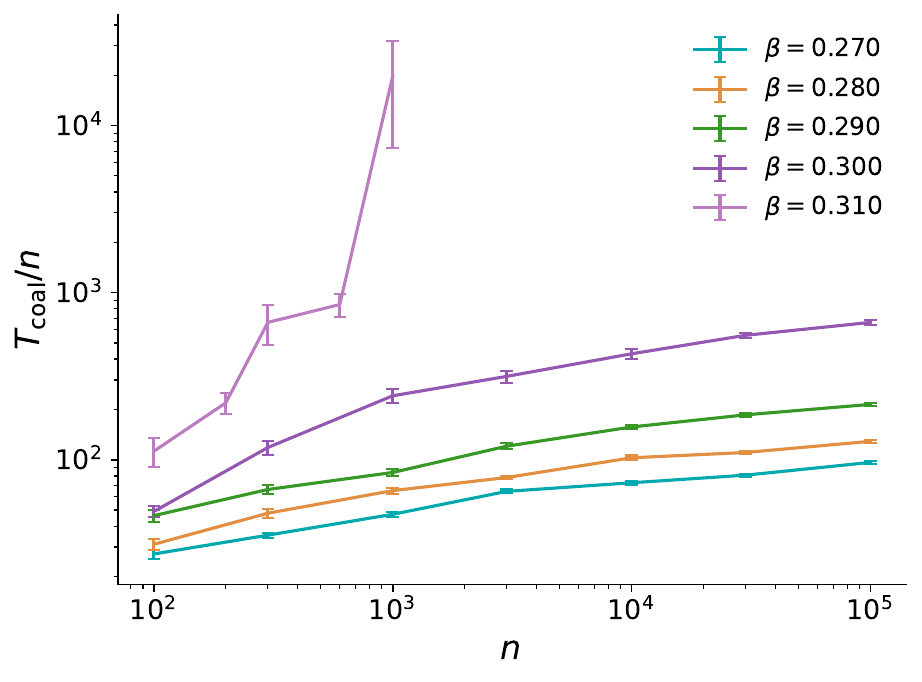}\\
	\includegraphics[width=0.45\textwidth]{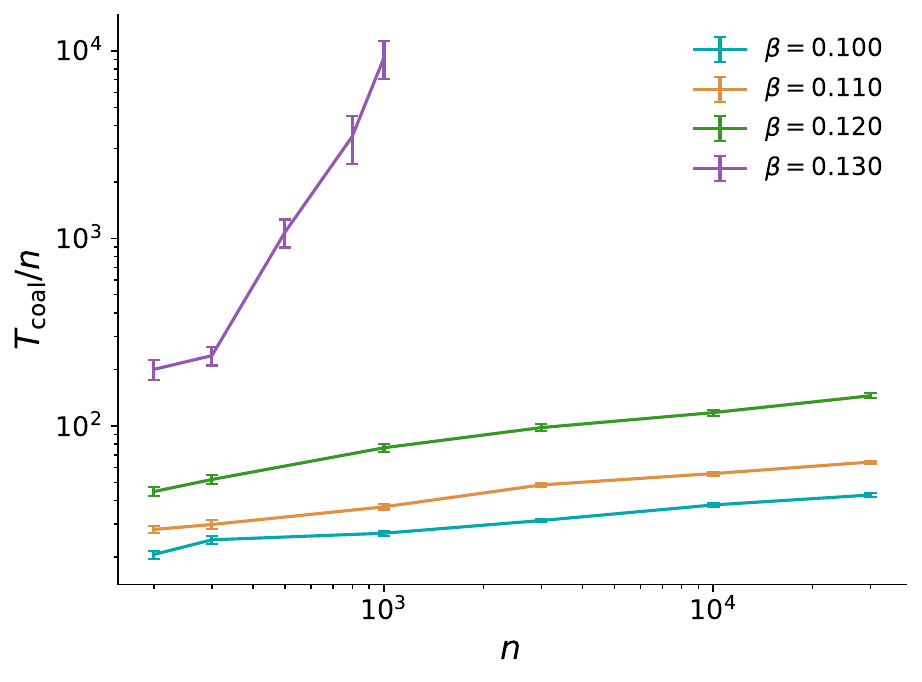}
	\includegraphics[width=0.45\textwidth]{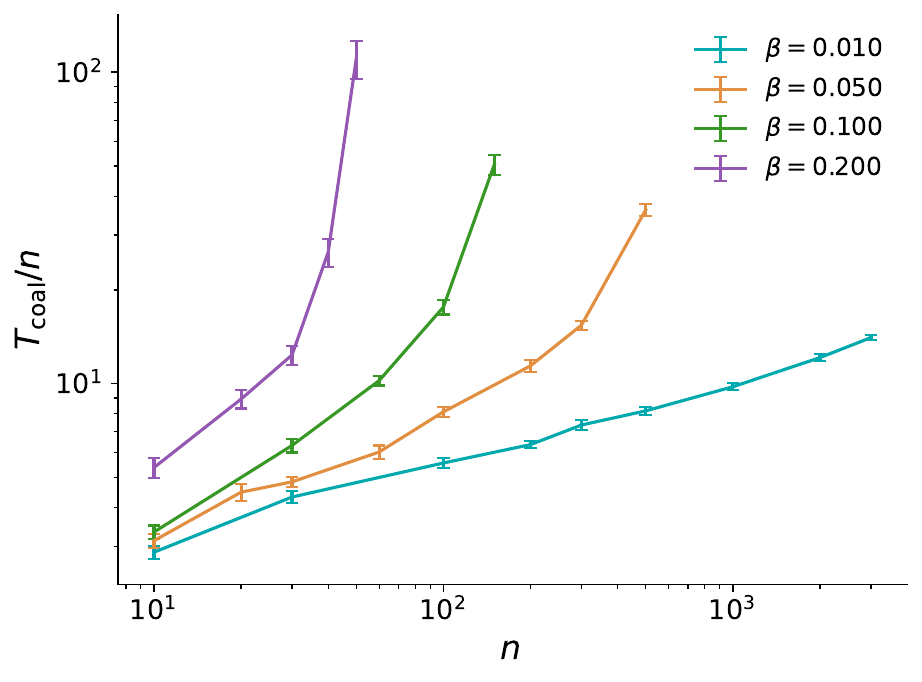}
	\caption{Empirical mean of the rescaled coalescence time $T_{\rm coal}/n$ as a function of $n$ on logarithmic axes. Top panels and bottom-left panel: on random $d$-regular graphs with $d=5$ (top-left, $\beta_{\rm uni}\simeq0.7477$) and $d=20$ (top-right, $\beta_{\rm uni}\simeq0.2958$) and $d=100$ (bottom left, $\beta_{\rm uni}\simeq0.1266$), with i.i.d. couplings in $\mathcal{N}(0,1/d)$.
	Bottom right panel: on the Sherrington-Kirkpatrick model (complete graph $K_n$ and i.i.d. couplings in $\mathcal{N}(0,1/n)$).}
	\label{fig_app:coal_loglog_SK}
\end{figure}

\begin{figure}[h!]
	\centering
	\includegraphics[width=0.9\textwidth]{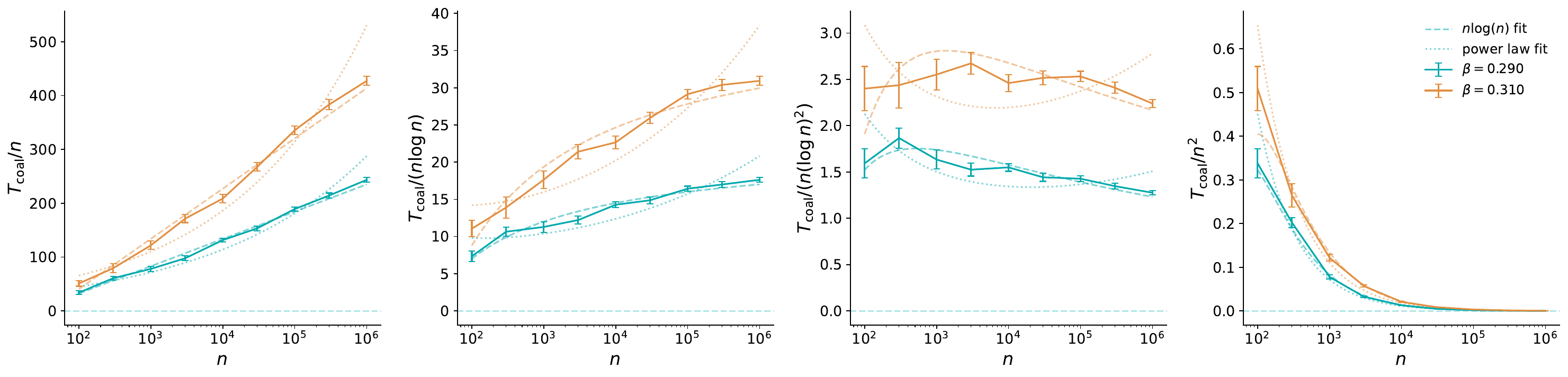}
	\includegraphics[width=0.9\textwidth]{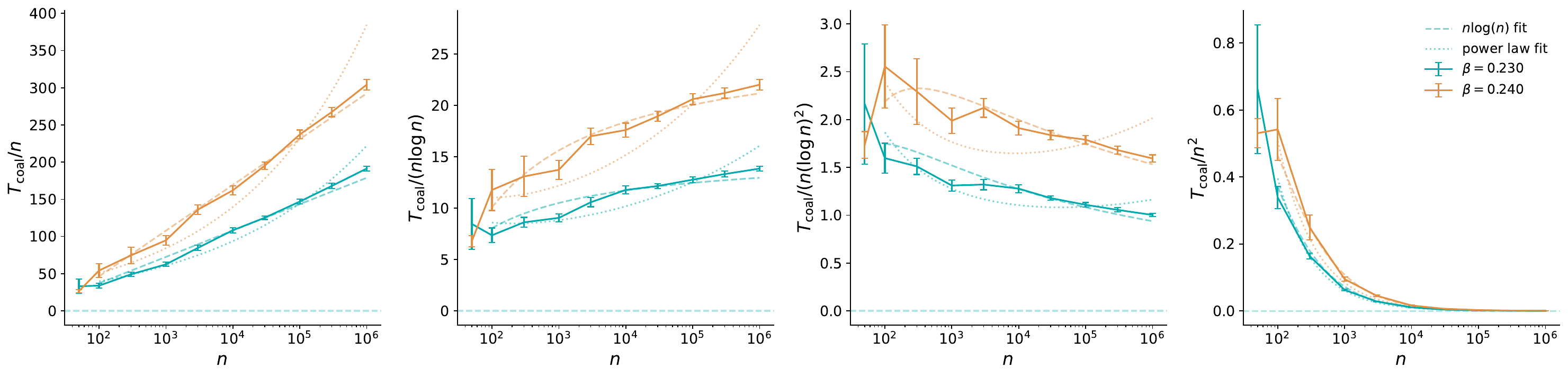}
	\includegraphics[width=0.9\textwidth]{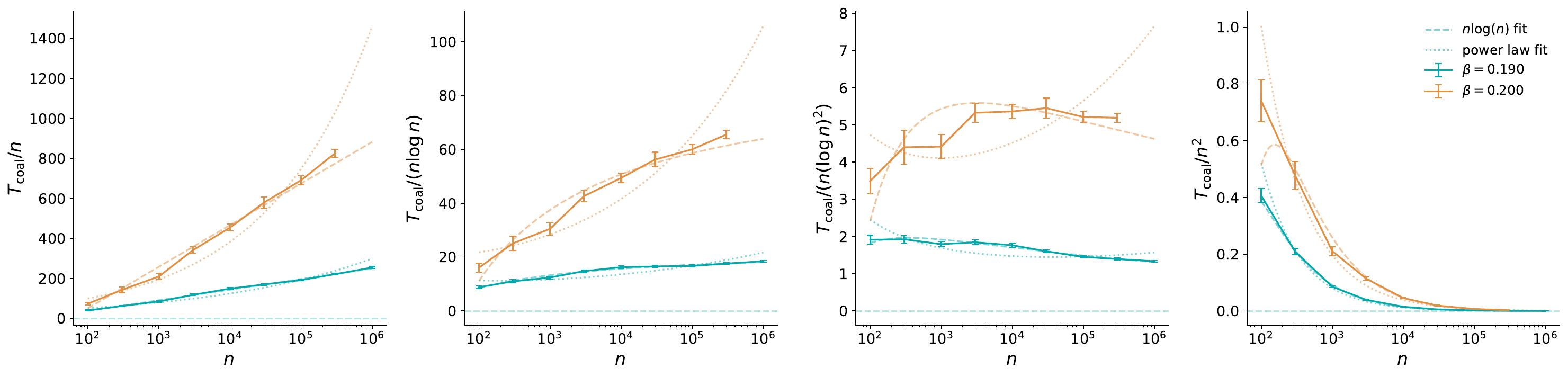}
	\includegraphics[width=0.9\textwidth]{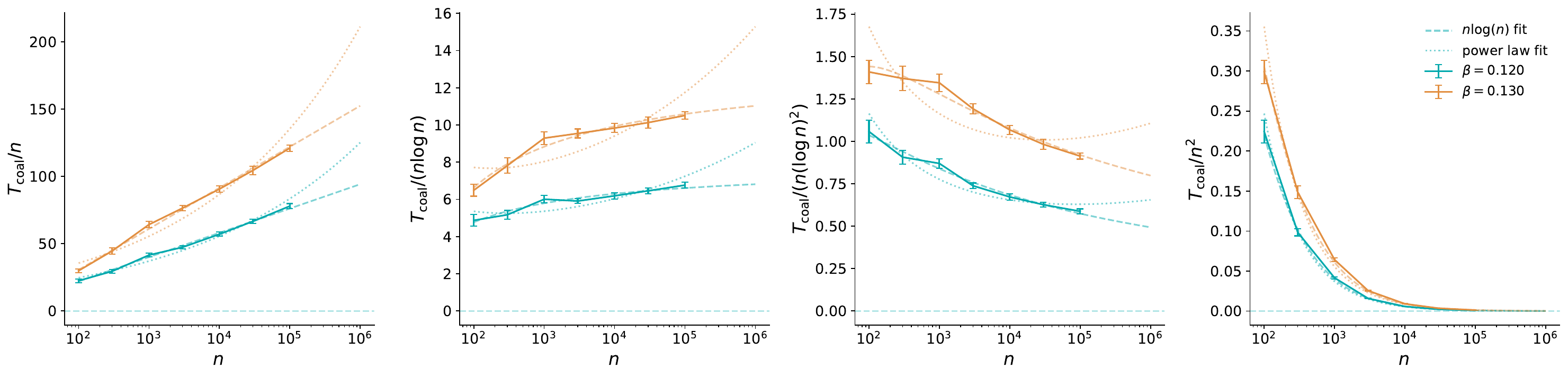}
	\includegraphics[width=0.9\textwidth]{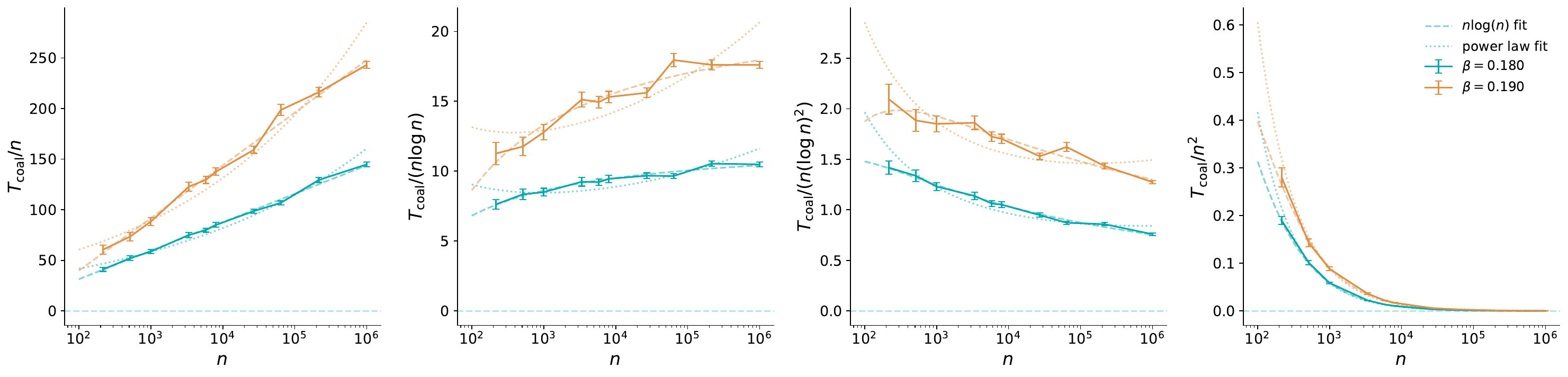}
	\includegraphics[width=0.9\textwidth]{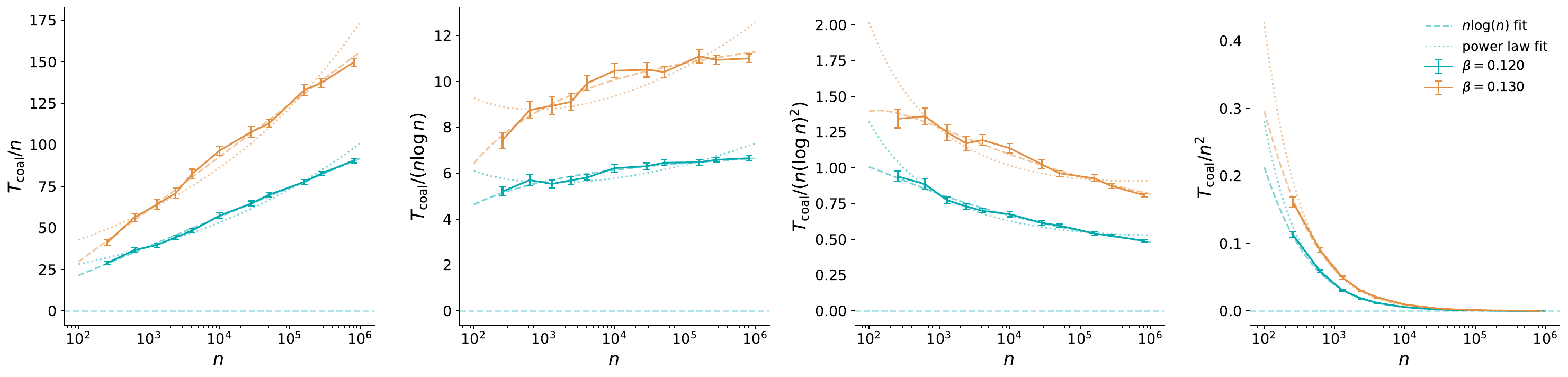}
	\caption{Rescaled empirical mean coalescence time as a function of $n$.
	The curves show the corresponding rescalings of two fits: $T_{\rm coal}=an\log n+bn$ (dashed line) and $T_{\rm coal}=cn^\delta$ (dotted line).
	On $\mathcal{G}(n,4/n)$ with i.i.d standard Gaussian couplings (1st row) and i.i.d uniform $\pm1$ couplings (2nd row), on random $d$-regular graphs with i.i.d. uniform $\pm1$ couplings with $d=6$ (3rd row) and $d=8$ (4th row), and on periodic $d$-dimensional lattices with $d=3$ (5th row) and $d=4$ (6th row).}
	\label{fig_app:coal_scaling}
\end{figure}

Additional plots of the scaling of $T_{\rm coal}/n$ against $n$ (log-log scale) are shown:
\begin{itemize}
	\item[--] in Figure~\ref{fig_app:coal_loglog_RR}: on random $d$-regular graphs and periodic $d$-lattices.
	\item[--] in Figure~\ref{fig_app:coal_loglog_SK}: on the Sherrington-Kirkpatrick model and on random $d$-regular graphs with rescaled couplings.
\end{itemize} 
Additional plots of $T_{\rm coal}/n$, $T_{\rm coal}/(n\log n)$, $T_{\rm coal}/(n(\log n)^2)$ and $T_{\rm coal}/n^2$ as functions of $n$ for several values of $\beta$ are shown in Figure~\ref{fig_app:coal_scaling}.

\subsection{Evolution of unresolved sites}
\label{subsec_app:numerics_unresolved_sites}
\begin{figure}[h]
	\centering
	\includegraphics[width=1.0\textwidth]{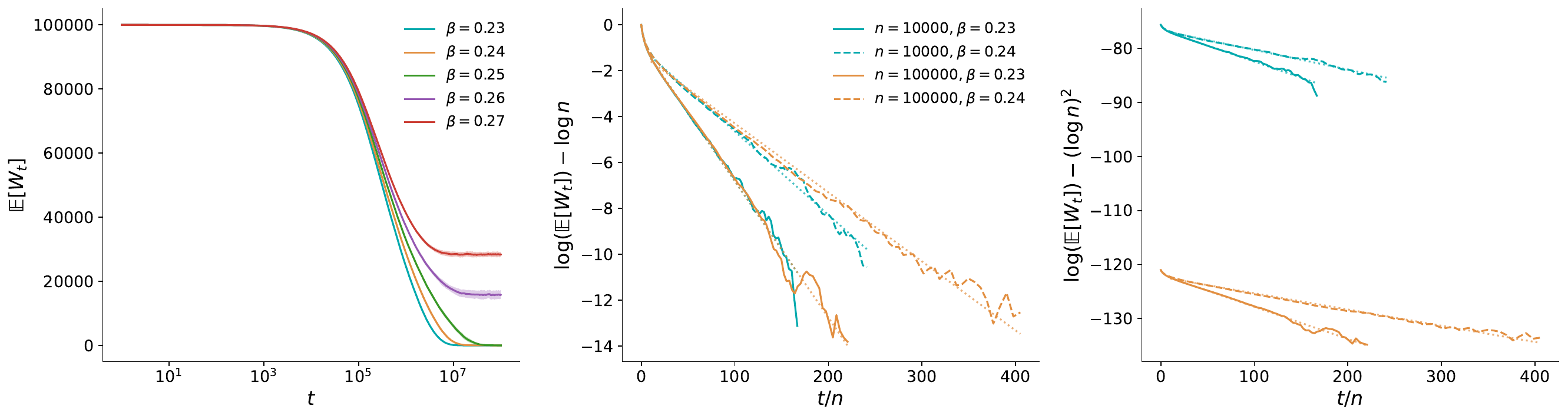}
	\includegraphics[width=1.0\textwidth]{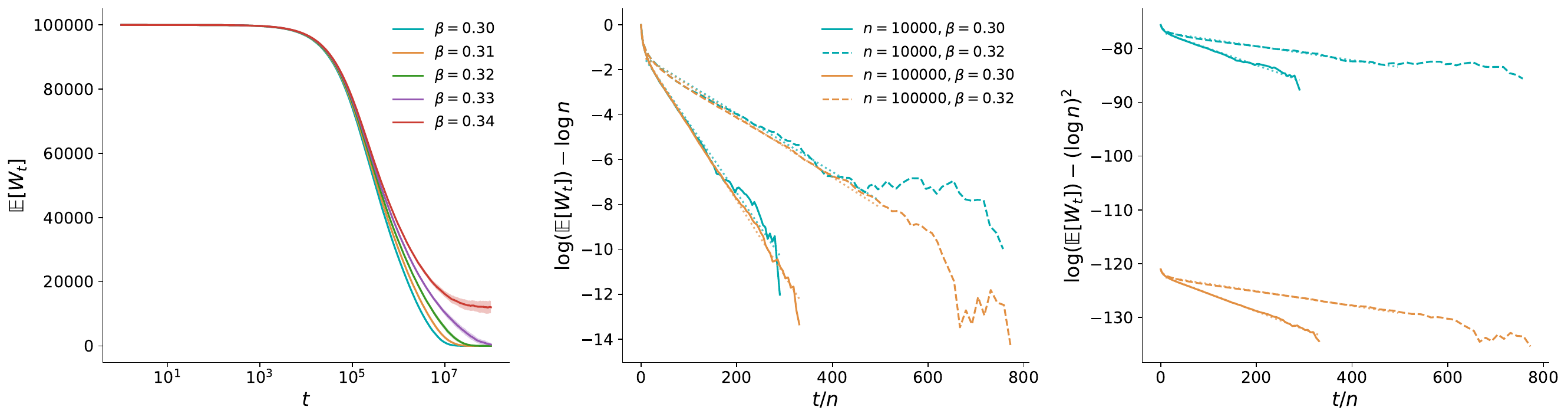}
	\includegraphics[width=1.0\textwidth]{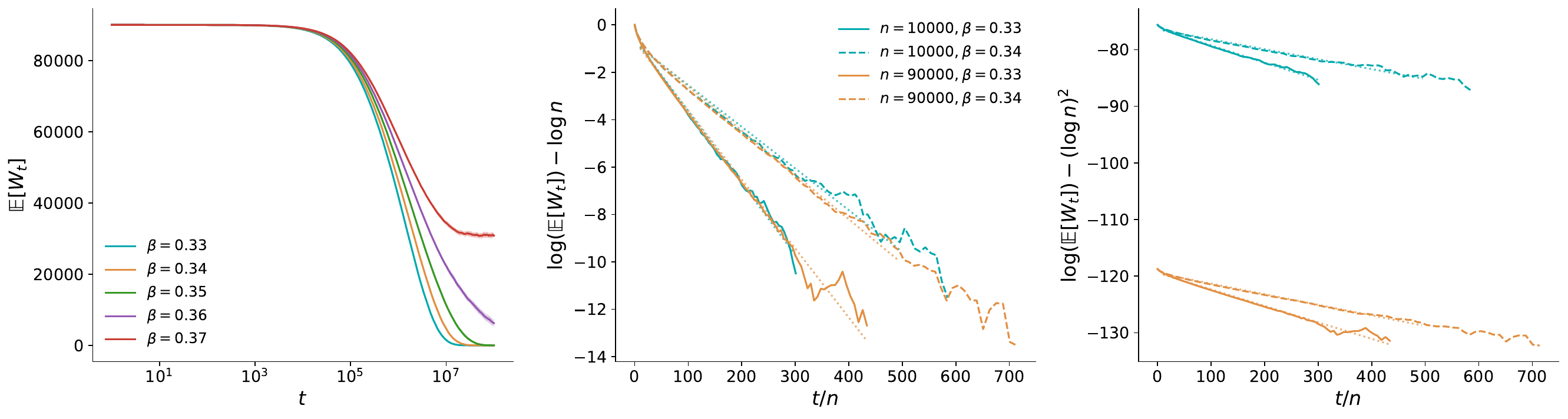}
	\includegraphics[width=1.0\textwidth]{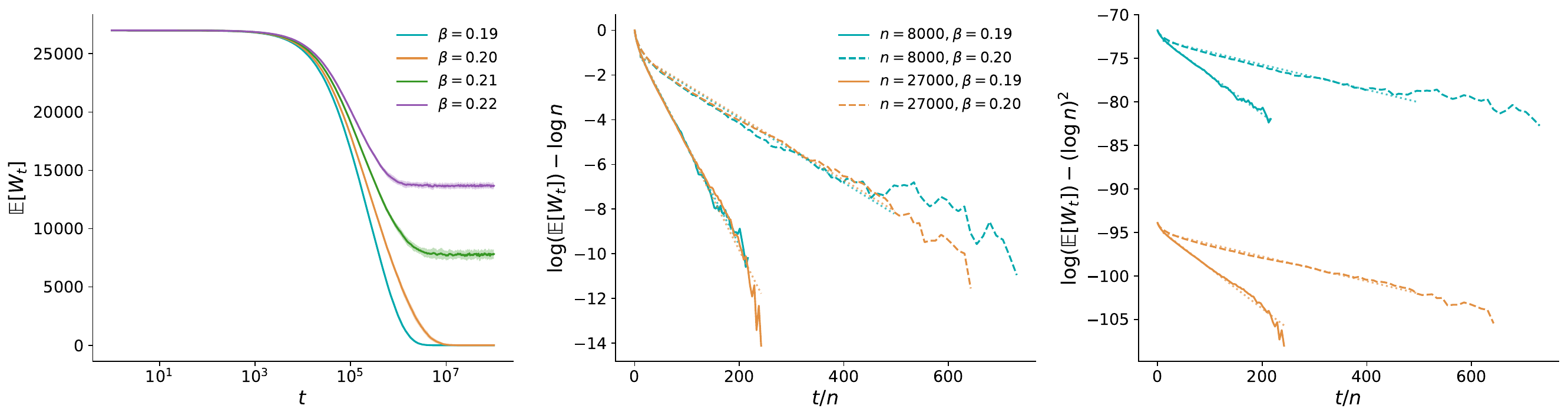}
	\caption{Evolution of the mean number of unresolved sites $\mathbb{E}[W_t]$, averaged over $50$ independent runs, horizon time $T_{\rm hor}=10^8$.
	Left panel: $\mathbb{E}[W_t]$ as a function of the number of single-site updates. 	
	Middle and right panels: linear fits to test the scalings $W_t=Kne^{-\delta t/n}$ and $W_t=Ke^{(\log n)^2-\delta t/n}$. 
	First row: on $\mathcal{G}(n,d/n)$ with $d=4$, size $n=10^5$ and i.i.d. uniform in $\{1,-1\}$ couplings.
	Second row: on $\mathcal{G}(n,d/n)$ with $d=4$, size $n=10^5$ and i.i.d. standard Gaussian couplings. 
	Third row: on periodic $2$-dimensional lattices, size $L=300$ and i.i.d. uniform in $\{1,-1\}$ couplings. 
	Fourth row: on periodic $3$-dimensional lattices, size $L=30$.}
	\label{fig_app:nb_star_evol}
\end{figure}

Additional plots of $W_t=\sum_{u\in V}\mathbbm{1}_{\{Y_t(u)=\star\}}$, the number of unresolved (star $\star$) sites as a function of $t$ (the number of single-site updates), averaged of $n_{\rm runs}=50$ independent runs, are shown in Figure~\ref{fig_app:nb_star_evol}.

\end{document}